\documentclass{aa}  

\usepackage{graphicx}
\usepackage{txfonts}
\usepackage{color}
\usepackage{caption}
\usepackage{subcaption}
\usepackage{tabularx}
\usepackage{array}
\usepackage{graphbox}
\usepackage{adjustbox}
\usepackage{makecell}
\usepackage{float}
\usepackage{multirow}
\newcommand*\rot{\rotatebox{90}}

\begin{document}

\title{Spatially Resolved Properties of Galaxies with a Kinematically Distinct Core}

   \author{Kiyoaki Christopher Omori
          \inst{1}
          \and
          Tsutomu T. Takeuchi\inst{1,2}
          }

   \institute{$^1$Division of Particle and Astrophysical Science, Nagoya University, 
Furo-cho, Chikusa-ku, Nagoya 464--8602, Japan\\
$^2$The Research Centre for Statistical Machine
Learning, the Institute of Statistical Mathematics, 10--3 Midori-cho, Tachikawa, Tokyo 190--8562, Japan
}

   \date{Received November 30, 2020; accepted December 12, 2020}

 
  \abstract
   {}
 {Interacting galaxies show unique irregularities in their kinematic structure. By investigating the spatially resolved kinematics and stellar population properties of galaxies that show irregularities, we can paint a detailed picture of the formation and evolutionary processes that took place during its lifetimes.}
   {In this work, we focus on galaxies with a specific kinematic irregularity, a kinematically distinct stellar core (KDC), in particular, counter-rotating galaxies where the core and main body of the galaxy are rotating in opposite directions. We visually identify eleven MaNGA galaxies with a KDC from their stellar kinematics, and investigate their spatially resolved stellar and gaseous kinematic properties, namely the two-dimensional stellar and gaseous velocity and velocity dispersion ($\sigma$) maps.  Additionally, we examine the stellar population properties, as well as spatially resolved recent star formation histories using the D$_{n}$4000 and H$\delta$ gradients.}
   {The galaxies display multiple off-centred, symmetrical peaks in the stellar $\sigma$ maps. 
   The gaseous velocity and $\sigma$ maps display regular properties.
   The stellar population properties and their respective gradients show differing properties depending on the results of the spatially resolved emission line diagnostics of the galaxies, with some galaxies showing inside-out quenching but others not.
   The star formation histories also largely differ based on the spatially resolved emission line diagnostics, but most galaxies show indications of recent star formation either in their outskirts or core. 
}
{We find a distinct difference in kinematic and stellar population properties in galaxies with a counter-rotating stellar core, depending on its classification using spatially resolved emission line diagnostics.}

    \titlerunning{Spatially Resolved Properties of Galaxies with Counter-Rotating Cores.}
   \maketitle
%

\section{Introduction}
\label{section:Introduction}

Hierarchical growth of galaxies via merging is a commonly accepted pathway of galaxy evolution in the current $\Lambda$ cold dark matter ($\Lambda$CDM) framework for structure formation in the Universe. 
There are small density fluctuations in the early homogeneous, expanding Universe. These perturbations grow over time, and in over-dense regions where the density difference becomes large compared to the under-dense regions, gravitational collapse occurs creating protogalaxies. Gas in these protogalaxies form stars to form the first galaxies, which will further evolve from internal star formation events or external accretion events.
One way galaxies undergo such accretion is through galaxy interaction and mergers.

Galaxy interactions are commonly associated with disturbances in galaxy kinematics. In interacting galaxies, we expect to find complex and disturbed kinematics, such as asymmetries and distortions in both the stellar and gaseous velocity fields \citep{2007MNRAS.376..997J}. By investigating the spatially resolved stellar and gaseous properties of galaxies that show disturbances in their kinematics, we can develop an understanding of the interaction processes that took place during the galaxy's existence and paint a picture of its formation and evolutionary pathways.

One type of disturbance that may form from galaxy interactions, and the main focus of this paper, is kinematic misalignments within a galaxy, where the galaxy has a core with a distinct rotation. Such feature is referred to as a kinematically distinct stellar core (KDC). 
In this work, we focused on a specific type of KDC.  Within KDCs, which have a broad definition of having a core and main body misalignment $>30\deg$ \citep{10.1111/j.1365-2966.2011.18560.x}, we selected galaxies with counter-rotating core, where the core and main body were rotating in opposite directions ($=180 \deg$). 

Such feature is thought to be a relic of an external gas accretion event \citep{1992ApJ...401L..79B}. 
A significant amount of angular momentum is required to change the orientation of the co-rotating gas, and external processes, for example a major merger \citep[e.g. ][]{1992A&A...258..250B, 1991ApJ...370L..65B}, are more likely to provide this angular momentum than internal ones.

Investigating the stellar and gaseous properties of both the core and outer region/main body of galaxies with a counter-rotating core will give us an understanding of their formation and evolutionary pathways, however there have been a number of limitations, mainly related to observational equipment, that have made studies of such properties difficult. Single fiber surveys such as the Sloan Digital Sky Survey (SDSS, \citealt{2000AJ....120.1579Y}) are limited to the central region of the galaxy, and cannot paint a picture of the behaviour of the entire galaxy. Similarly, long-slit surveys also have limitations because while they are able to spatially resolve a target galaxy, it is limited to an elongated region.
Recent advances in integral field spectroscopy have given us access to spatially resolved data which will help us identify and analyze galaxies and their properties. In addition, these advancements have made mapping of galaxies extending to their outer regions possible, allowing for a more thorough understanding of the galaxy.
For example, projects such as SAURON \citep{Bacon_2001} and ATLAS3D \citep{Cappellari_2011} have been successful at spatially resolving local early type galaxies out to 1 effective radius.
This paper uses data from the integral field spectroscopy survey Mapping Nearby Galaxies at Apache Point Observatory (MaNGA, \citealt{2015ApJ...798....7B}) to search for and analyze KDCs. In comparison to the projects above, the MaNGA survey can map out effective information out to 1.5 effective radius for two-thirds of its sample and 2.5 effective radius for one-third of its sample \citep{Yan_2016}, which allows for coverage of the majority of light as well as allowing observations of properties such as gradients at galaxy outskirts and observations of accretion events occurring in the outskirts. MaNGA also has a larger number of galaxies observed over a wider wavelength range compared to the above surveys.
This paper is organised as follows: We briefly describe the MaNGA survey, its instrumentation, and sample selection criteria in Sect. \ref{section:Data}. In Sect. \ref{section:Method}, we describe our data analysis method. We show the physical properties of our sample in Sect. \ref{section:Results}. We discuss our results in Sect. \ref{section:Discussion}.

\section{Data}
\label{section:Data}

MaNGA, an integral field spectroscopic survey, is one of the three core projects of Sloan Digital Sky Survey IV (SDSS-IV, \citealt{2017AJ....154...28B}). 
It uses the 2.5 meters telescope at the Apache Point Observatory \citep{2006AJ....131.2332G}. 
MaNGA aims to map and acquire spatially resolved spectroscopic observations of ~10,000 local galaxies, in a redshift range of 0.01 $<$ \textit{z} $<$ 0.15, at an average redshift of 0.037 \citep{Law_2016} by 2020.
MaNGA spectra cover a wavelength range of 3,600{\AA}\mbox{--}10,000{\AA}, at a resolution of R $\sim$ 2,000. 

The MaNGA target selection is optimised in a way where galaxies are selected based on only their SDSS $i$-band absolute magnitude and redshift, and the sample is unbiased based on their sizes or environments. 
The methodology and extensive efforts taken for this optimization are highlighted in \citet{2017AJ....154...86W}. 
We use data from SDSS Data Release 16 (DR16), which includes 4675 unique MaNGA galaxies.

We have selected our galaxy sample from our DR16 data by visual classification of 2D galaxy kinematic maps provided by MaNGA. We considered galaxies to be counter-rotating if their stellar velocity maps \textbf{a)} had a rotational misalignment between the inner and outer galactic regions and \textbf{b)} the misalignment was large enough such that the inner and outer regions were rotating in opposite directions, as in the core and main body were counter-rotating with respect to each other ($=180 \deg$).  We investigated the stellar velocity maps of 4671 of 4675 MaNGA galaxies, and identified eleven galaxies with a kinematically distinct core in their stellar maps. Of these eleven galaxies, we used the spatially resolved BPT diagrams \citep{1981PASP...93....5B} to identify three galaxies as star-forming (SF), five as AGN-hosting, two non-classified, and one ambiguous. We investigate the spatially resolved kinematics, stellar population properties and star formation histories of these galaxies.
Table \ref{tab:Samples} shows our galaxy sample.
\begin{table*}[htp]
\centering
\begin{tabular}{||c c c c c c||} 
 \hline
 \thead{Plate-IFU \\ \textbf{(a)}} &  \thead{\textit{z} \\ \textbf{(b)}} &  \thead{$M_*$  \\ \textbf{(c)}} & \thead{$R_e$ \\ \textbf{(d)}} & \thead{Age\\ \textbf{(e)}} & \thead{Type \\ \textbf{(f)}}\\ [0.5ex] 
 \hline\hline
  8143-3702 & 0.025 & 6.384x$10^9$ & 4.454 & 0.852 & AGN \\
 \hline
 8155-3702 & 0.023 & 1.111x$10^{10}$ & 3.516 & 0.553 & AGN \\
 \hline
 8606-3702 & 0.024 & 1.596x$10^{10}$ & 8.042 & N/A & AGN \\
 \hline
 8989-9101 & 0.033 & 2.318x$10^{10}$ & 7.961 & 0.814 & AGN \\
 \hline
 8995-3703 & 0.055 & 2.174x$10^{10}$ & 3.178 & 0.976 & AGN \\
 \hline
  8615-1902 & 0.020 & 5.974x$10^9$ & 3.679 & 0.775 & starforming \\
 \hline
 9027-3703 & 0.021 & 2.443x$10^9$ & 4.904 & 0.335 & starforming \\ 
 \hline
  9872-3701 & 0.020 & 5.556x$10^9$ & 3.849 & 0.743 & starforming \\ 
 \hline
  8143-1902 & 0.041 & 7.967x$10^9$ & 1.901 & 0.984 & unclassified \\
 \hline
   8335-1901 & 0.055 & 2.174x$10^{10}$ & 3.178 & 0.976 & unclassified \\
 \hline
  9027-1902 & 0.022 & 3.657x$10^9$ & 2.253 & 0.8958 & ambiguous \\ [1ex] 
 \hline
\end{tabular}
\caption{\label{tab:Samples}List of galaxy sample. Column \textbf{(a)}: MaNGA plate-ifu identification. Column \textbf{(b)}: Redshift Column \textbf{(c)}: Galaxy stellar mass, given in $M_{\odot}$ Column \textbf{(d)}: Elliptical Petrosian 50 percent light radius, given in arcsec. Column \textbf{(e)}: Age, given in Gyr, from \citet{2017MNRAS.466.4731G}. Estimated with 1-sigma errors for a central 3-arcsecond aperture and for an elliptical radius of one effective radius $R_e$. Column \textbf{(f)}: Galaxy classification from spatially resolved BPT diagram. }
\end{table*}

\section{Data analysis}
\label{section:Method}
For this work, we focused on investigating the spatially resolved \textbf{a)} stellar and gaseous kinematics, \textbf{b)} stellar population properties and \textbf{c)} star formation histories of our sample.
By investigating \textbf{a)}, we can check for consistency with previous works that suggest galaxies that have underwent an external accretion event are likely to have misaligned stellar and gaseous kinematics \citep{1992ApJ...401L..79B}. 
We investigate \textbf{b)} to understand the process in which the KDC may have formed.
We investigate \textbf{c)} to check if the galaxy has recently experienced a star formation episode, and if so, where it occurred.
\subsection{Stellar and gas kinematics}

The kinematic maps of both stellar and ionised gas were obtained from the output of the data analysis pipeline (DAP) in MaNGA \citep{Westfall_2019}. Obtaining accurate stellar kinematics requires a maximum S/N. 
The DAP bins adjacent spaxels using the Voronoi binning method of \citet{2003MNRAS.342..345C}, and the spectra of these spaxels are stacked and averaged to meet such minimum, for this case greater than 10. 
The stellar continuum of each binned spectra is fitted using the penalised pixel-fitting (pPXF) method by \citet{2017MNRAS.466..798C} and hierarchically clustered MILES templates (MILES-HC) (MILES stellar library: \citealt{2006MNRAS.371..703S}). 
The stellar kinematic information, being velocity and velocity dispersion, is obtained through this fitting process. Once the fitting is finished, emission line analysis is conducted. Emission line measurements are made in two ways, both from emission line moments and Gaussian emission-models, and provide us with best fit continuum models and fluxes and equivalent widths of emission lines. 
These measurements also give us the information on the velocity and velocity dispersion, $\sigma$.
For this work, we utilise H$\alpha$ (6564~{\AA}) emission lines to investigate gaseous kinematics.

\subsection{Stellar population properties}

To obtain the stellar population properties, such as metallicity and age, we refer to the MaNGA FIREFLY Value Added Catalogue (MaNGA FIREFLY VAC, \citealt{2017MNRAS.466.4731G, 2018MNRAS.477.3954P}), a catalogue providing us with spatially resolved stellar population properties for MaNGA galaxies. The MaNGA FIREFLY VAC uses the full spectral fitting code FIREFLY \citep{2017MNRAS.472.4297W} to obtain the stellar population properties. 
FIREFLY is a $\chi^2$ minimization fitting code that best fits an spectral energy distribution (SED) with a linear combination of simple stellar populations (SSPs).
The best-fit combination will provide us with light and mass-weighted ages \textit{log (Gyr)} and metallicities $[Z/H]$. This code is applied onto the spatially resolved spectra of MaNGA galaxies to give us spatially resolved properties and gradients. Details on how the ages and metallicities are obtained from spectra is written in \citet{2017MNRAS.466.4731G}.
FIREFLY is applied to spectra binned using the Voronoi method with a minimum signal-to-noise ratio of 10 per pixel. The on-sky position of each bin (relative to the galactic centre) is used to obtain the effective radius of each bin, and gradients are measured using least-squares linear regression for data points out to $1.5R_e$.

\subsection{Star formation histories}

To investigate the star formation history of the galaxies with a counter-rotating core, we use the D$_{n}$4000 versus H$\delta_{A}$ diagnostic diagram developed by \citet{Kauffmann_2003}. 
This diagnostic is based on two spectral indices, the 4000~{\AA}~break and the strength of the H$\delta$ absorption line. 
The combination of these two indices can tell us about the recent star formation history of galaxies. 
The two indices are inversely correlated. 
Galaxies with recent star formation activity have a weaker D$_n$4000 value and deeper H$\delta_{A}$ absorption index. 

The 4000~{\AA}~break is seen due to the accumulation of multiple spectral lines in a narrow wavelength region. 
This is due to two reasons, one being the lack of hot blue stars, and the other due to metals in the stellar atmosphere absorbing high energy radiation.  
Old stellar populations have a large 4000~{\AA}~break, and young stellar populations a small one. 
The D$_{n}$4000 index is a parameter defined by \citet{1983ApJ...273..105B} as the ratio of the average flux density $F_{\nu}$ in the bands 3750--3950 and 4050--4250~{\AA}.   

Strong H$\delta$ absorption lines are an indicator of recently ceased starburst activity, with no star formation in the last 0.1--1~Gyr \citep{10.1093/pasj/55.4.771}. 
The stellar population is dominated by A stars, which have a stronger H$\delta$ absorption than O and B stars. 
The H$\delta_{A}$ index is defined as the equivalent width of the H$\delta$ absorption feature in the bandpass 4083--4122~{\AA}. 

These indices are available in spatially resolved form for MaNGA galaxies in the MaNGA FIREFLY VAC.
For the D$_{n}$4000 index, MaNGA FIREFLY VAC uses a narrower continuum band than defined by \citet{1983ApJ...273..105B}, 3850--3950 and 4000--4100~{\AA}, introduced by \citet{1999ApJ...527...54B} to calculate the index.
Thus we followed their method and adopted that for this work.
For H$\delta_{A}$, MaNGA FIREFLY VAC calculates this index using Equation (2) of \citet{1994ApJS...94..687W}
\begin{align}
        EW = \int_{\lambda_1}^{\lambda_2} \left(1-\frac{F_{I\lambda}}{F_{C\lambda}}\right)
\end{align}
where $F_{I\lambda}$ is the flux per unit wavelength in the index passband, $F_{C\lambda}$ the straight-line continuum flux in the index passband, and $\lambda_1$ and $\lambda_2$ the wavelength range. 

The D$_{n}$4000--H$\delta_{A}$ plane introduced by \citet{Kauffmann_2003} is a very powerful diagnostic to estimate star formation activity in a galaxy over the past 1--2 Gyr. 
In addition, the two indices are largely insensitive to dust attenuation, which can increase the chance of inaccuracies. 
Whereas \cite{Kauffmann_2003} uses this plane to identify whether or not a galaxy as a whole has had recent star formation, we use this plane on the spatially resolved indices of each galaxy in our sample to examine if and where the galaxy has experienced recent star forming activity.

\section{Results}
\label{section:Results}

In the following sections we discuss our findings in the spatially resolved stellar and gaseous properties, and star formation histories. 

\renewcommand{\thefigure}{1}
\begin{figure*}[htp]
    \centering
    \begin{tabular} {m{0cm} m{5.1cm} m{0cm} m{5.1cm} m{0cm} m{5.1cm}}
     \rotatebox[origin=c]{90}{8155-3702 (AGN)}  & 
       \includegraphics[width=0.3\textwidth]{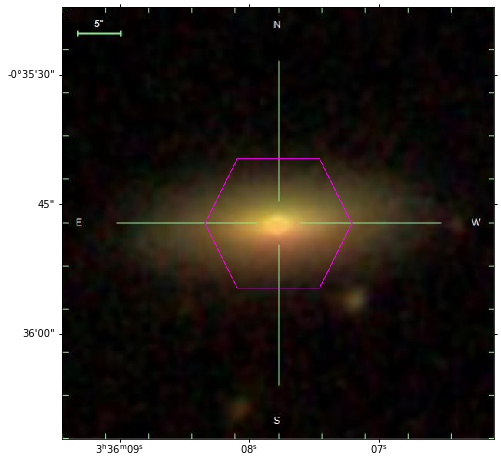} &
     \rotatebox[origin=c]{90}{8143-3702 (AGN)}  &  
       \includegraphics[width=0.3\textwidth]{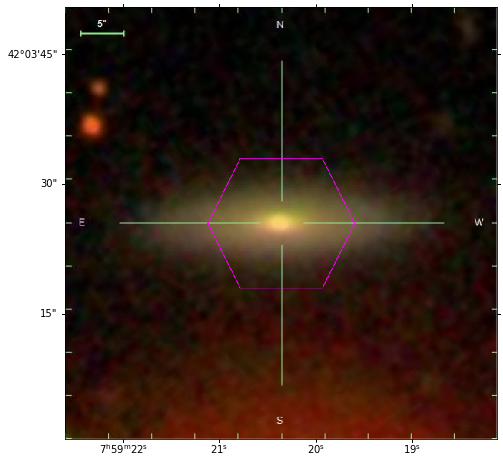} &
     \rotatebox[origin=c]{90}{8606-3702 (AGN)}  &  
       \includegraphics[width=0.3\textwidth]{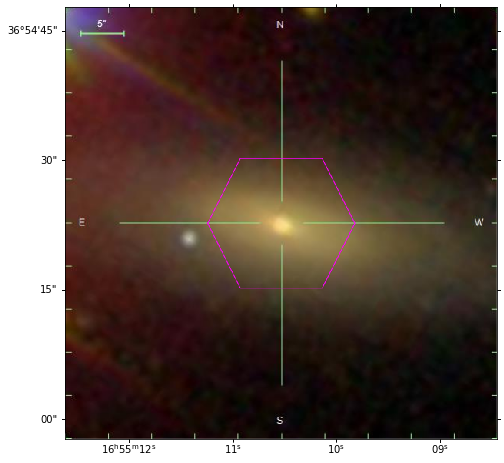} \\
    \rotatebox[origin=c]{90}{8995-3703 (AGN)}  &  
       \includegraphics[width=0.3\textwidth]{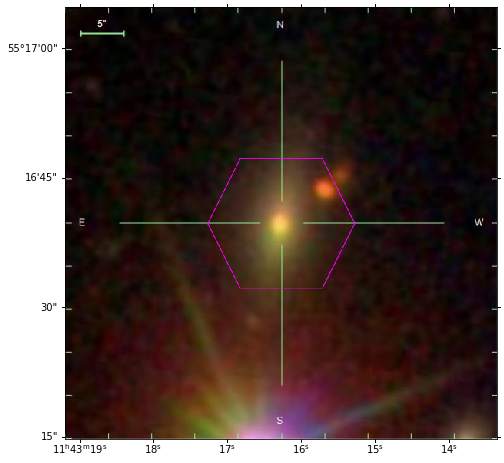} &
    \rotatebox[origin=c]{90}{8989-9101 (AGN)}  &  
       \includegraphics[width=0.3\textwidth]{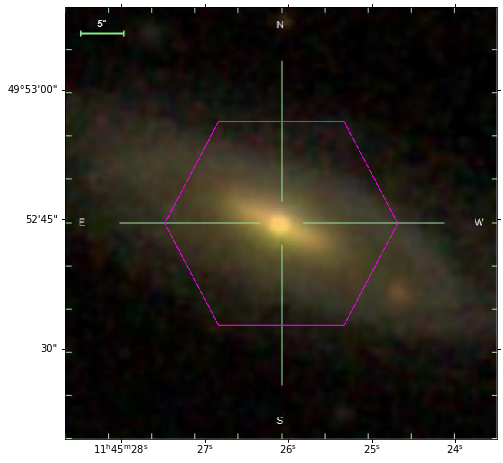} &
      \rotatebox[origin=c]{90}{8615-1902 (SF)}  &  
       \includegraphics[width=0.3\textwidth]{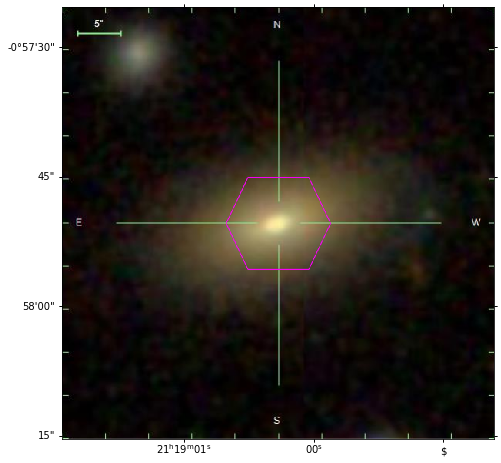} \\
       \rotatebox[origin=c]{90}{9027-3703 (SF)}  &  
       \includegraphics[width=0.3\textwidth]{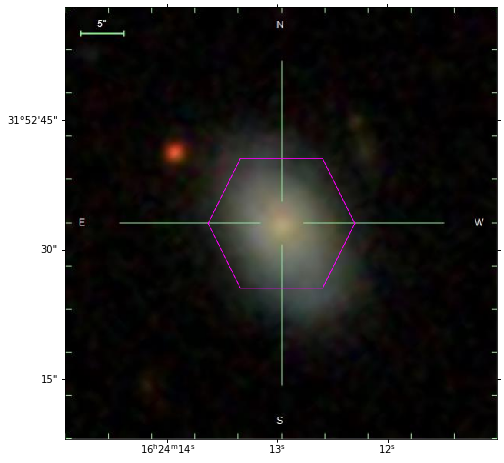} &
        \rotatebox[origin=c]{90}{9872-3701 (SF)}  &  
       \includegraphics[width=0.3\textwidth]{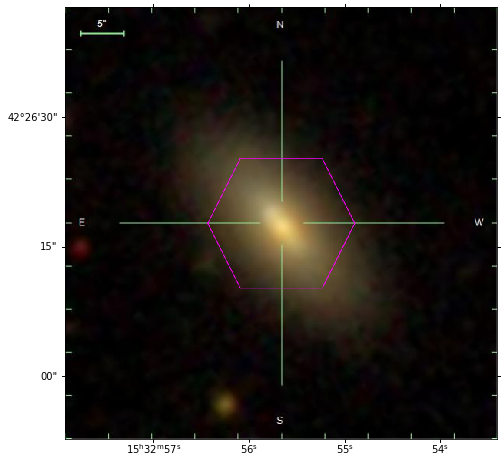} &
     \rotatebox[origin=c]{90}{8143-1902}  &  
       \includegraphics[width=0.3\textwidth]{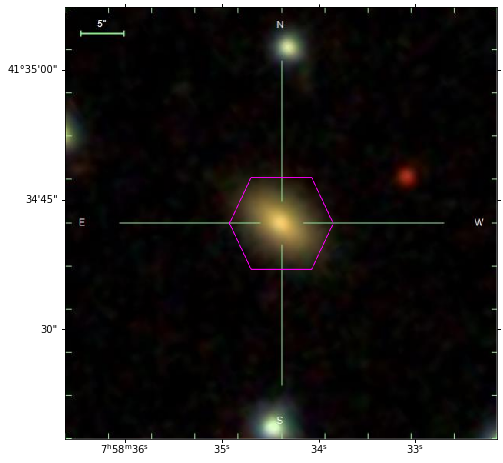} \\
     \rotatebox[origin=c]{90}{8335-1901}  &  
       \includegraphics[width=0.3\textwidth]{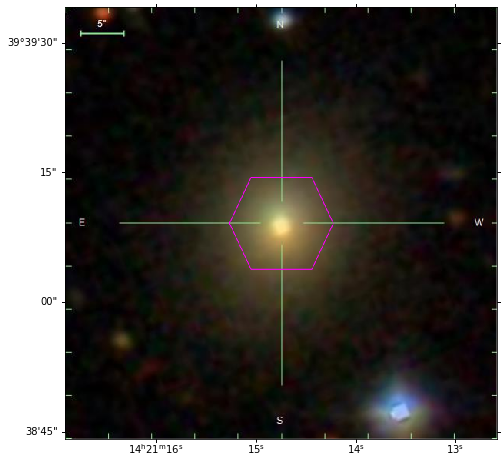} &
         \rotatebox[origin=c]{90}{9027-1902 (amb)}  &  
       \includegraphics[width=0.3\textwidth]{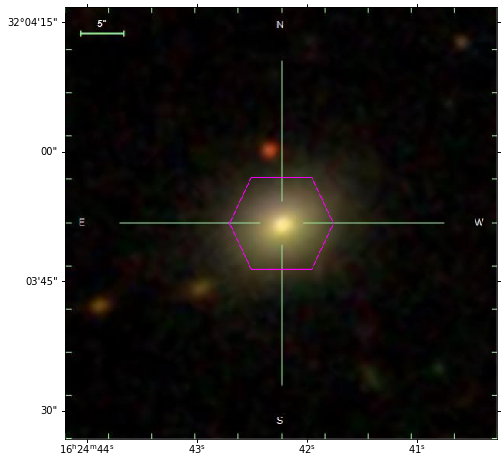} \\
       
    \end{tabular}
    \caption{SDSS optical images of the galaxies sampled in this work. The magenta hexagons indicate the surveyed area of the galaxy.}
    \label{fig:galimg}
\end{figure*}

\renewcommand{\thefigure}{2}
\begin{figure*}[htp]
    \centering
    \begin{tabular}{m{0cm} m{22cm}}

        \rotatebox[origin=c]{90}{8143-3702 (AGN)}  & 
       \includegraphics[width=0.9\textwidth]{8143-3702/kinematics.png} \\
        \rotatebox[origin=c]{90}{8155-3702 (AGN)}  &  
       \includegraphics[width=0.9\textwidth]{8155-3702/kinematics.png} \\
        \rotatebox[origin=c]{90}{8606-3702 (AGN)}  &  
       \includegraphics[width=0.9\textwidth]{8606-3702/kinematics.png} \\
    \rotatebox[origin=c]{90}{8995-3703 (AGN)}  &  
       \includegraphics[width=0.9\textwidth]{8995-3703/kinematics.png} \\
        \rotatebox[origin=c]{90}{8989-9101 (AGN)}  &  
       \includegraphics[width=0.9\textwidth]{8989-9101/kinematics.png} \\
        \rotatebox[origin=c]{90}{8615-1902 (SF)}  &  
       \includegraphics[width=0.9\textwidth]{8615-1902/kinematics.png} \\
     \end{tabular}
    \caption{}
\end{figure*}
\begin{figure*}[!h]
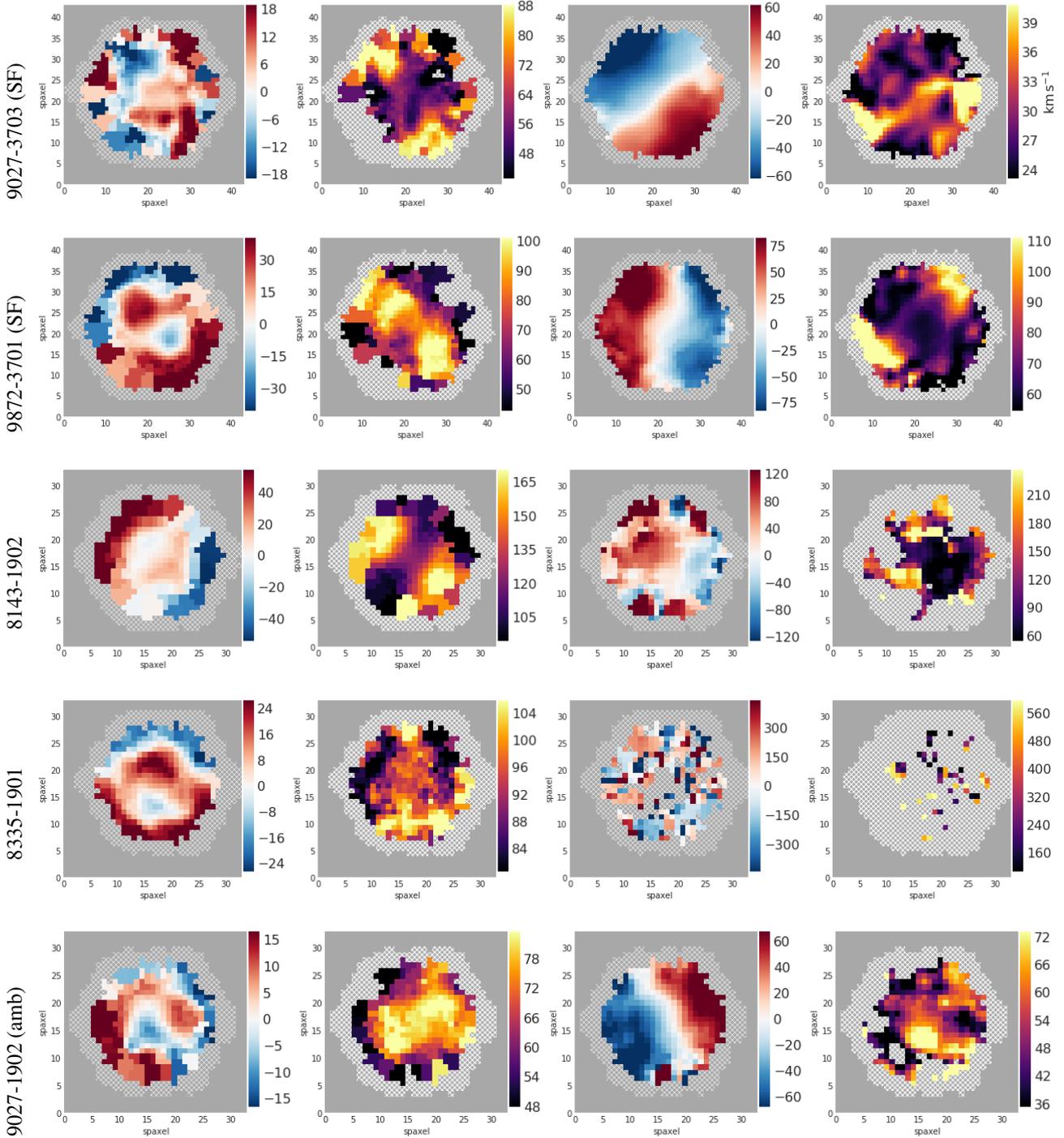

    \ContinuedFloat
    \centering
    \begin{tabular}{m{0cm} m{22cm}} 
        \rotatebox[]{90}{9027-3703 (SF)}  &  
       \includegraphics[width=0.9\textwidth]{9027-3703/kinematics.png} \\
        \rotatebox[origin=c]{90}{9872-3701 (SF)}  &  
       \includegraphics[width=0.9\textwidth]{9872-3701/kinematics.png} \\
     \rotatebox[origin=c]{90}{8143-1902}  &  
       \includegraphics[width=0.9\textwidth]{8143-1902/kinematics.png} \\
     \rotatebox[origin=c]{90}{8335-1901}  &  
       \includegraphics[width=0.9\textwidth]{8335-1901/kinematics.png} \\
         \rotatebox[origin=c]{90}{9027-1902 (amb)}  &  
       \includegraphics[width=0.9\textwidth]{9027-1902/kinematics.png} \\
       
    \end{tabular}
    \caption{Kinematic maps for our galaxy sample. The first column shows the MaNGA plate-ifu identification, the second the line-of-sight stellar velocity (km s$^{-1}$), third the line-of-sight stellar velocity dispersion (km s$^{-1}$), the fourth the line-of-sight H$\alpha$ velocity (km s$^{-1}$), and the fifth the line-of-sight H$\alpha$ velocity dispersion (km s$^{-1}$). In the maps, 1 spaxel $=$ 0.5 arcsec.}
    \label{fig:galkin}
\end{figure*}

\subsection{Stellar and gaseous properties}

    Figure \ref{fig:galimg} shows the SDSS optical images, and Fig. \ref{fig:galkin} shows the kinematic maps of our entire galaxy sample.  
    From left to right on Fig. \ref{fig:galkin}, the columns show the MaNGA plate-IFU identification, the stellar velocity, stellar $\sigma$, H$\alpha$ velocity, and H$\alpha$ $\sigma$ maps. 
    We find that most of our samples have a decoupling between the stellar and gaseous velocity maps. 
    Whereas the stellar velocity maps show the feature of a KDC, such feature is absent in most of the gaseous velocity maps.  
    
    We also investigated the rotational direction of the stars and gas, and noted the differences. 
    The three star-forming galaxies and one of the AGN-hosting galaxies showed that the rotational direction of gas was opposite to that of the main stellar body i.e., the gas co-rotates with respect to the KDC of these galaxies. 
    The four remaining AGNs showed that the rotational direction of the gas was the same as that of the main stellar body, i.e. the gas counter-rotates with respect to the KDC of these galaxies. 
    Of the two unclassified galaxies, the gas is co-rotating with the main body in 8143\mbox{--}1902. 8335\mbox{--}1901 does not have sufficient measurements to make any statement.
    Ambiguous galaxy 9027\mbox{-}1902 has gas co-rotating with the KDC.
    
    Looking at the velocity dispersion maps, we find that many of the galaxies show two distinct peaks in the stellar $\sigma$ maps, which decoupled with the H$\alpha$ $\sigma$ maps. These two peaks have previously been referred to as "2-$\sigma$ galaxies" \citep{2011MNRAS.414.2923K}, with the two peaks in the stellar velocity dispersion maps lying off-centred symmetrically along the major axis of the galaxy.
    The H$\alpha$ $\sigma$ maps show varying behaviour, with the star-forming galaxies showing peaks close to perpendicular to the stellar $\sigma$, and the AGN galaxies showing a central peak.
    The two unclassified galaxies lack sufficient data to make any statement about their H$\alpha$ $\sigma$ maps. The ambiguous galaxy has a slightly off-centred peak.

\subsection{Stellar population properties}
\renewcommand{\thefigure}{3}
\begin{figure*}[htp]
    \centering
    \begin{tabularx}{1.2\textwidth}{X X X X}
\hspace{-1.5cm}\includegraphics[width=0.28\textwidth, keepaspectratio]{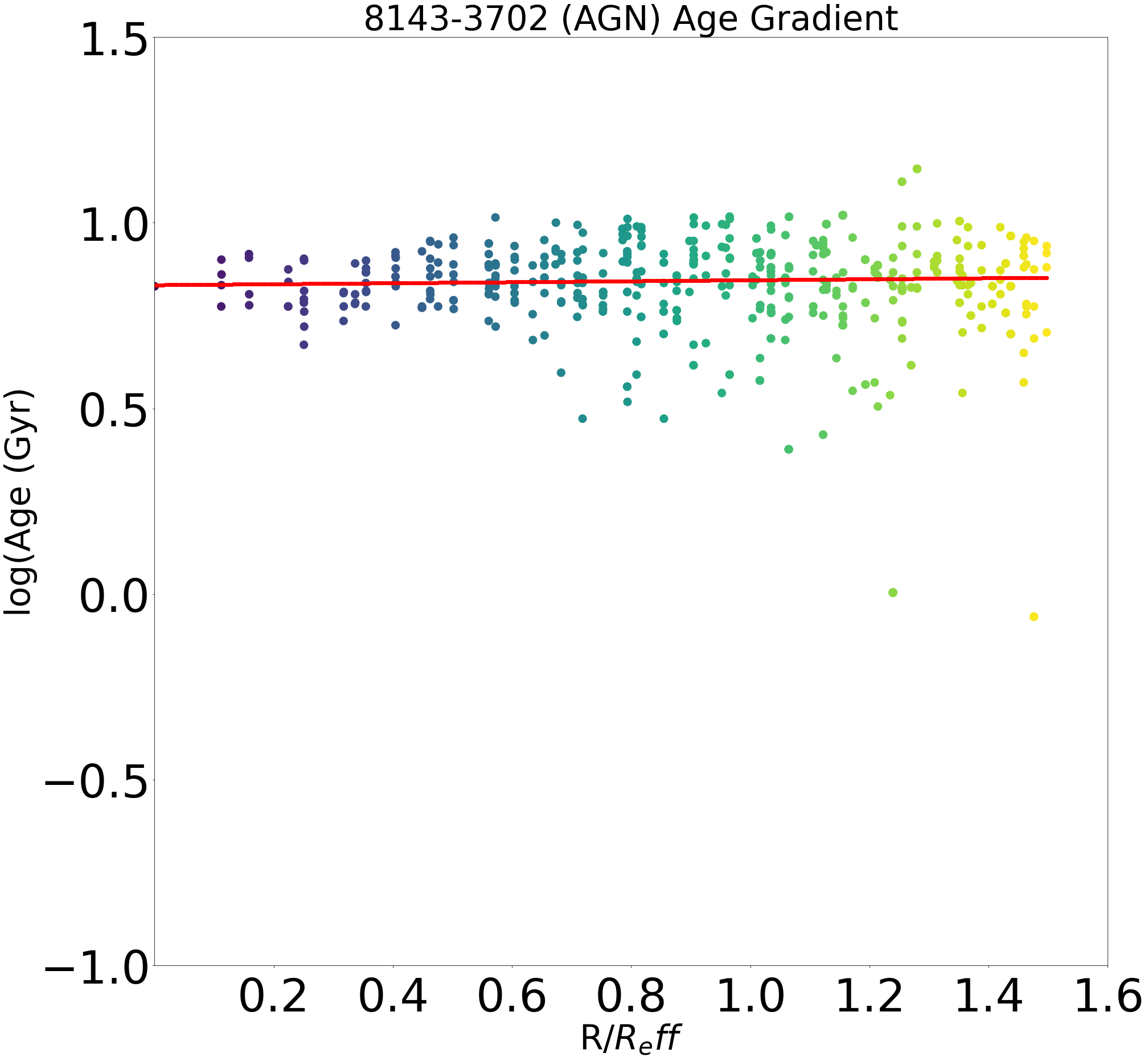}  &
\hspace{-2.0cm}\includegraphics[width=0.28\textwidth, keepaspectratio]{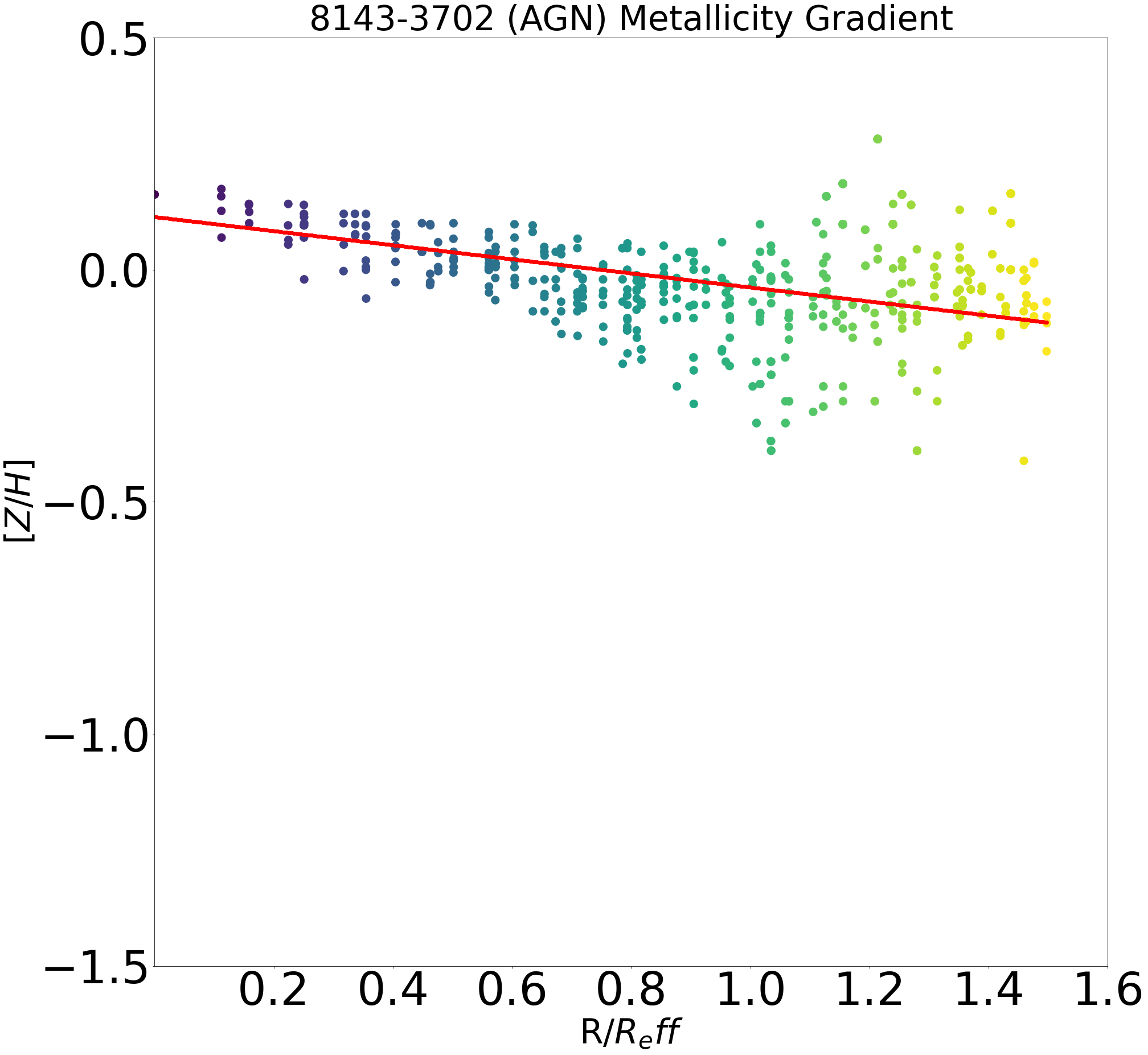} &
\hspace{-2.0cm}\includegraphics[width=0.28\textwidth, keepaspectratio]{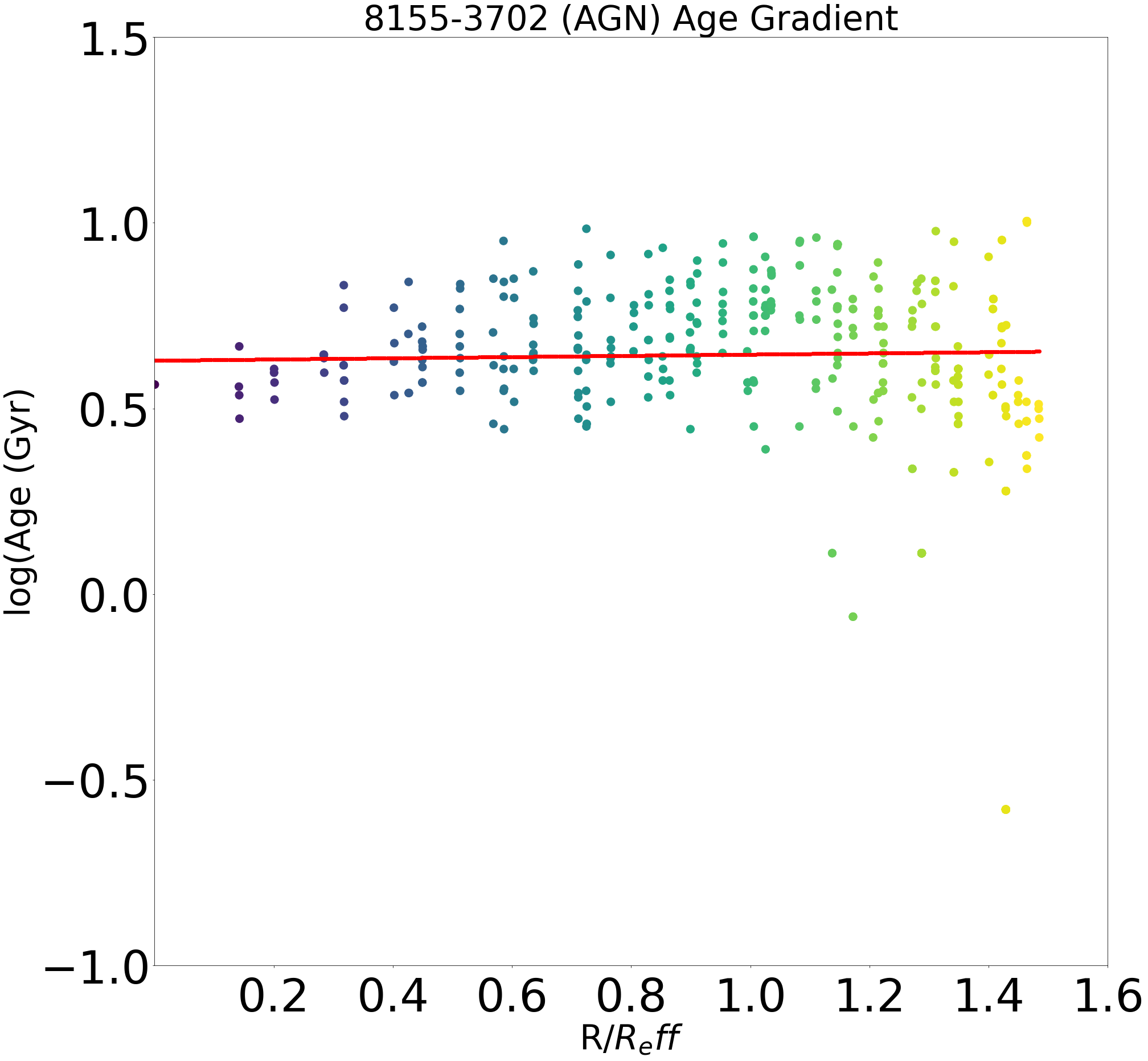} & 
\hspace{-2.3cm}\includegraphics[width=0.28\textwidth, keepaspectratio]{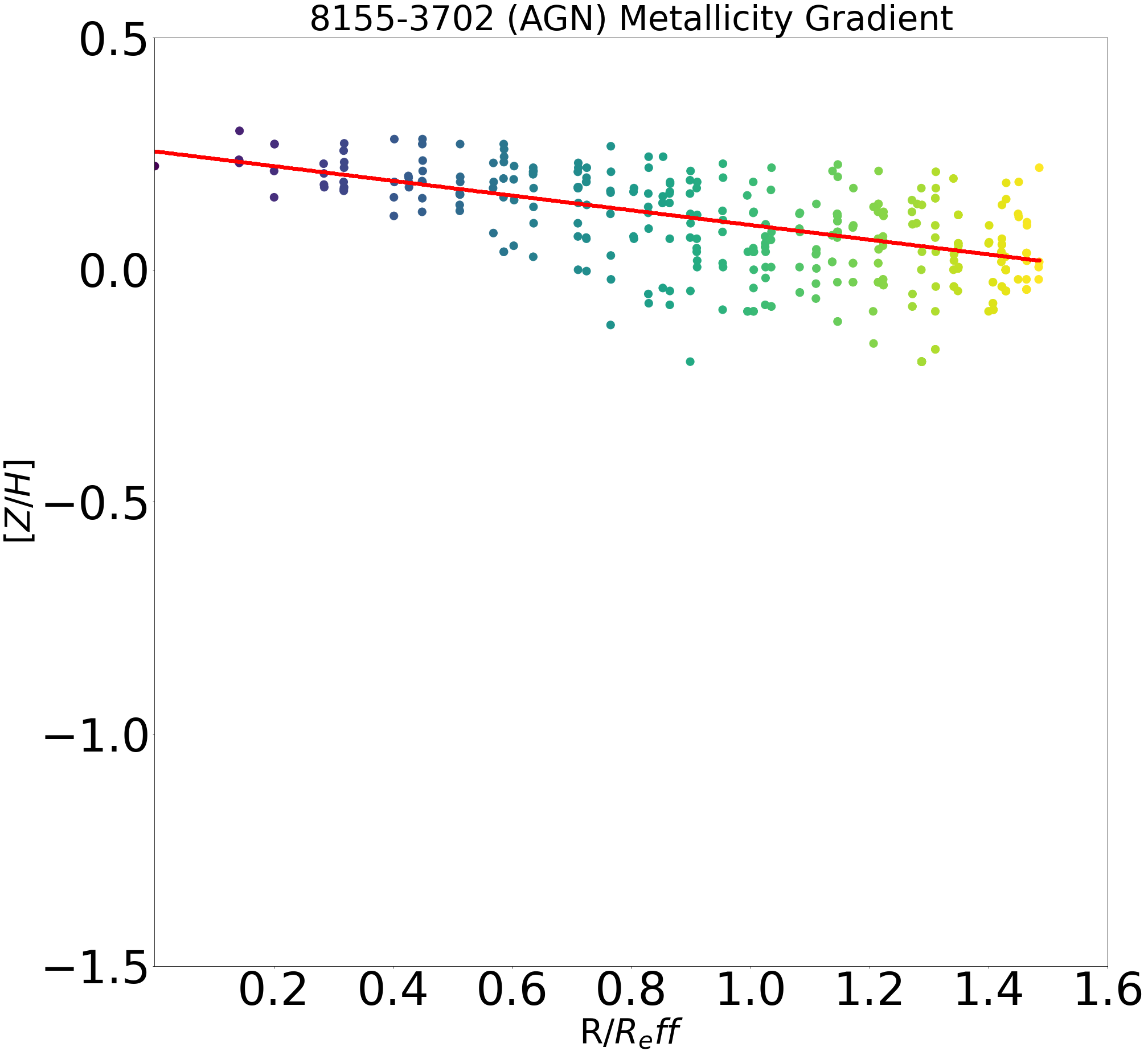}\\
\hspace{-1.5cm}\includegraphics[width=0.28\textwidth, keepaspectratio]{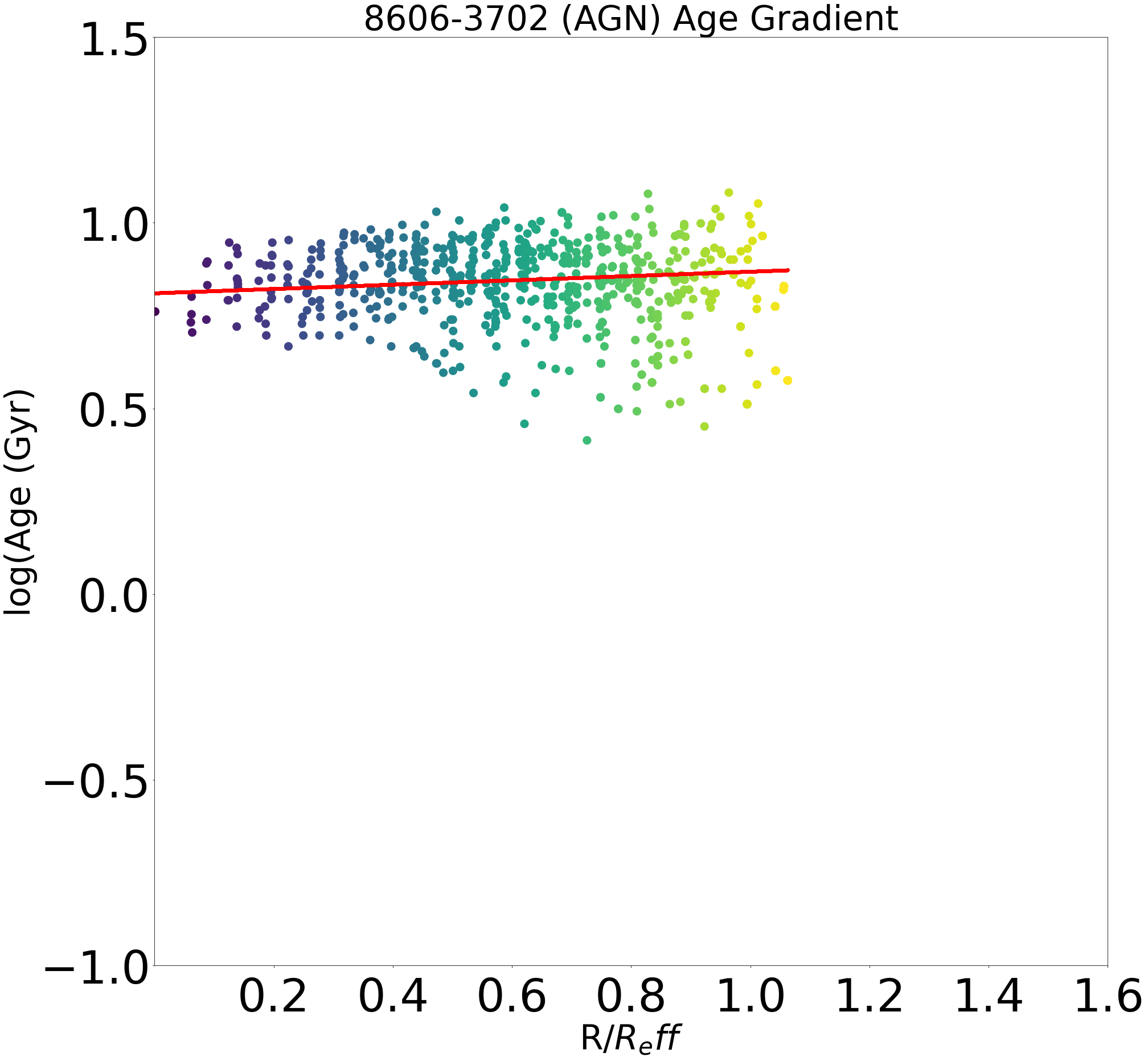}  &
\hspace{-2.0cm}\includegraphics[width=0.28\textwidth, keepaspectratio]{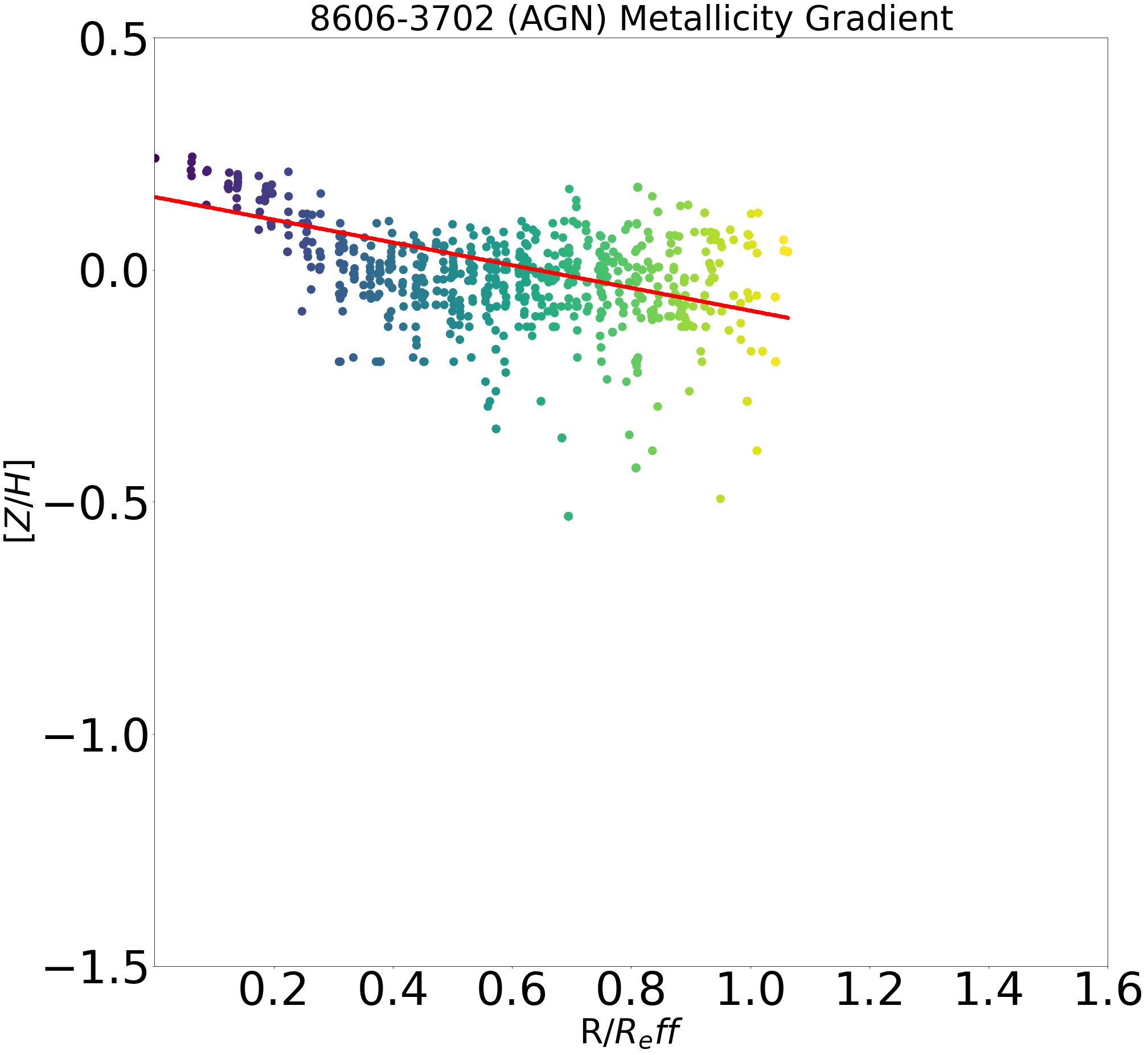} &
\hspace{-2.0cm}\includegraphics[width=0.28\textwidth, keepaspectratio]{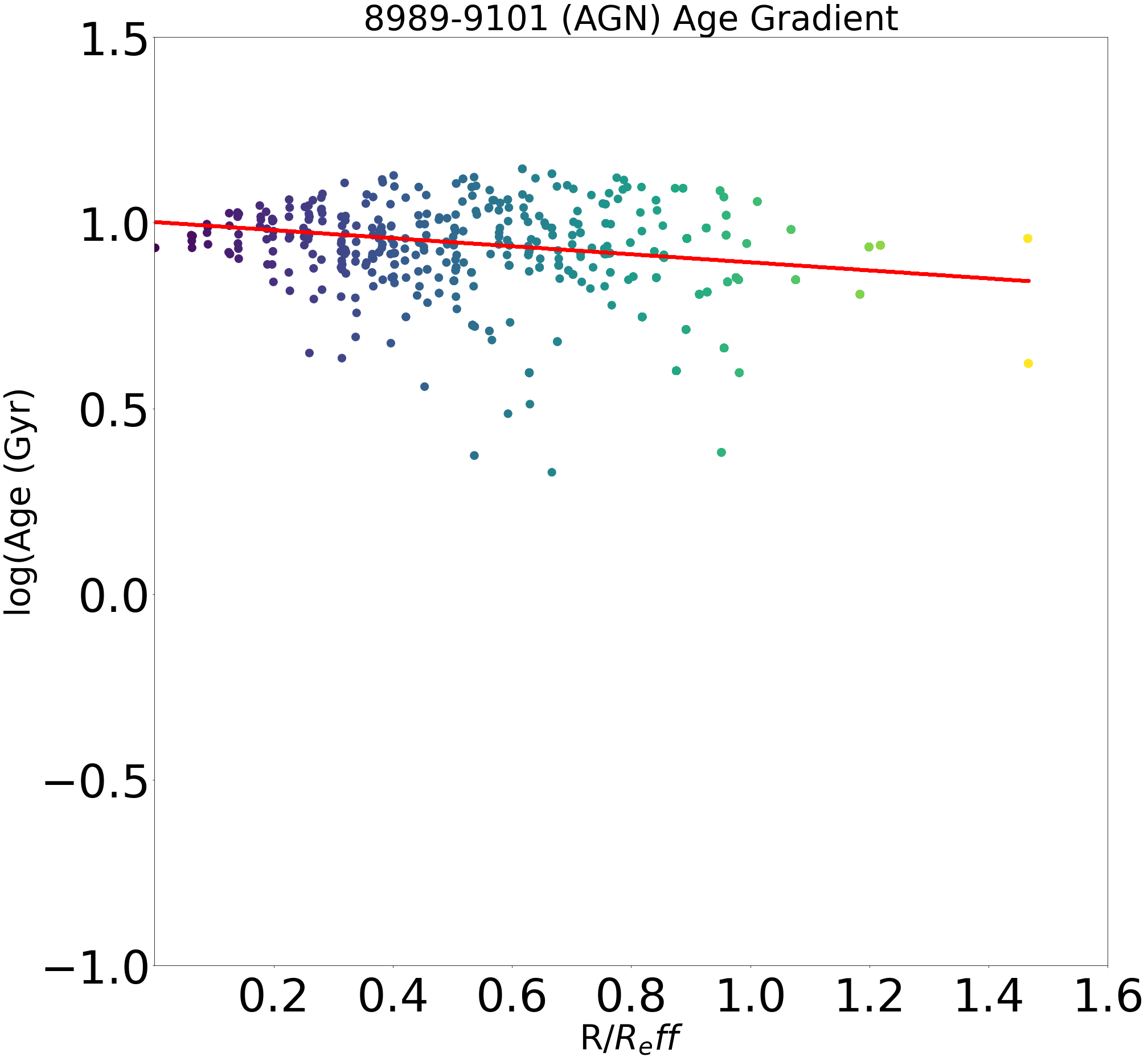} & 
\hspace{-2.3cm}\includegraphics[width=0.28\textwidth, keepaspectratio]{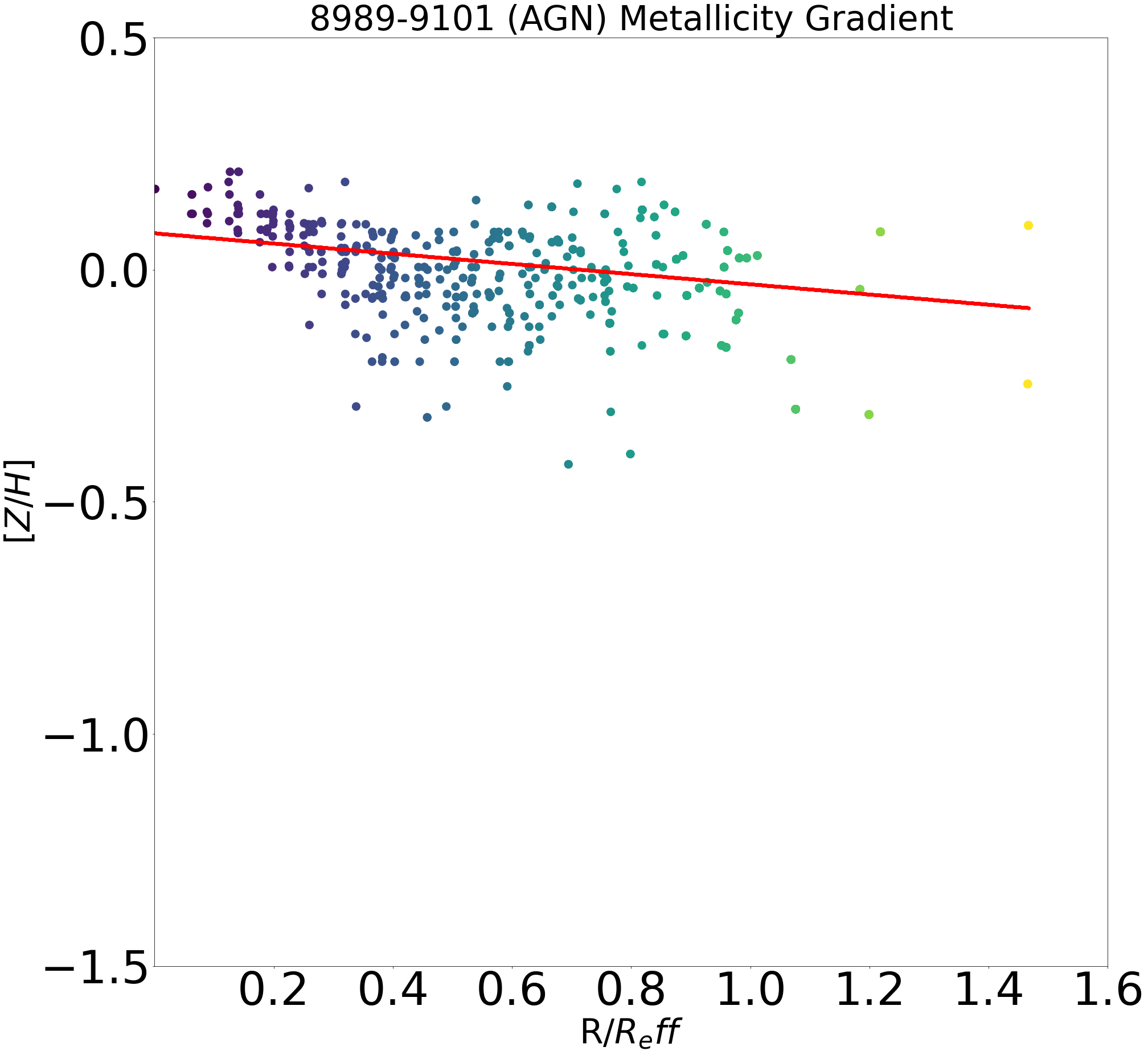}\\
\hspace{-1.5cm}\includegraphics[width=0.28\textwidth, keepaspectratio]{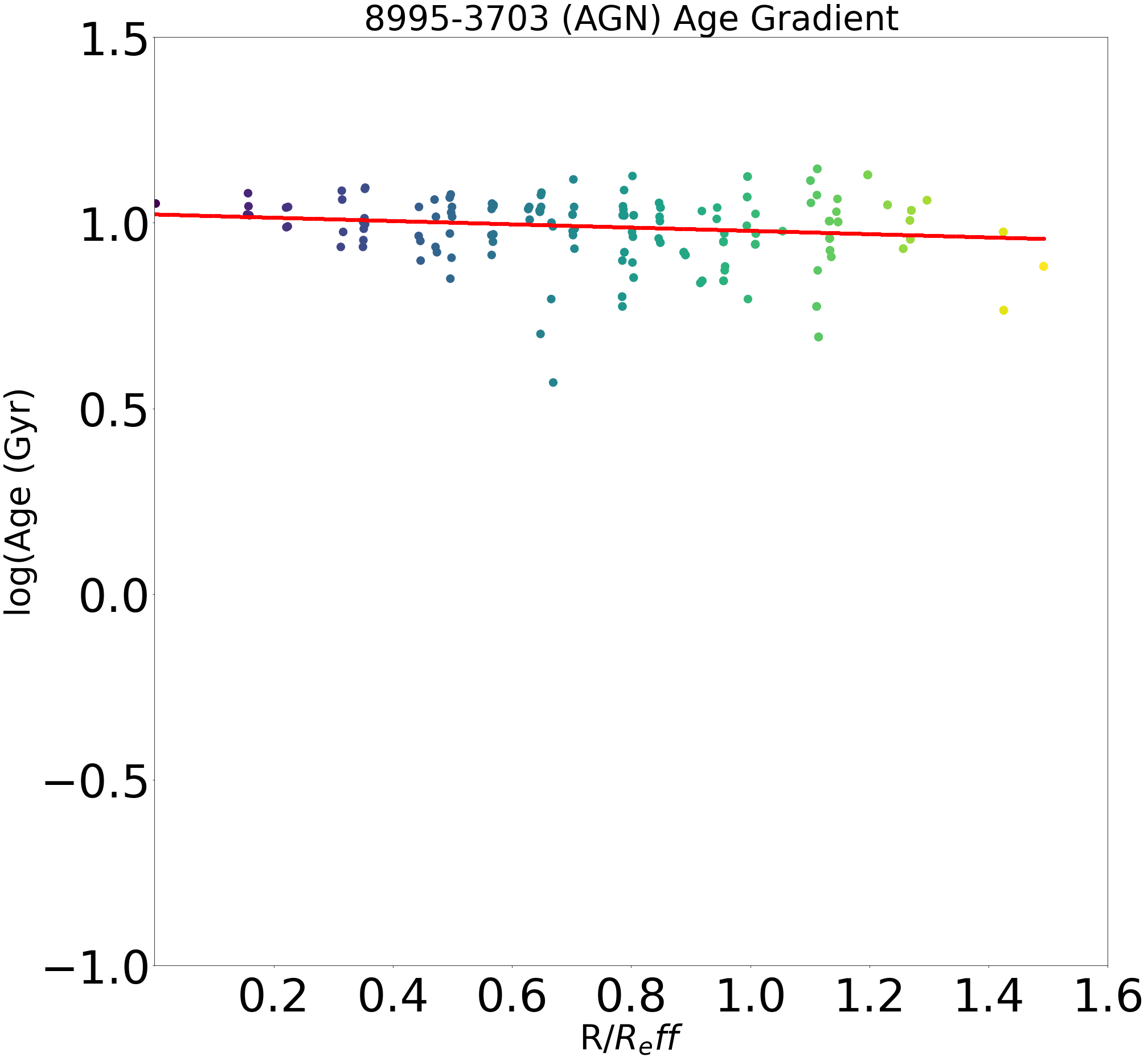}  &
\hspace{-2.0cm}\includegraphics[width=0.28\textwidth, keepaspectratio]{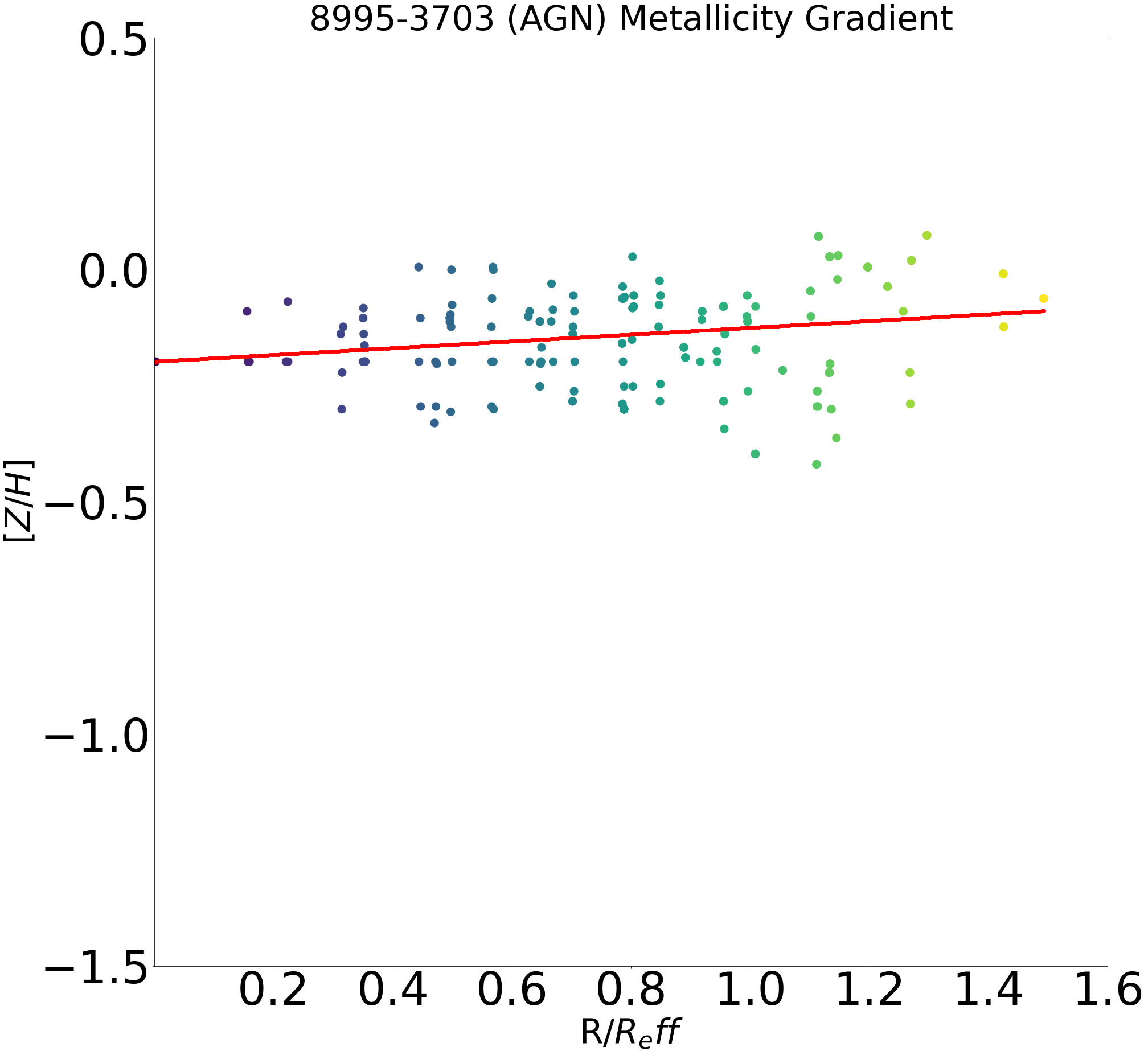} &
\hspace{-2.0cm}\includegraphics[width=0.28\textwidth, keepaspectratio]{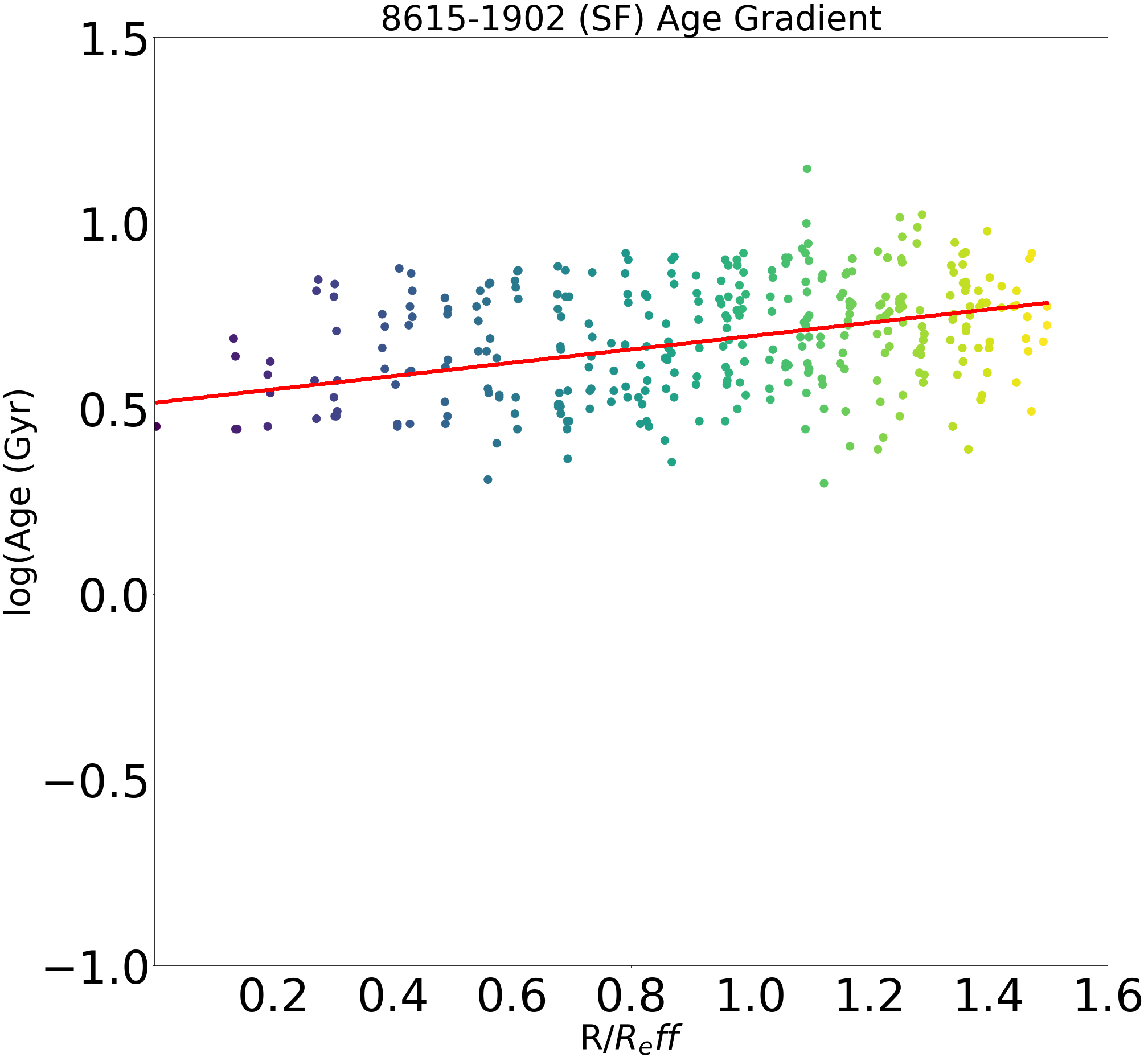} & 
\hspace{-2.3cm}\includegraphics[width=0.28\textwidth, keepaspectratio]{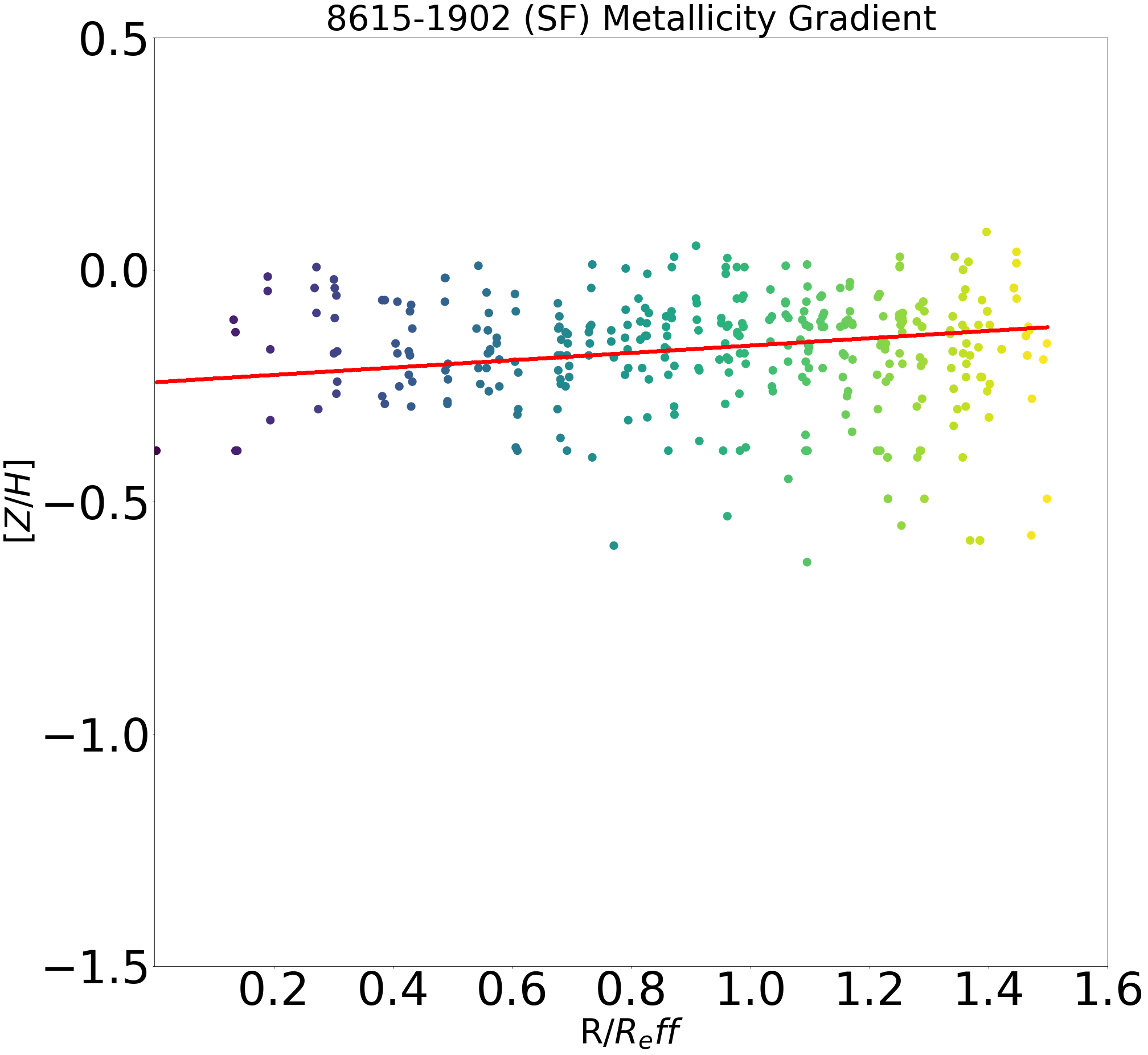}\\
\hspace{-1.5cm}\includegraphics[width=0.28\textwidth, keepaspectratio]{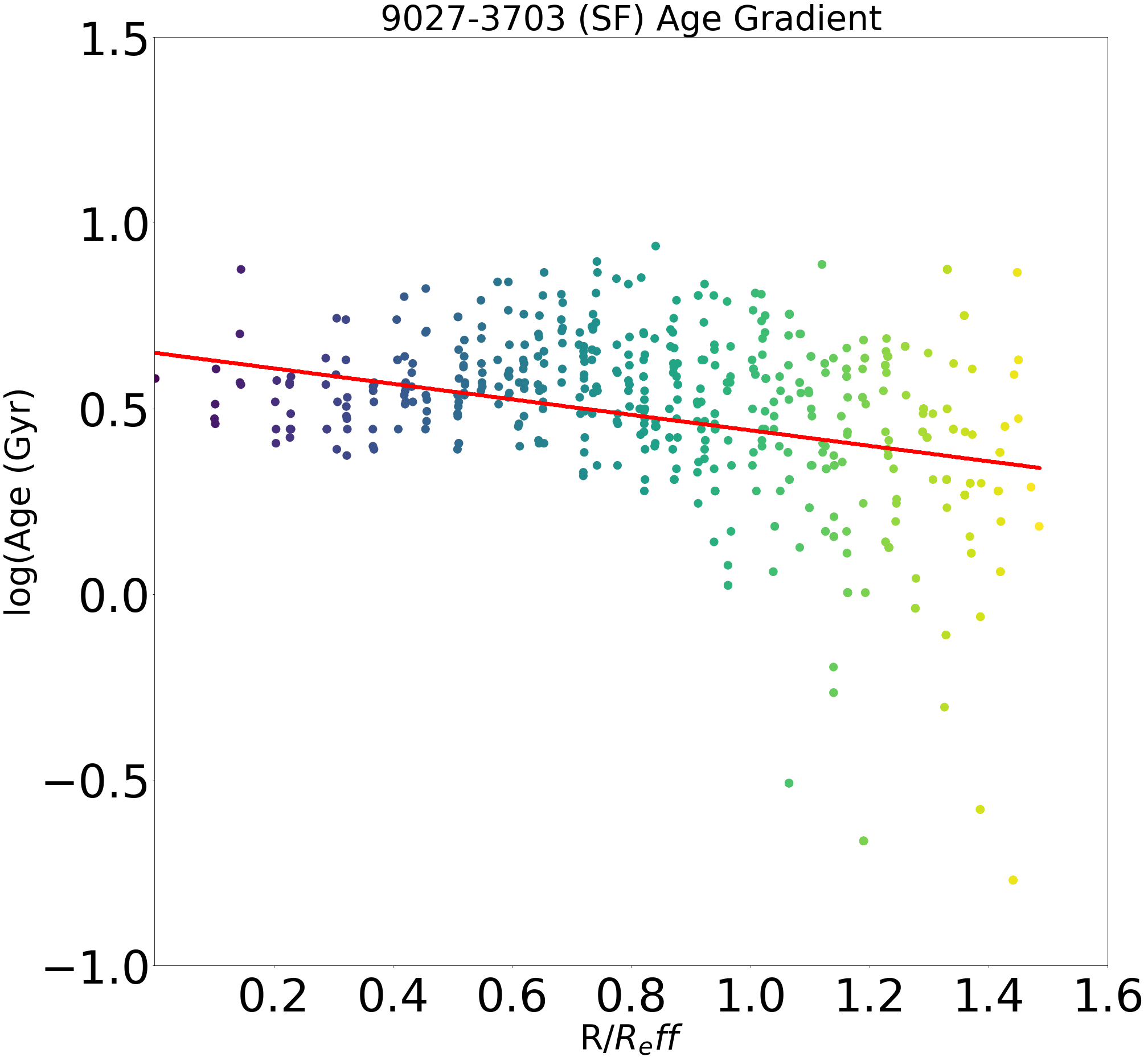}  &
\hspace{-2.0cm}\includegraphics[width=0.28\textwidth, keepaspectratio]{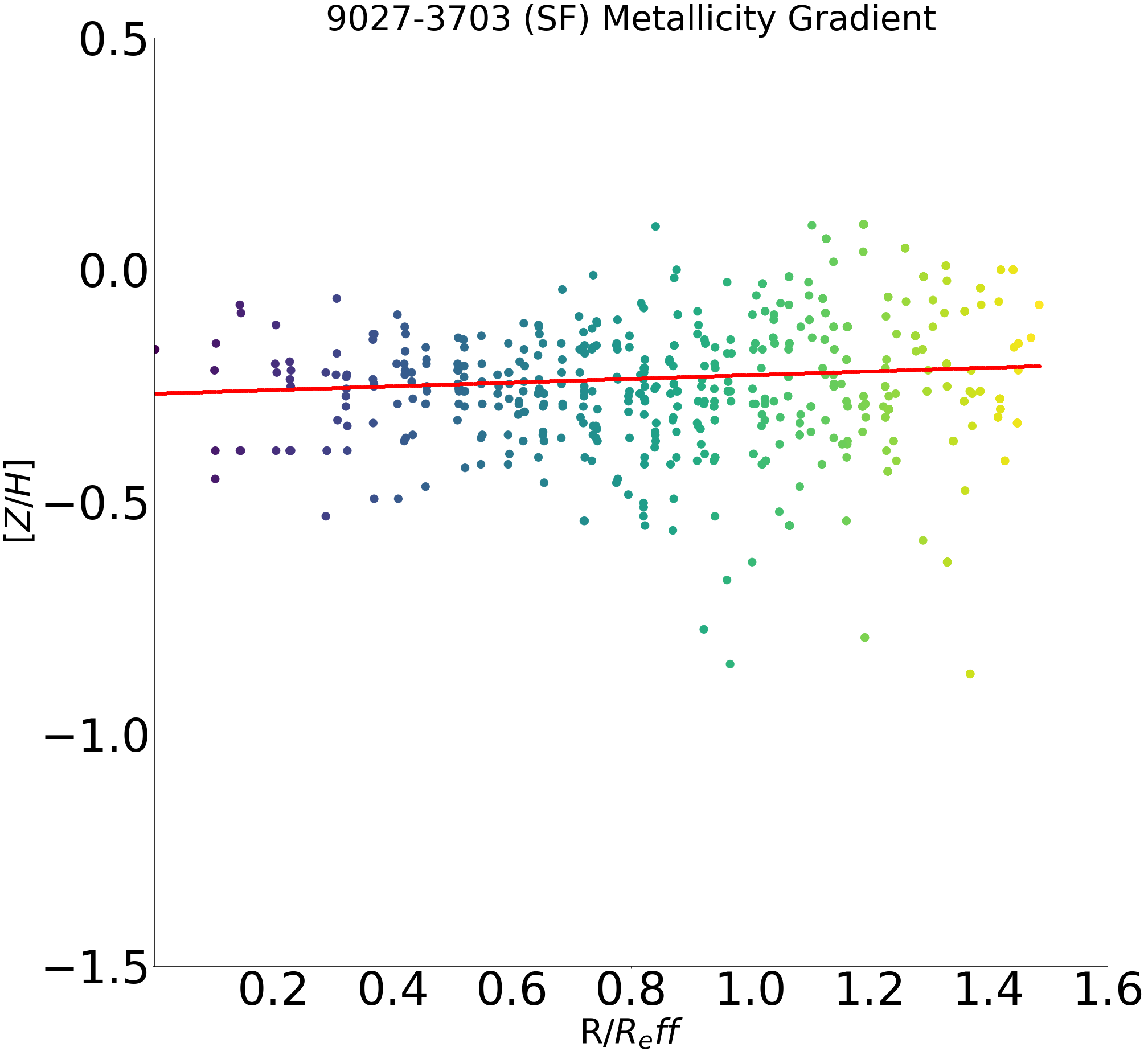} &
\hspace{-2.0cm}\includegraphics[width=0.28\textwidth, keepaspectratio]{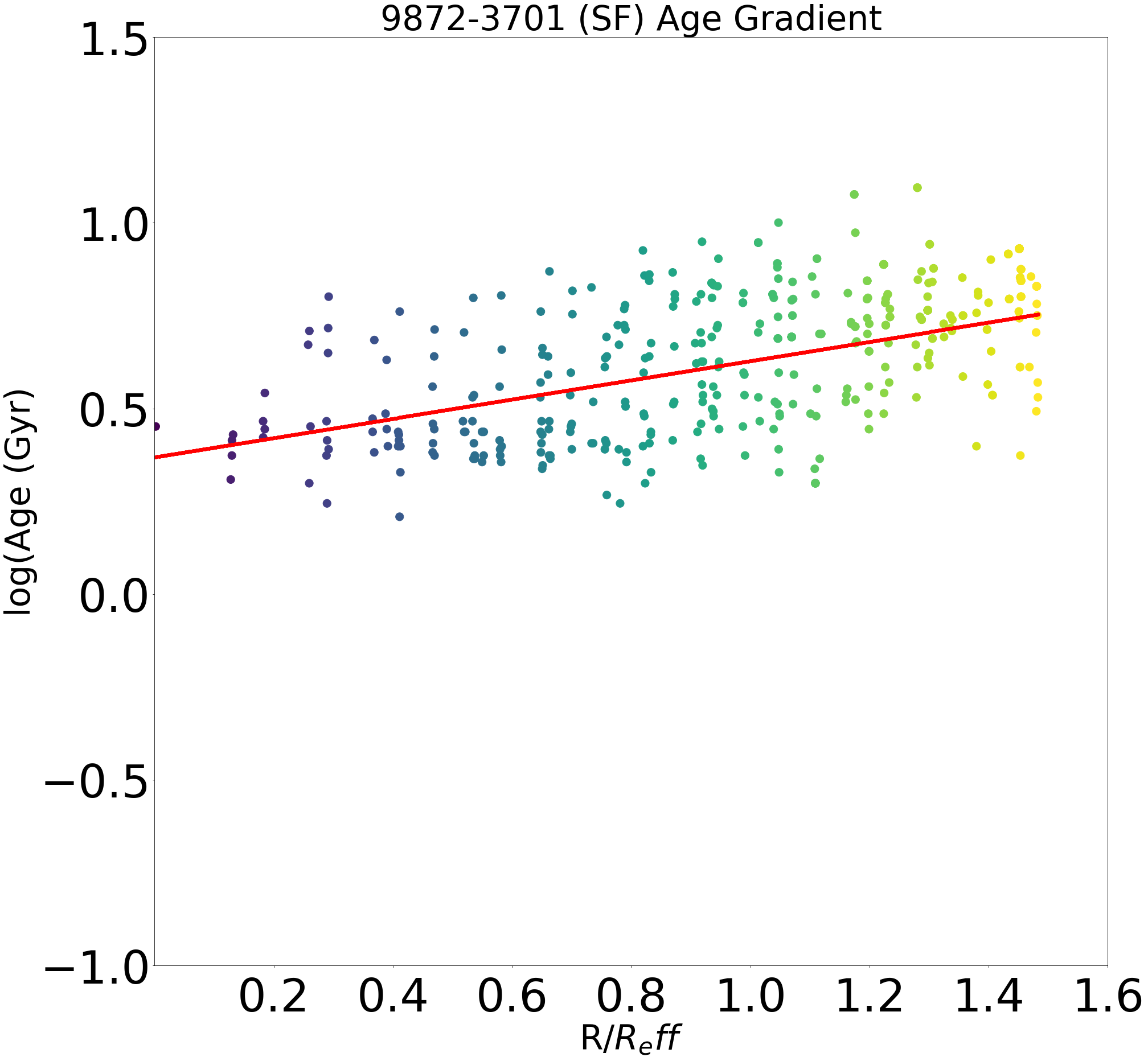} & 
\hspace{-2.3cm}\includegraphics[width=0.28\textwidth, keepaspectratio]{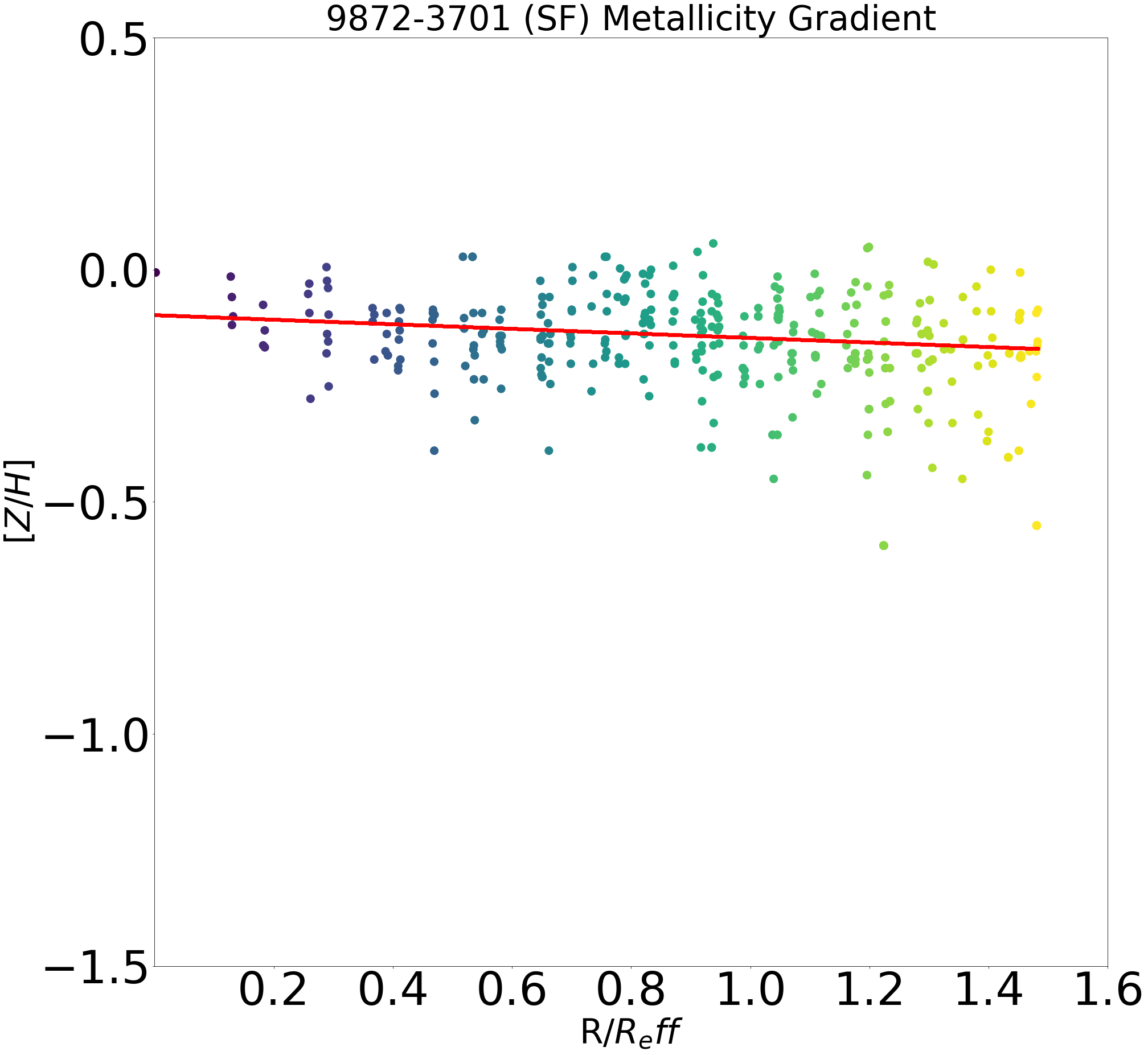}\\
\end{tabularx}
\end{figure*}
\begin{figure*}
\ContinuedFloat
\centering
\begin{tabularx}{1.2\textwidth}{X X X X}
\hspace{-1.5cm}\includegraphics[width=0.28\textwidth, keepaspectratio]{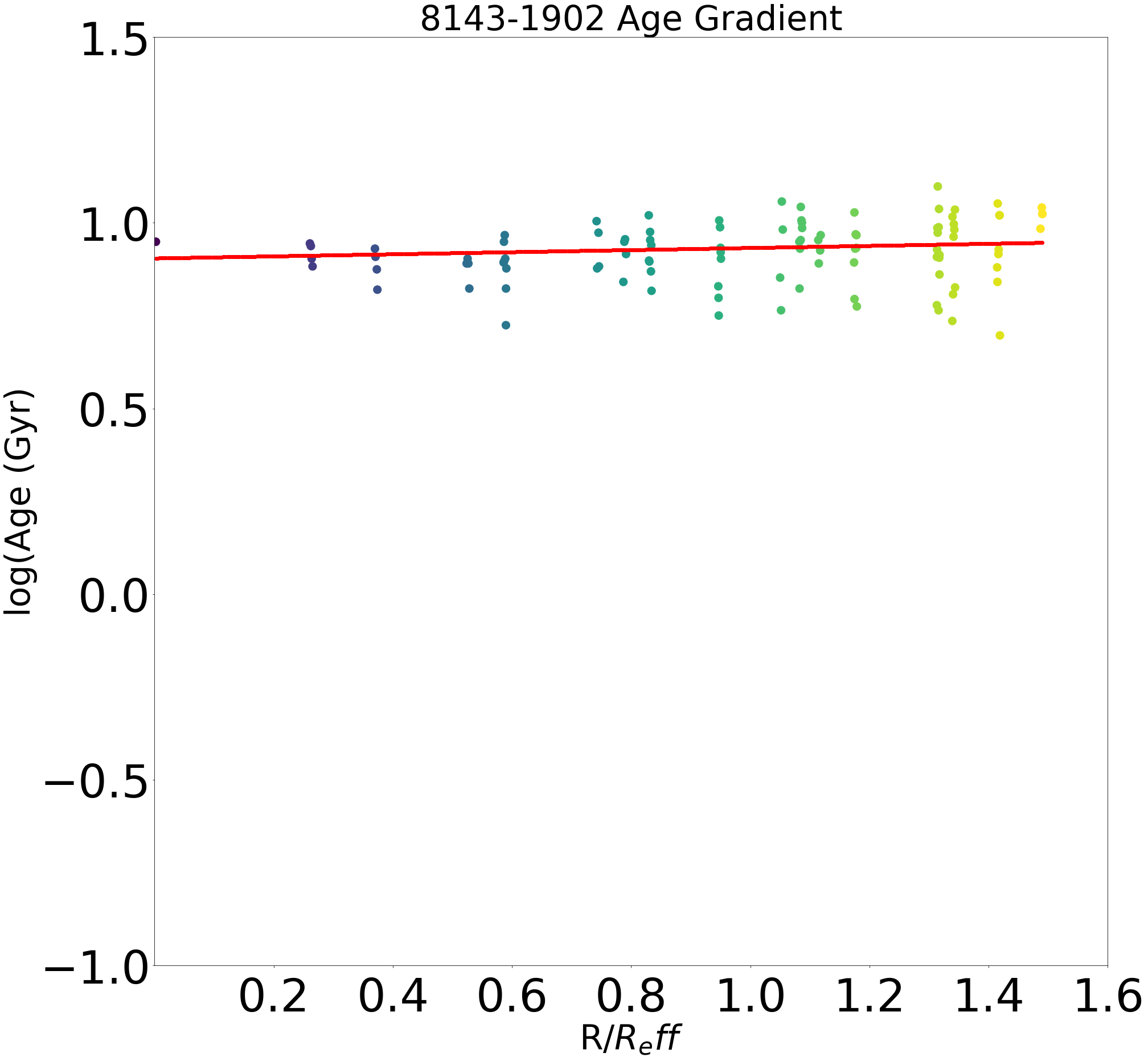} & 
\hspace{-2.3cm}\includegraphics[width=0.28\textwidth, keepaspectratio]{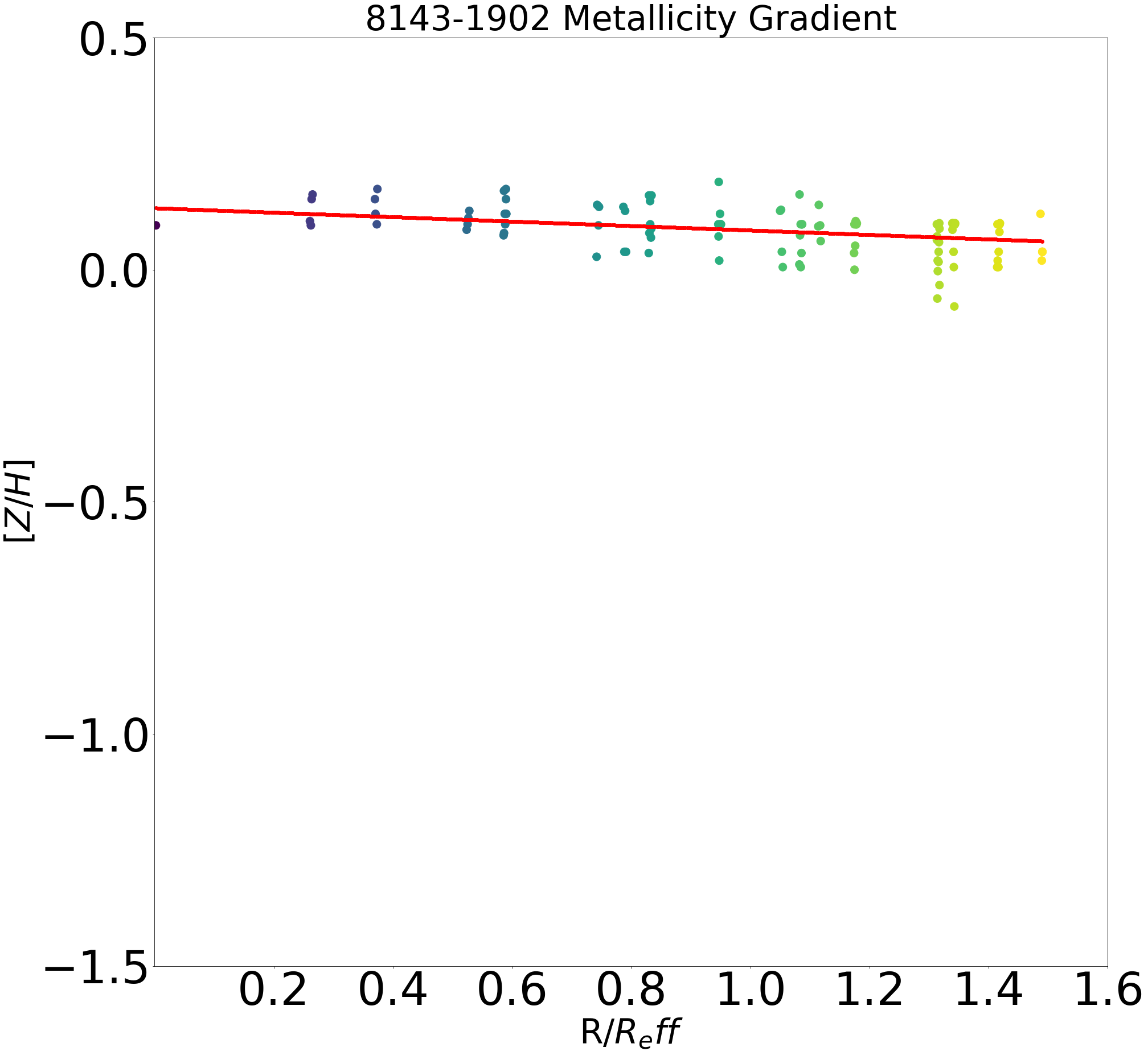}&
\hspace{-2.0cm}\includegraphics[width=0.28\textwidth, keepaspectratio]{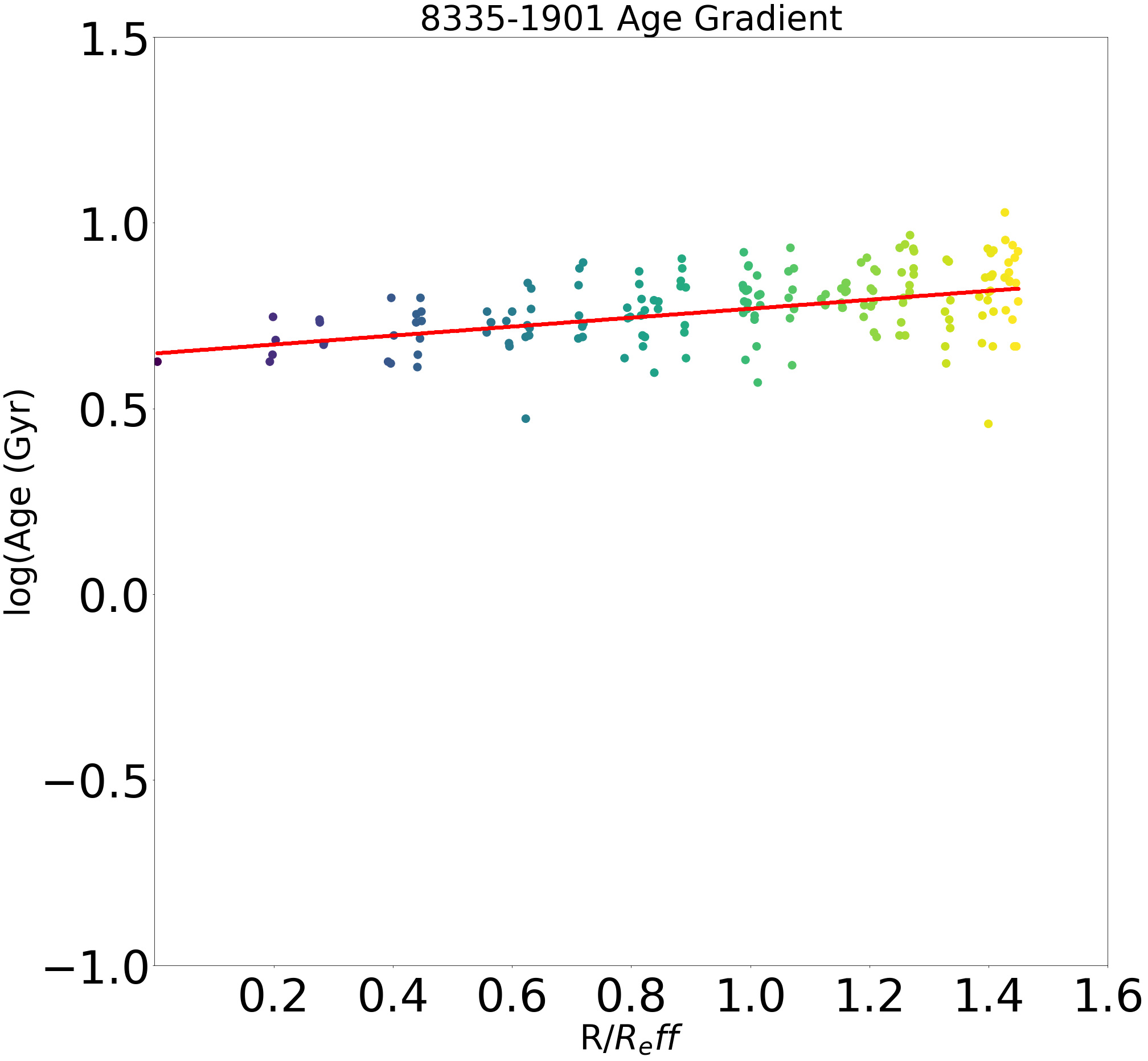} & 
\hspace{-2.3cm}\includegraphics[width=0.28\textwidth, keepaspectratio]{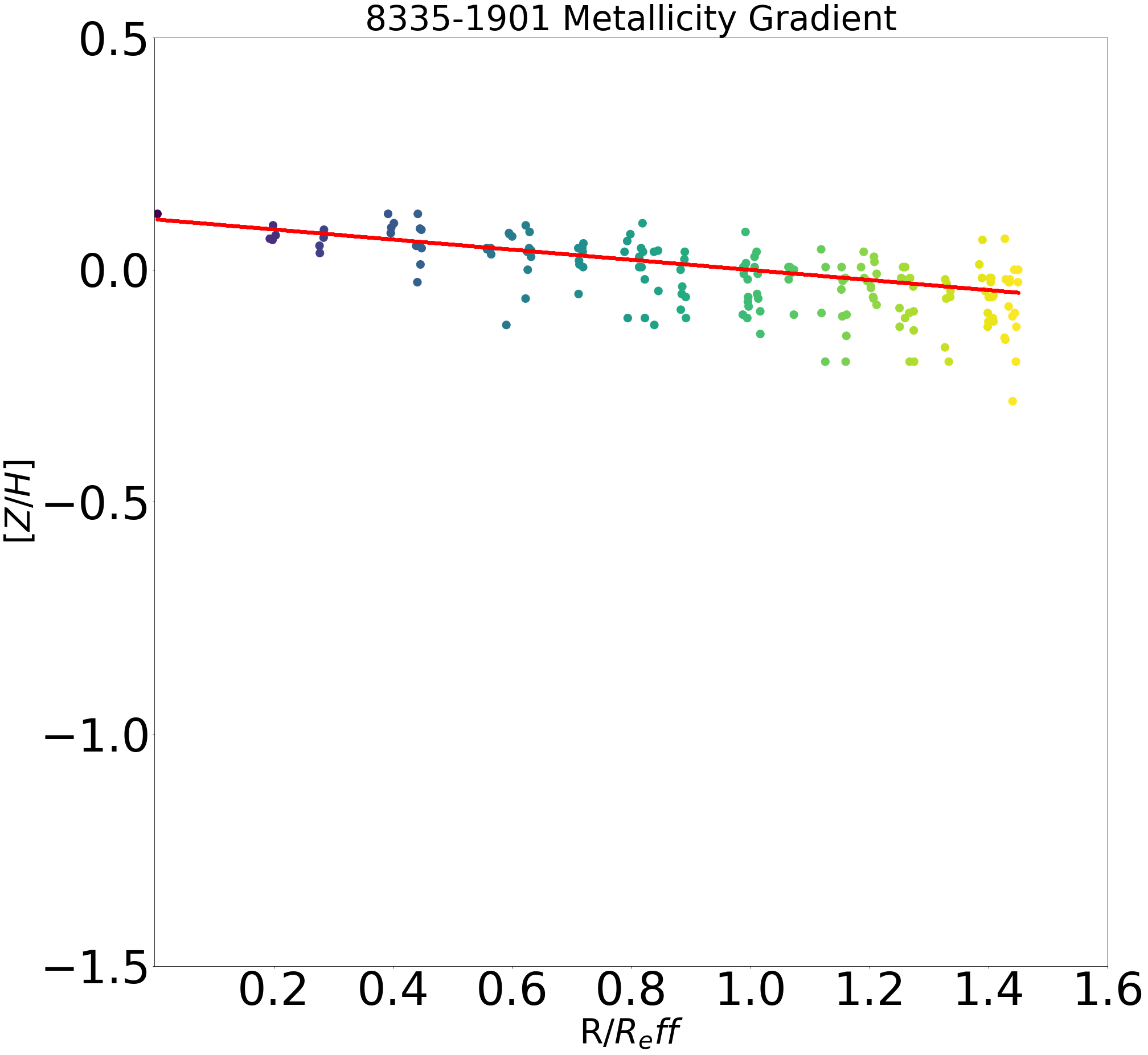}\\
\hspace{-1.5cm}\includegraphics[width=0.28\textwidth, keepaspectratio]{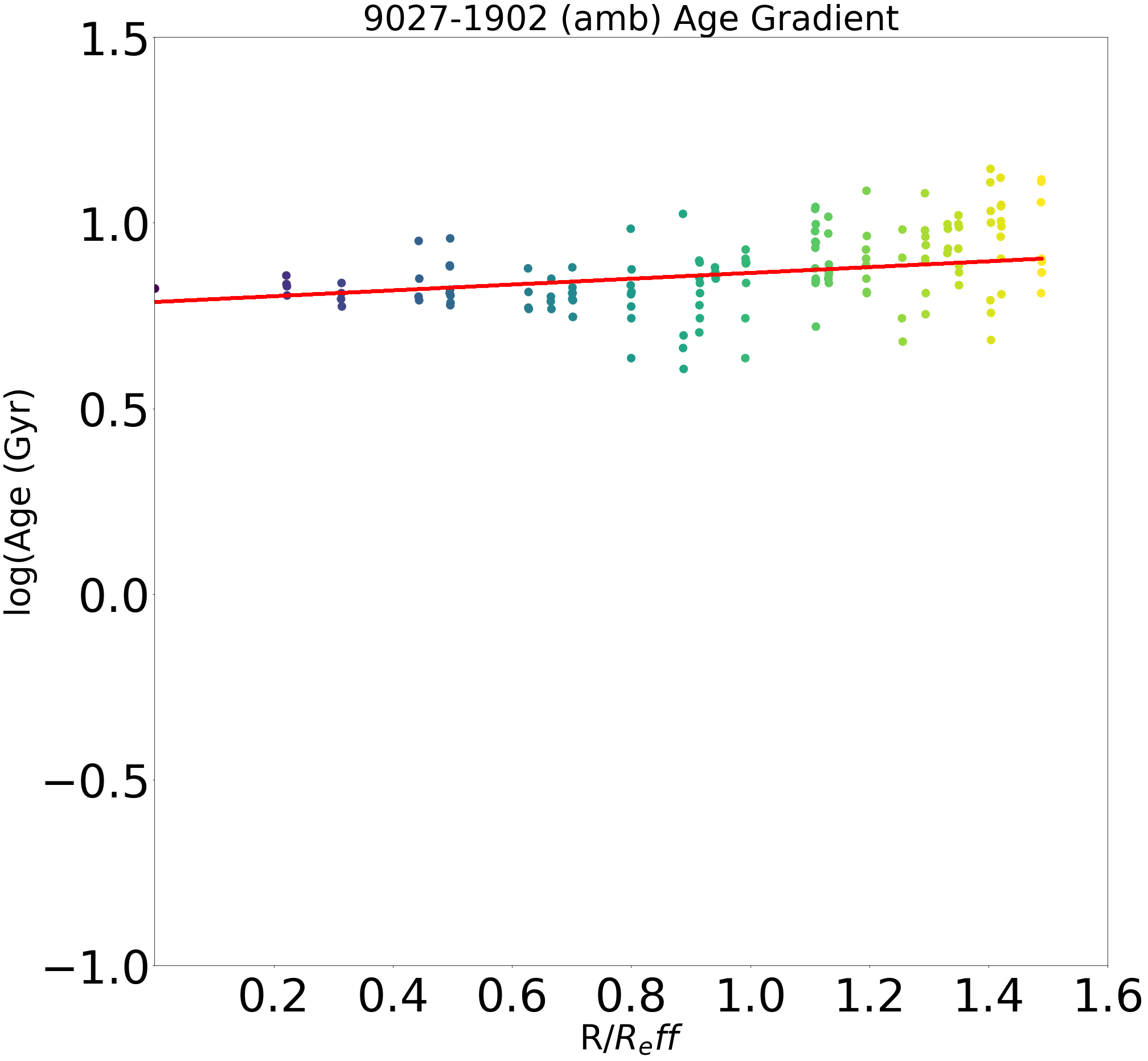} & 
\hspace{-2.3cm}\includegraphics[width=0.28\textwidth, keepaspectratio]{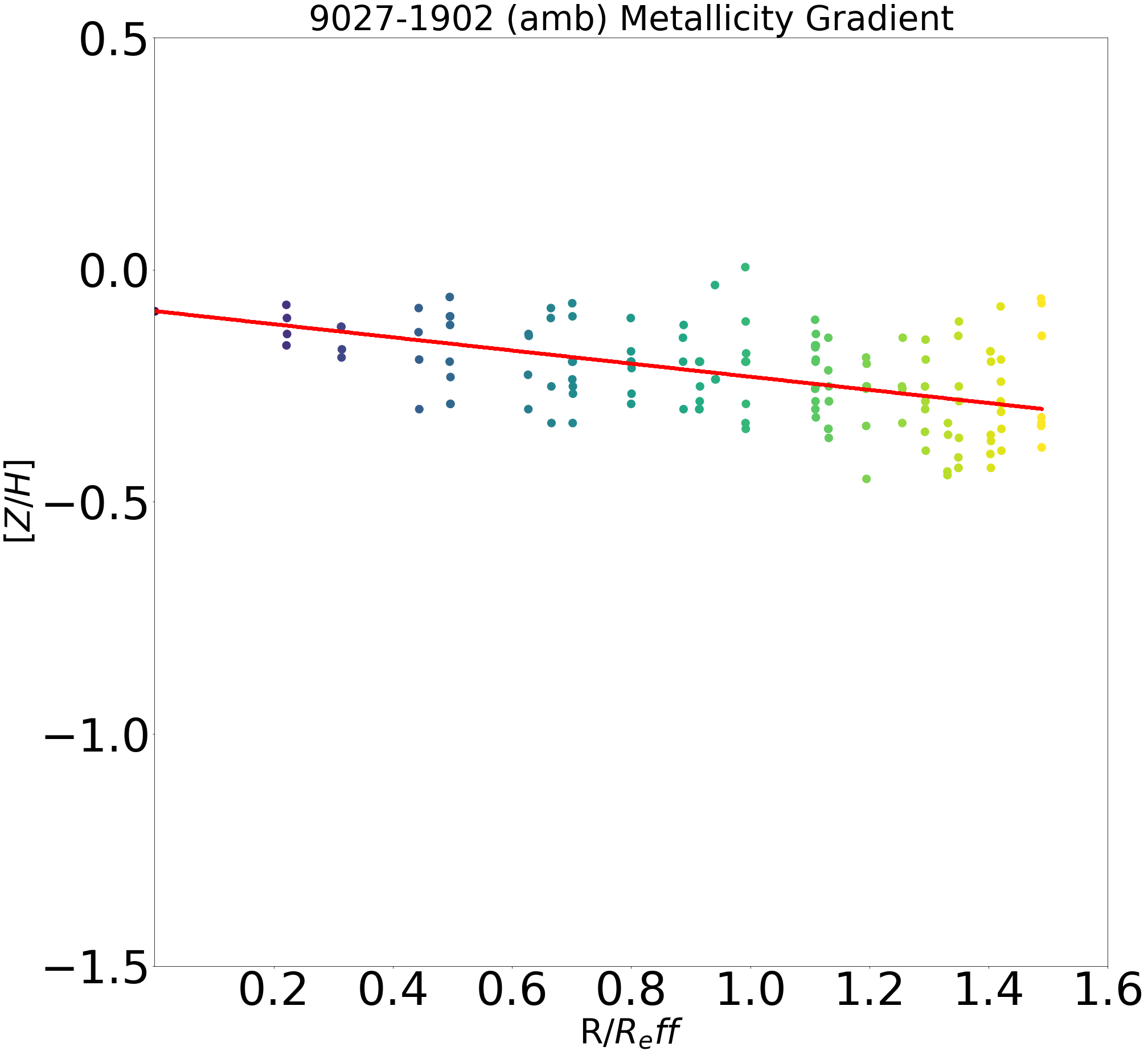}\\
    \end{tabularx}
        \caption{The age (in units of log(Age(Gyr))). and metallicity (in units of [Z/H]) gradients of our sample. Two galaxies are shown per row. The first and third columns show the age gradients, and the second and fourth the metallicity gradients. Red lines indicate our obtained gradients, and blue lines indicate gradients from the MaNGA FIREFLY VAC}
    \label{fig:graddiagram}
\end{figure*}
\begin{table*}[htp]
\centering
\begin{tabular}{||c c c c c||} 
 \hline
 \thead{Plate-IFU \\ \textbf{(a)}} &  \thead{Age Gradient \\ \textbf{(b)}} &  \thead{Age Gradient Zeropoint \\ \textbf{(c)}} & \thead{Metallicity Gradient \\ \textbf{(d)}} & \thead{Metallicity Gradient Zeropoint\\ \textbf{(e)}}\\ [0.5ex] 
 \hline\hline
  8143-3702 (AGN)& 0.013$\pm$0.013  & \textbf{0.832} & -0.152$\pm$0.027 & 0.114\\
 \hline
 8155-3702 (AGN) & 0.016$\pm$0.071 & 0.629 & -0.159$\pm$0.012 & 0.114\\
 \hline
8606-3702 (AGN) & 0.058$\pm$0.005 & 0.811 & -0.245$\pm$0.142 & 0.157\\
 \hline
  8989-9101 (AGN) & -0.108$\pm$0.028 & 1.003 & -0.110$\pm$0.079 & 0.079\\
 \hline
  8995-3703 (AGN) & -0.044$\pm$0.030 & 1.023 & 0.073$\pm$0.027 & -0.198 \\
 \hline
   8615-1902 (starforming) & 0.179$\pm$0.005 & 0.516 & 0.079$\pm$0.070 & -0.242 \\
 \hline
 9027-3703 (starforming) & -0.209$\pm$0.052 & 0.650 & 0.040$\pm$0.049 & -0.267 \\  
 \hline
  9872-3701 (starforming) & 0.259$\pm$0.032 & 0.369 & -0.049$\pm$0.011 & -0.098 \\ 
\hline
 8143-1902 (unclassified) & 0.028$\pm$0.021 & 0.905 & -0.048$\pm$0.017 & 0.133\\ 
 \hline
8335-1901 (unclassified) & 0.121$\pm$0.024 & 0.649 & -0.109$\pm$0.008 & 0.109\\ 
\hline
9027-1902 (ambiguous) & 0.078$\pm$0.030 & 0.787 & -0.142$\pm$0.014 & -0.089\\ 
  [1ex]
 \hline
\end{tabular}
\caption{\label{tab:lwam}Stellar population gradients given for our galaxy sample within 1 $R_e$, obtained from the MaNGA FIREFLY VAC.}
\end{table*}

Table \ref{tab:lwam} shows the age and metallicity gradients of our sample, with values provided both from the MaNGA FIREFLY VAC, obtained by linearly fitting the datapoints within 1.5 $R_e$. Fig. \ref{fig:graddiagram} graphs the gradients.
Figure \ref{fig:stellarpop} shows the spatially resolved age and metallicity maps of our sample. The left diagram shows the ages, and the right the metallicities.
Our results show that in the AGN-hosting galaxies, the ages showed relatively shallow gradients compared to the star-forming galaxies, and the metallicities showed more steep gradients in the AGN-hosts compared to the star-forming galaxies. Two of the three star-forming galaxies, 9872\mbox{--}3701 and 8615\mbox{--}1902 showed positive gradients, whereas 9027\mbox{--}3702 has a negative gradient.
The two unclassified galaxies have positive age and negative age gradients, however 8143\mbox{--}1902 has shallower gradients compared to 8335\mbox{--}1901. The ambiguous galaxy showed gradient steepnesses similar to those of the AGN-hosts.
\subsection{Star formation histories}

\renewcommand{\thefigure}{4}
\begin{figure*}[htp]
    \centering
    \begin{tabularx}{\textwidth}{X X}
\hspace{-1.5cm}\includegraphics[width=0.55\textwidth, keepaspectratio]{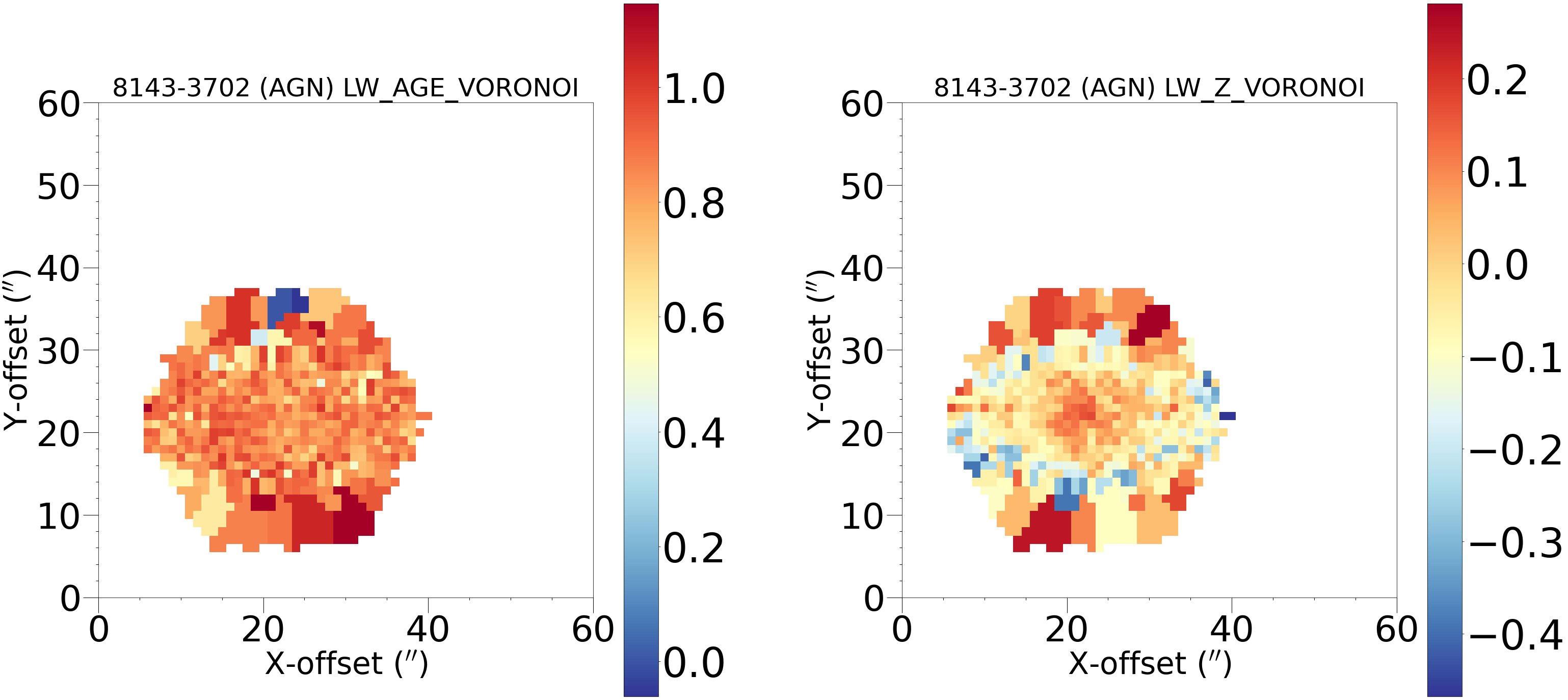}  &
\hspace{0cm}\includegraphics[width=0.55\textwidth, keepaspectratio]{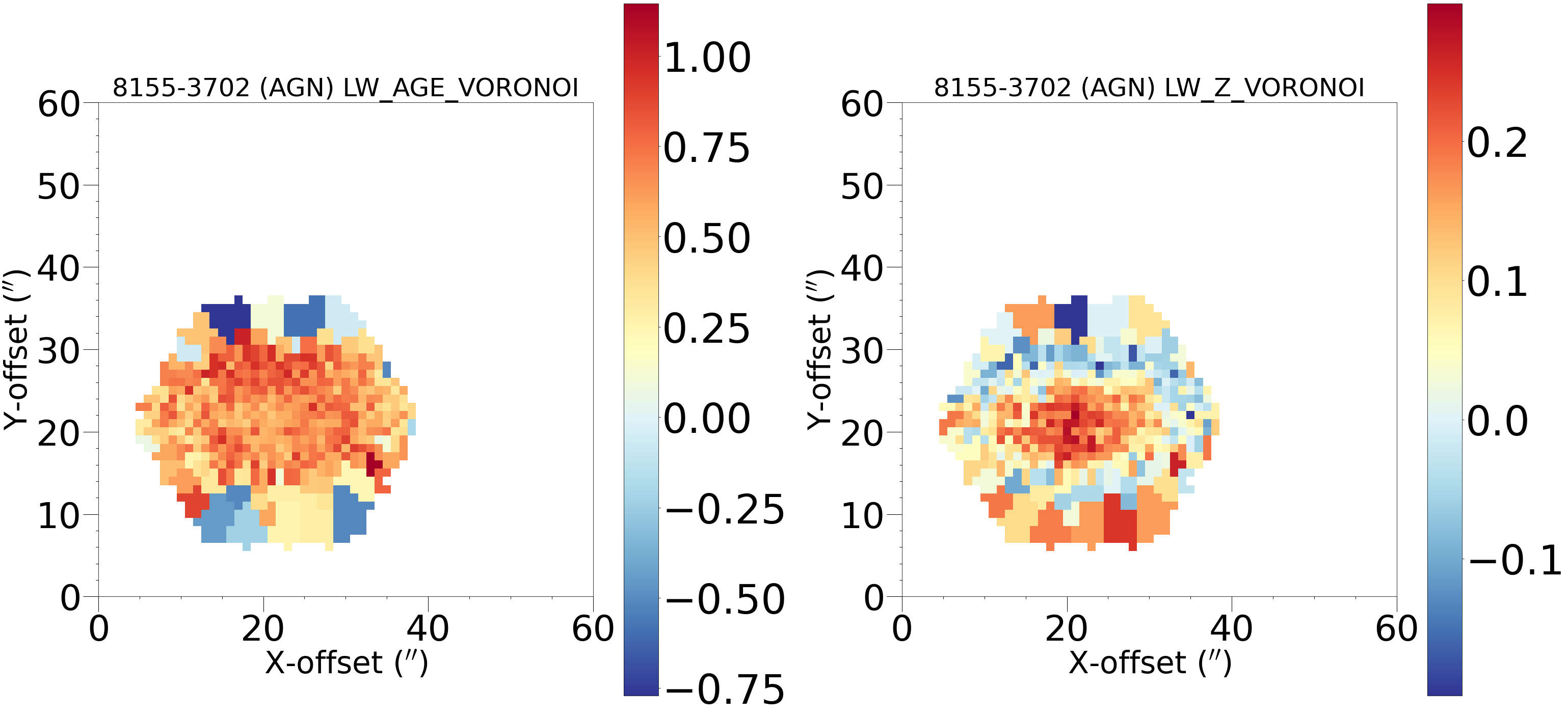} \\
\hspace{-1.5cm}\includegraphics[width=0.55\textwidth, keepaspectratio]{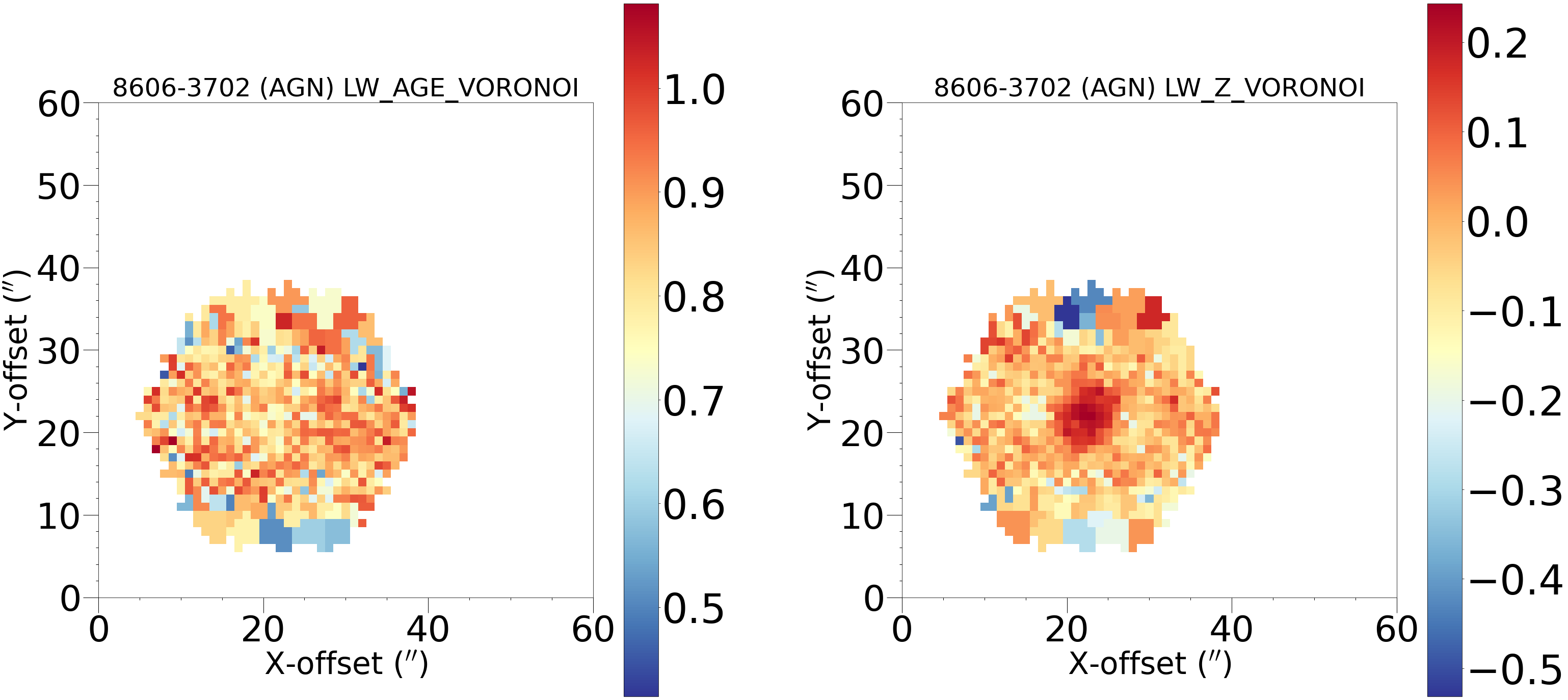} &
\hspace{0cm}\includegraphics[width=0.55\textwidth,keepaspectratio]{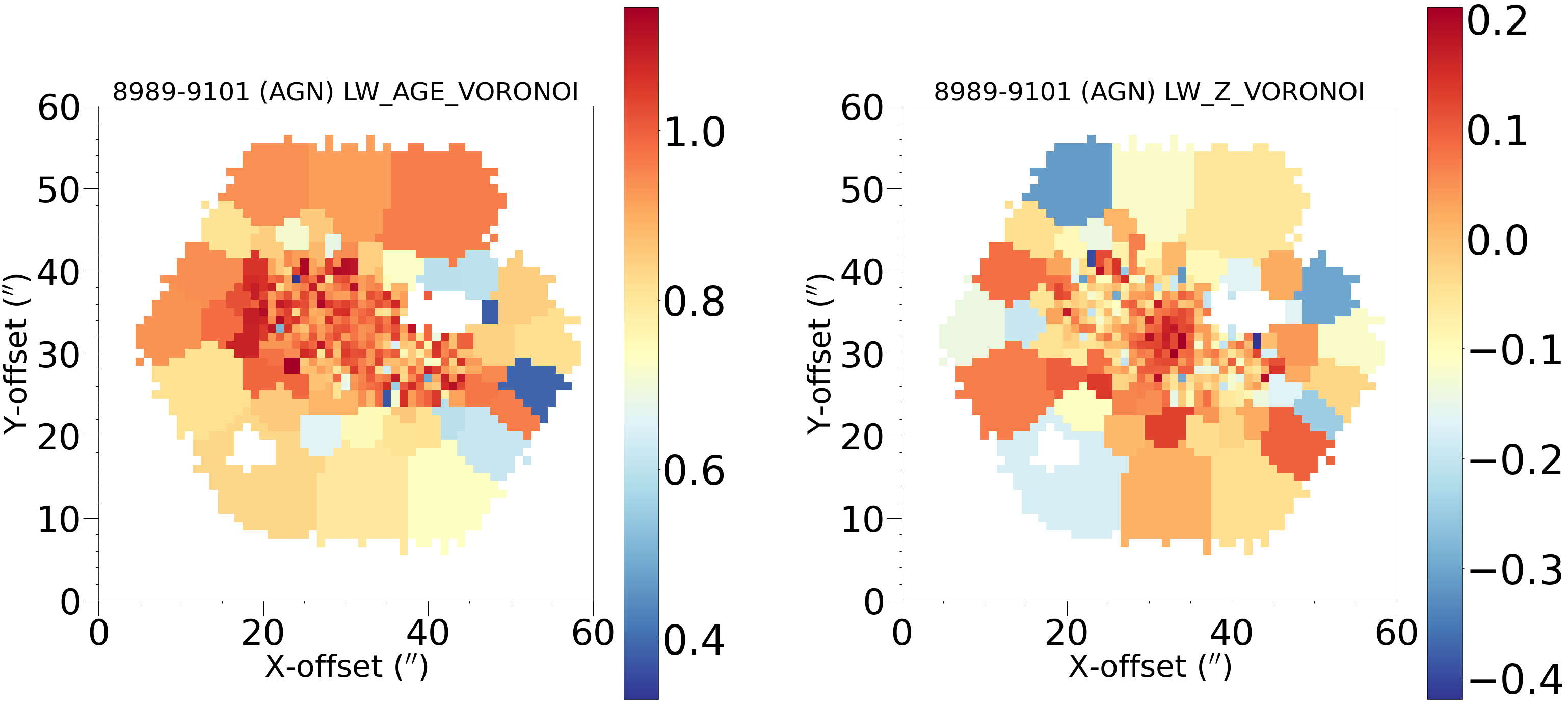} \\
\hspace{-1.5cm}\includegraphics[width=0.55\textwidth,keepaspectratio]{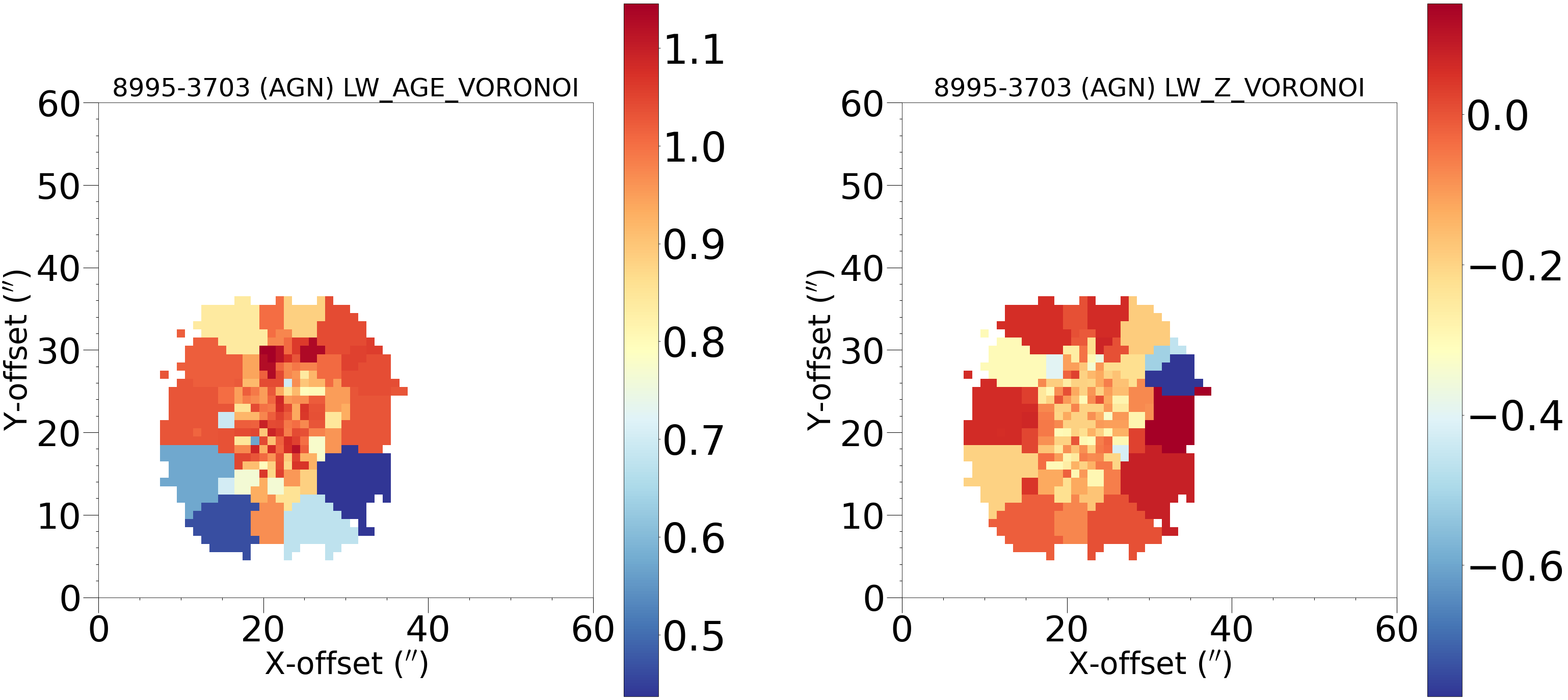}  &
\hspace{0cm}\includegraphics[width=0.55\textwidth, keepaspectratio]{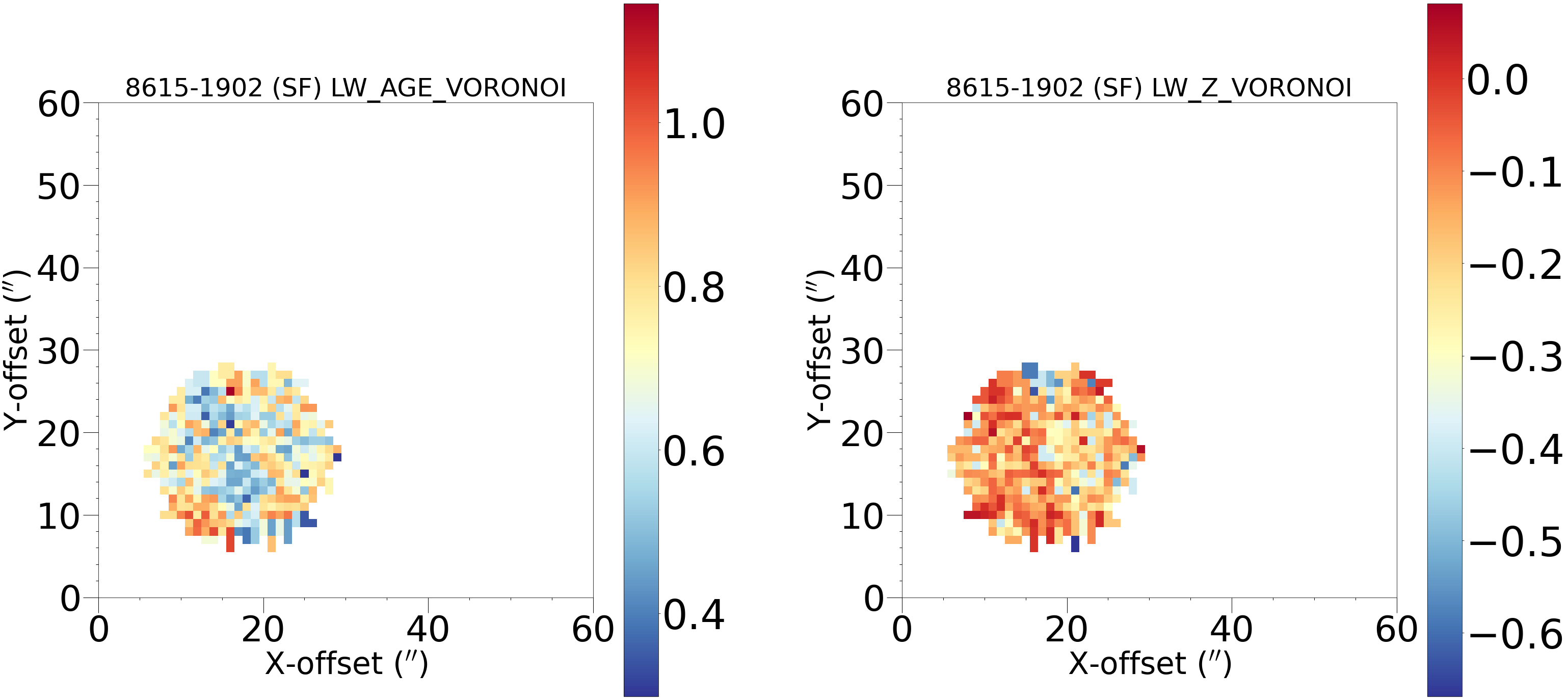}  \\
\hspace{-1.5cm}\includegraphics[width=0.55\textwidth, keepaspectratio]{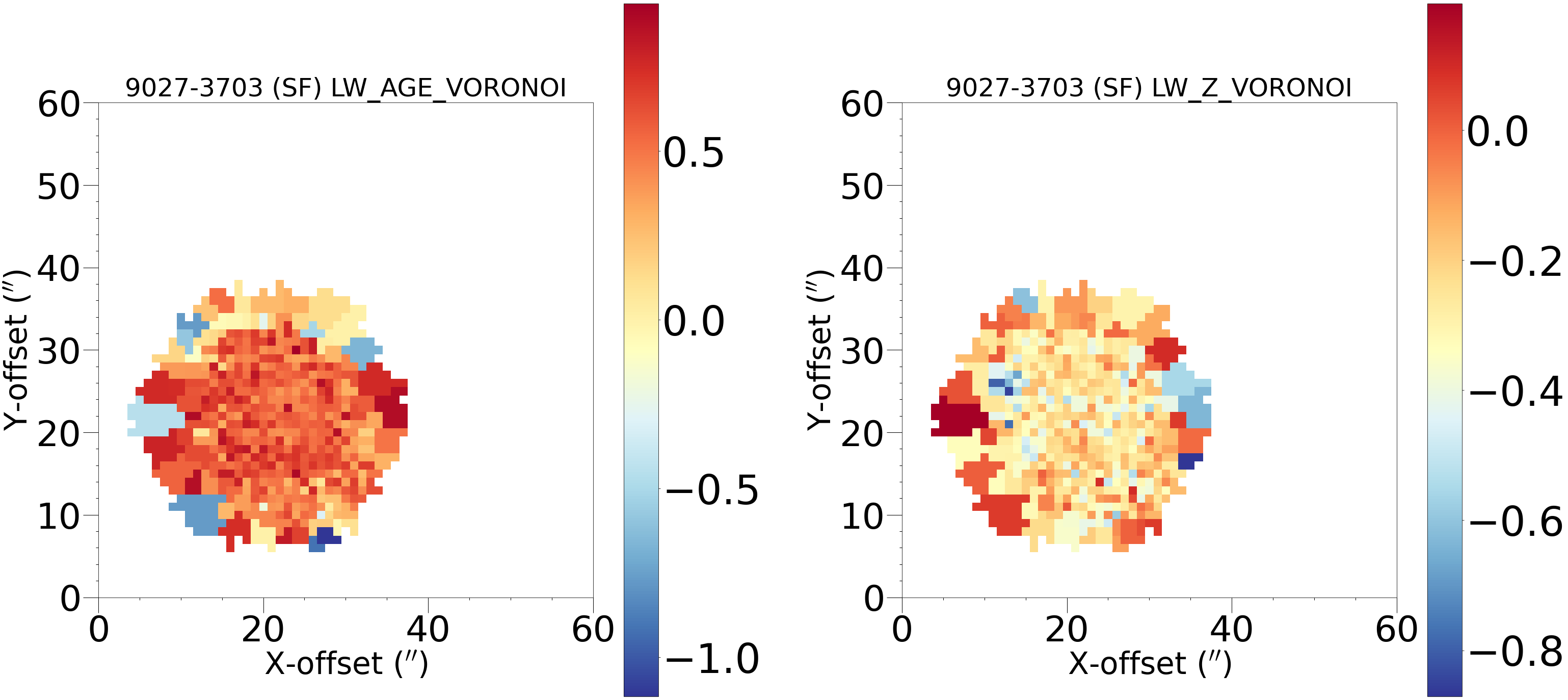}  &
\hspace{0cm}\includegraphics[width=0.55\textwidth, keepaspectratio]{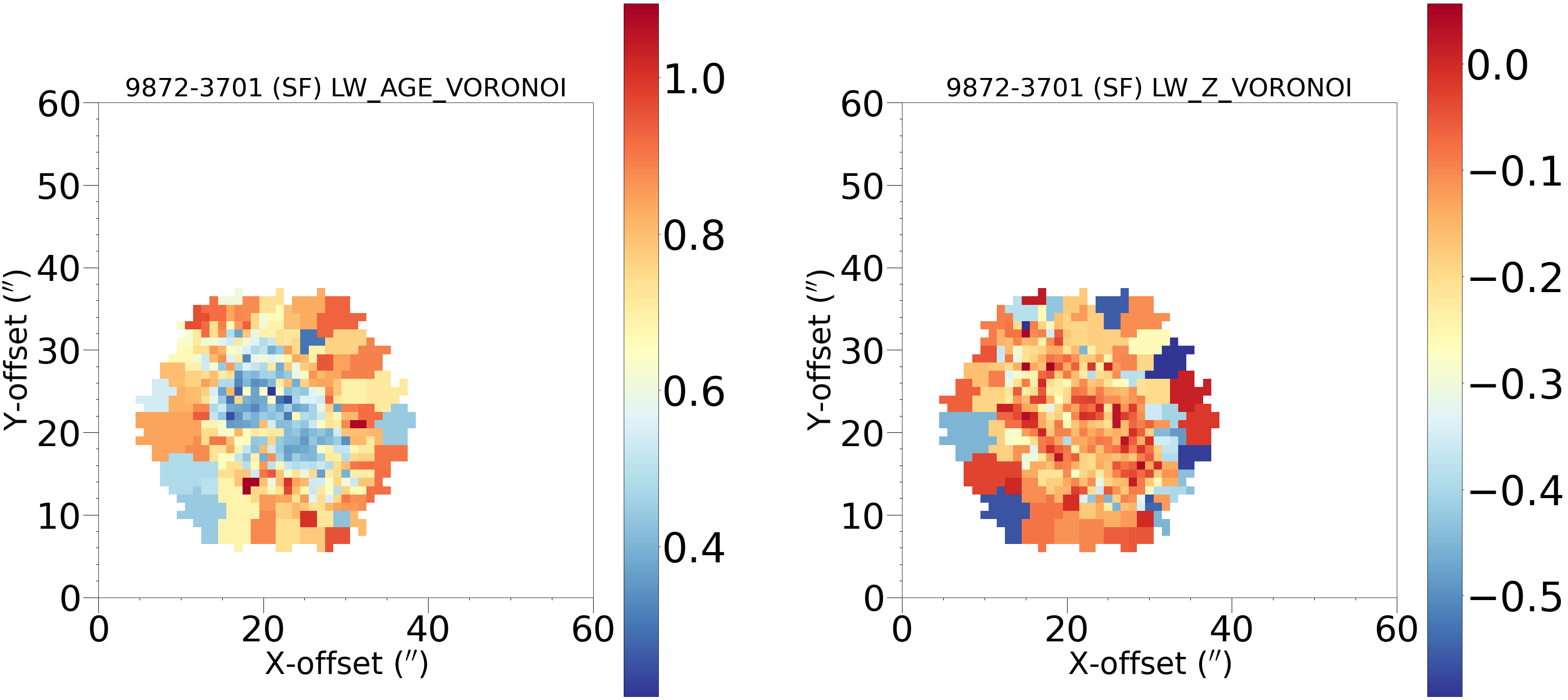}   \\

    \end{tabularx}
\end{figure*}
\newpage
\begin{figure*}
\centering
\begin{tabularx}{\textwidth}{X X}
\hspace{-1.5cm}\includegraphics[width=0.55\textwidth, keepaspectratio]{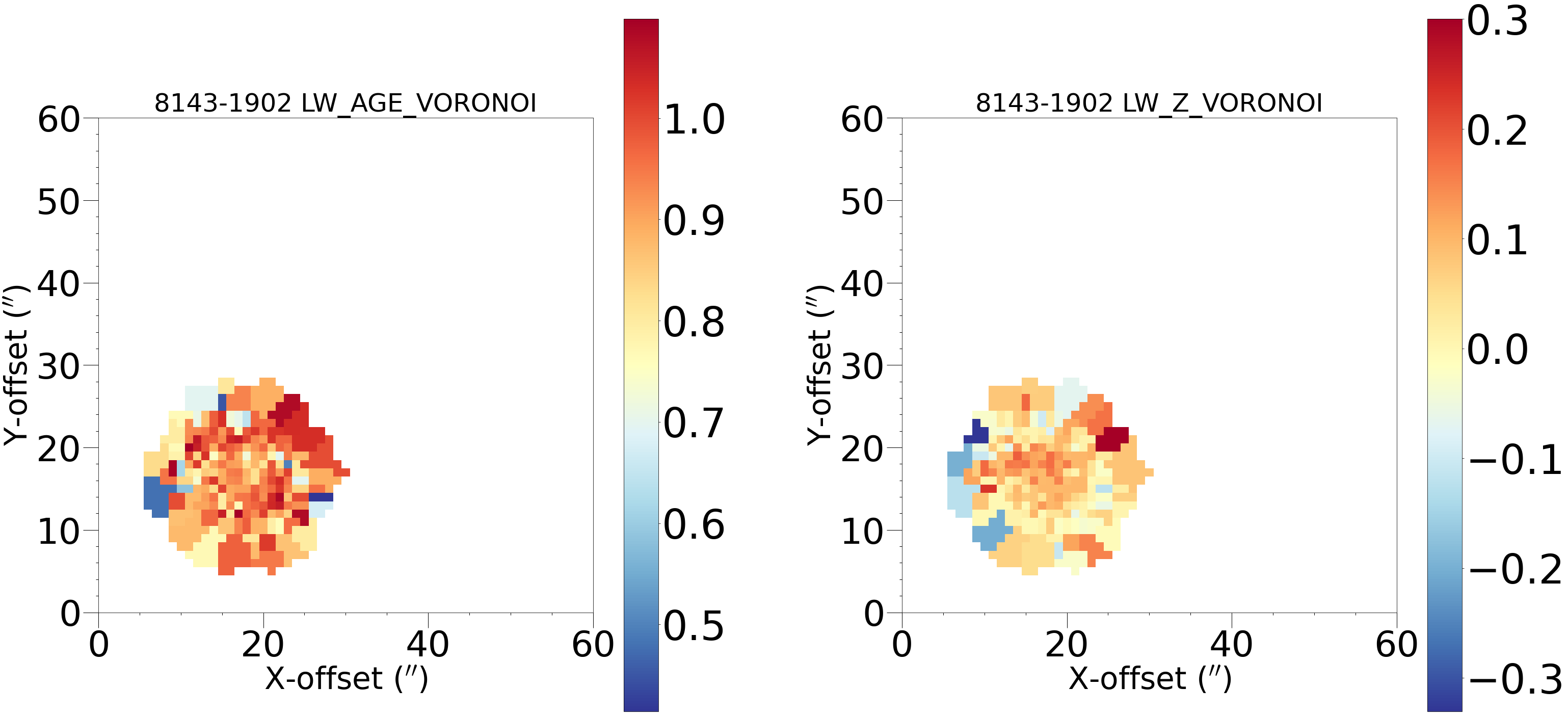}  &
\hspace{0cm}\includegraphics[width=0.55\textwidth, keepaspectratio]{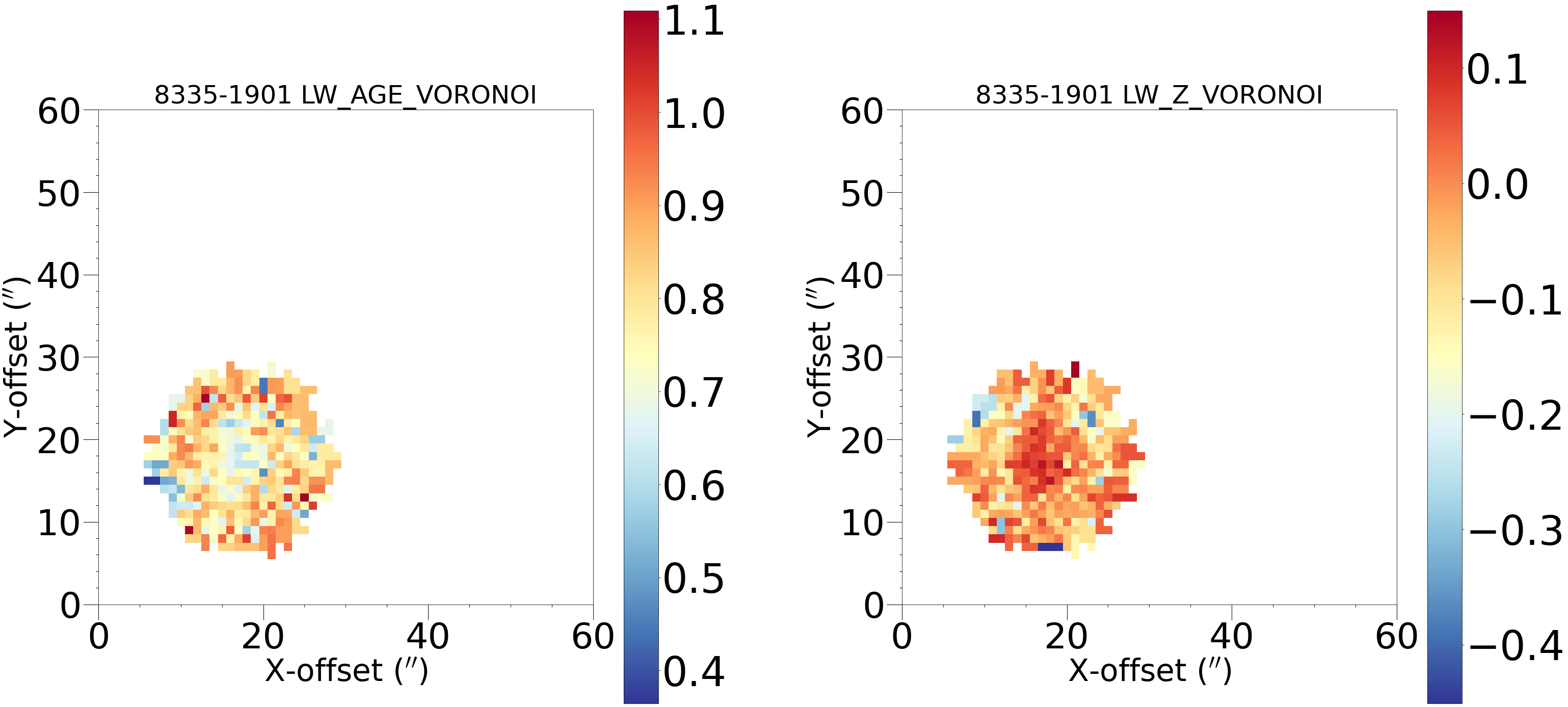}   \\
\hspace{-1.5cm}\includegraphics[width=0.55\textwidth, keepaspectratio]{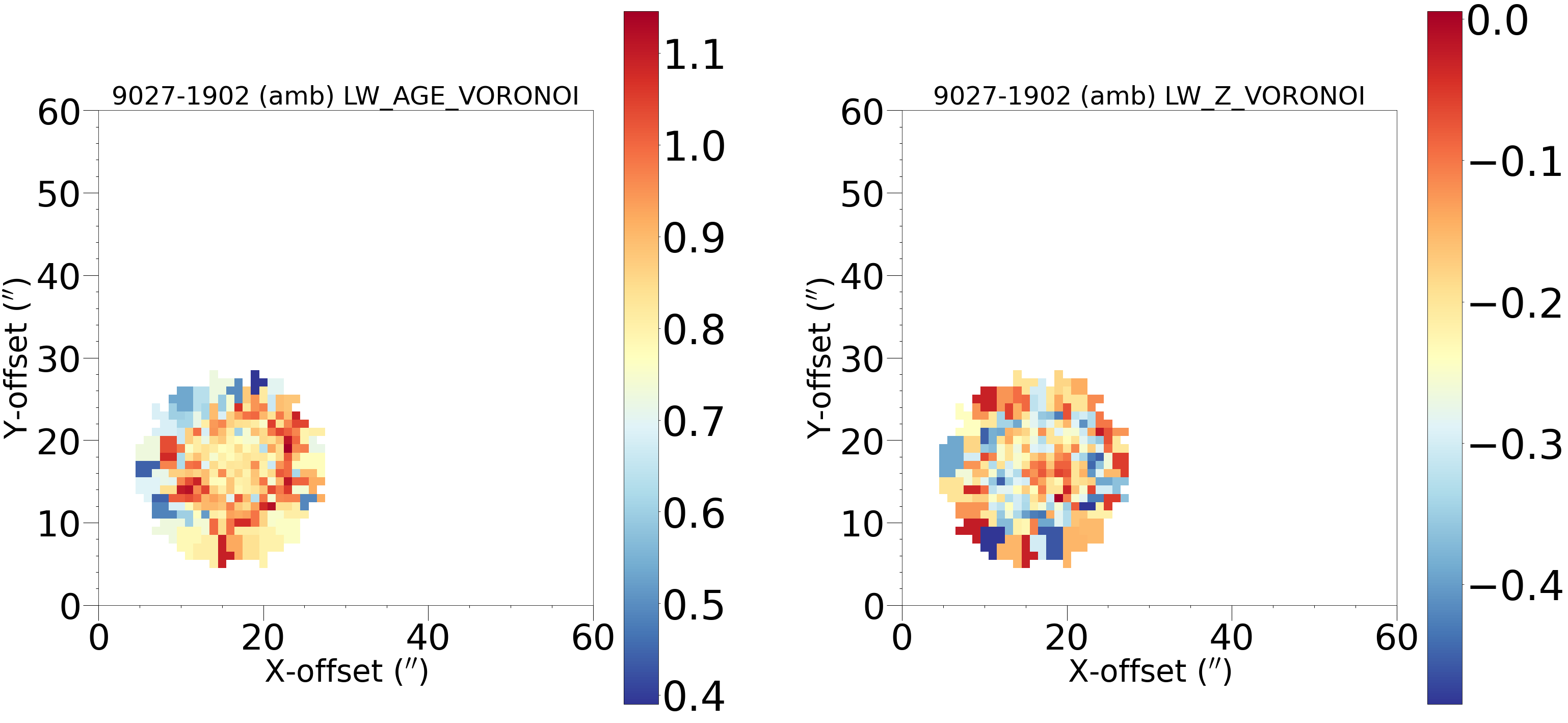}   \\
\end{tabularx}
        \caption{The spatially resolved age (in units of log(Age(Gyr))). and metallicity (in units of [Z/H]) distributions of our sample. Two galaxies are shown per row. The first and third columns show the spatially resolved age distribution, and the second and fourth the spatially resolved metallicity distribution.}
    \label{fig:stellarpop}
\end{figure*}

\renewcommand{\thefigure}{5}
\begin{figure*}[htp]
    \centering
    \begin{tabularx}{\textwidth}{X X}
\hspace{-1.5cm}\includegraphics[height=0.19\textheight, keepaspectratio]{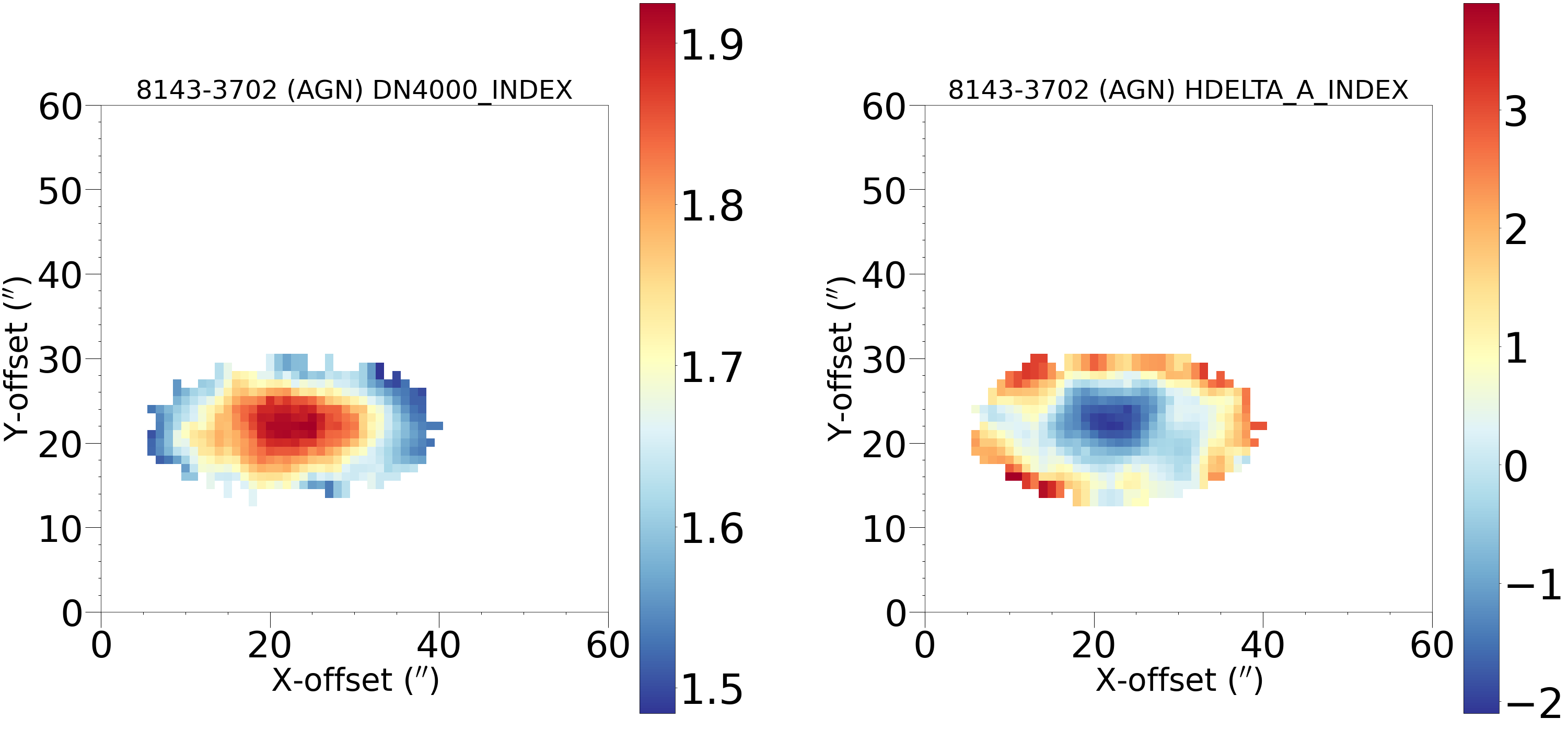}  &
\hspace{-.5cm}\includegraphics[height=0.19\textheight, keepaspectratio]{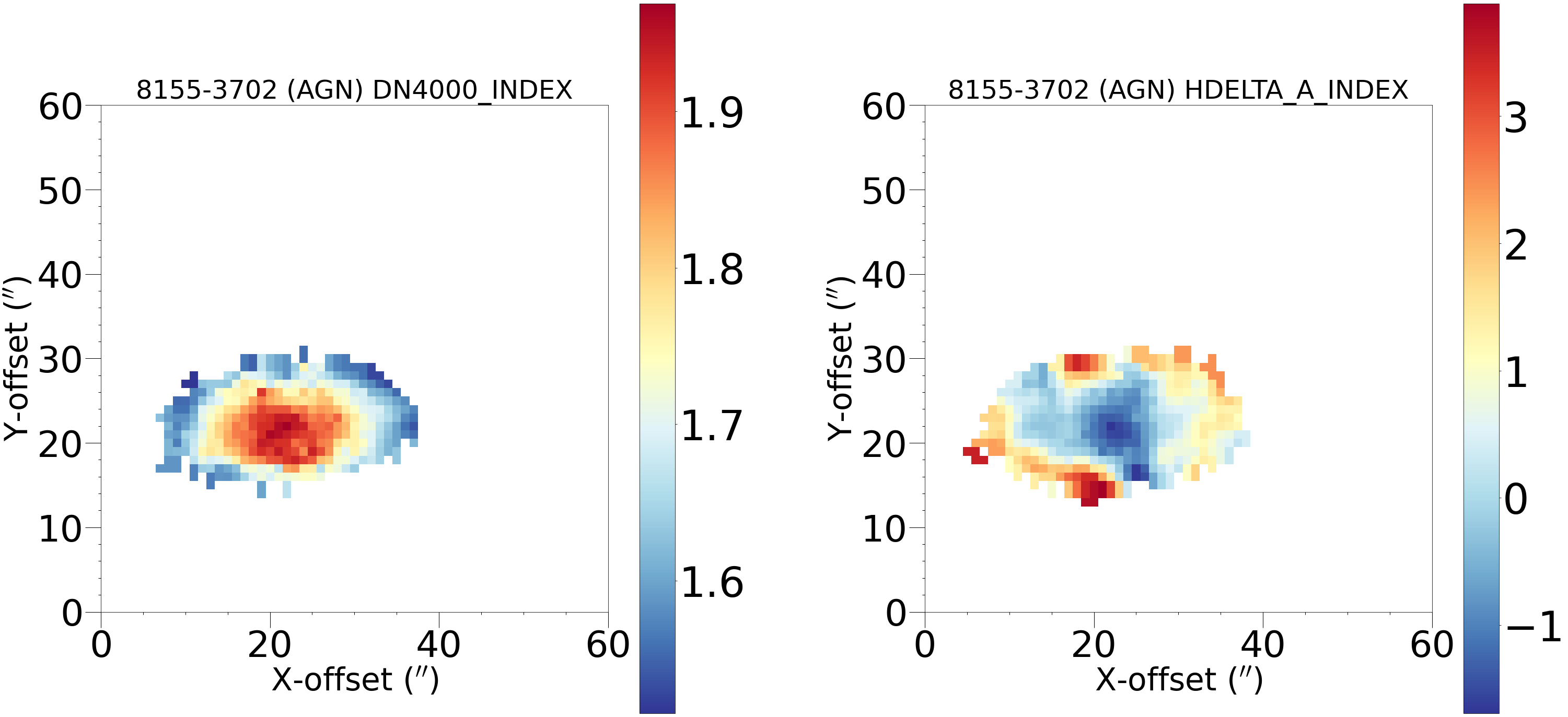} \\
\hspace{-1.5cm}\includegraphics[height=0.19\textheight, keepaspectratio]{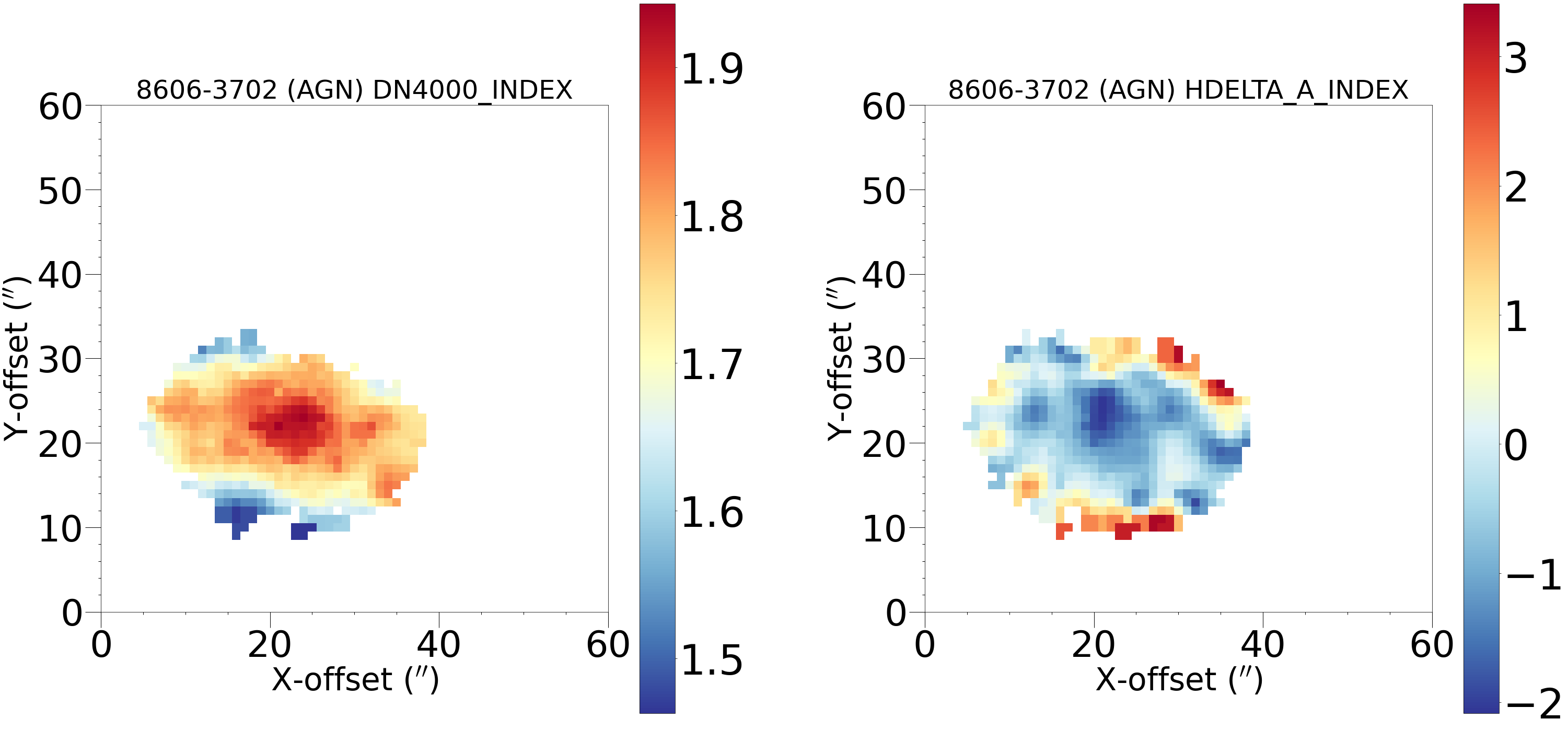} &
\hspace{-.5cm}\includegraphics[height=0.19\textheight,keepaspectratio]{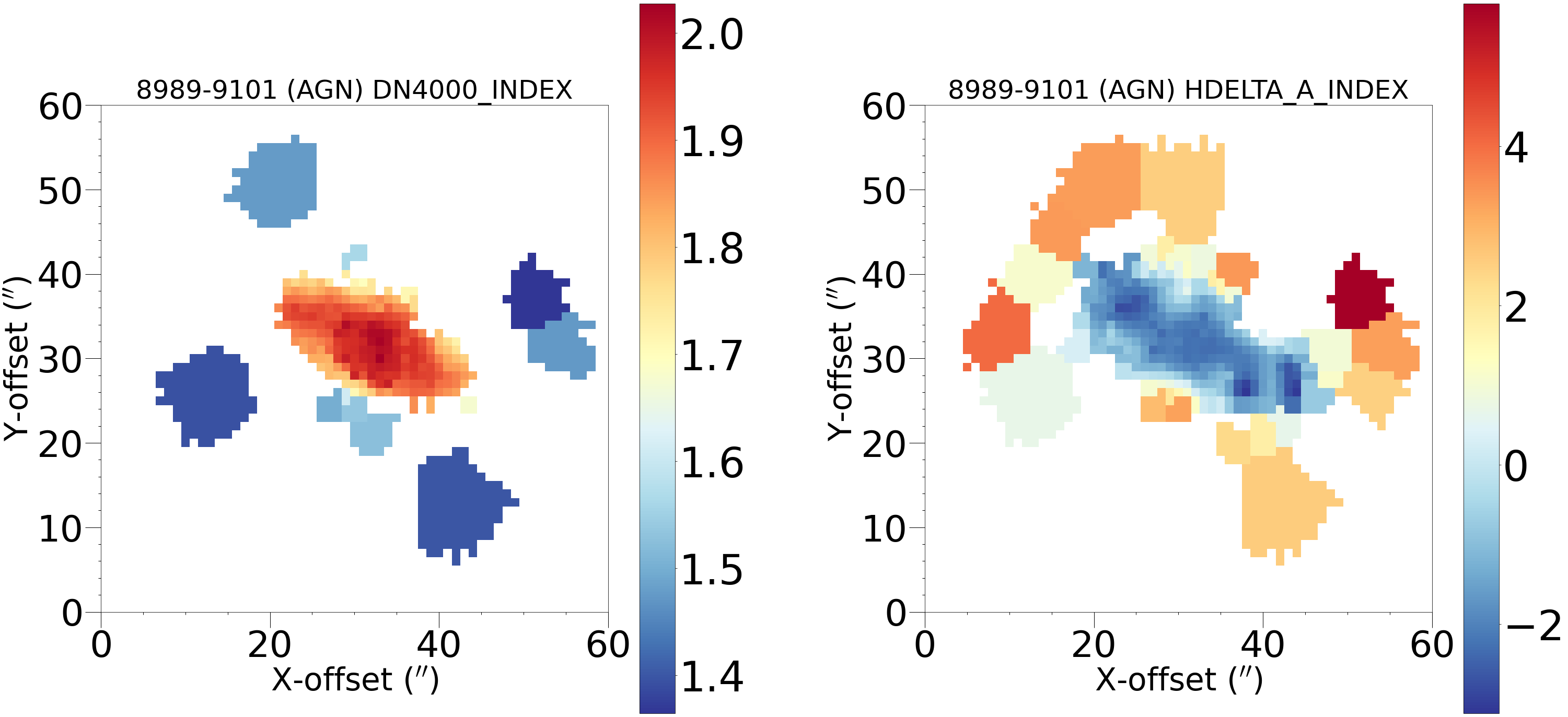} \\
\hspace{-1.5cm}\includegraphics[height=0.19\textheight,keepaspectratio]{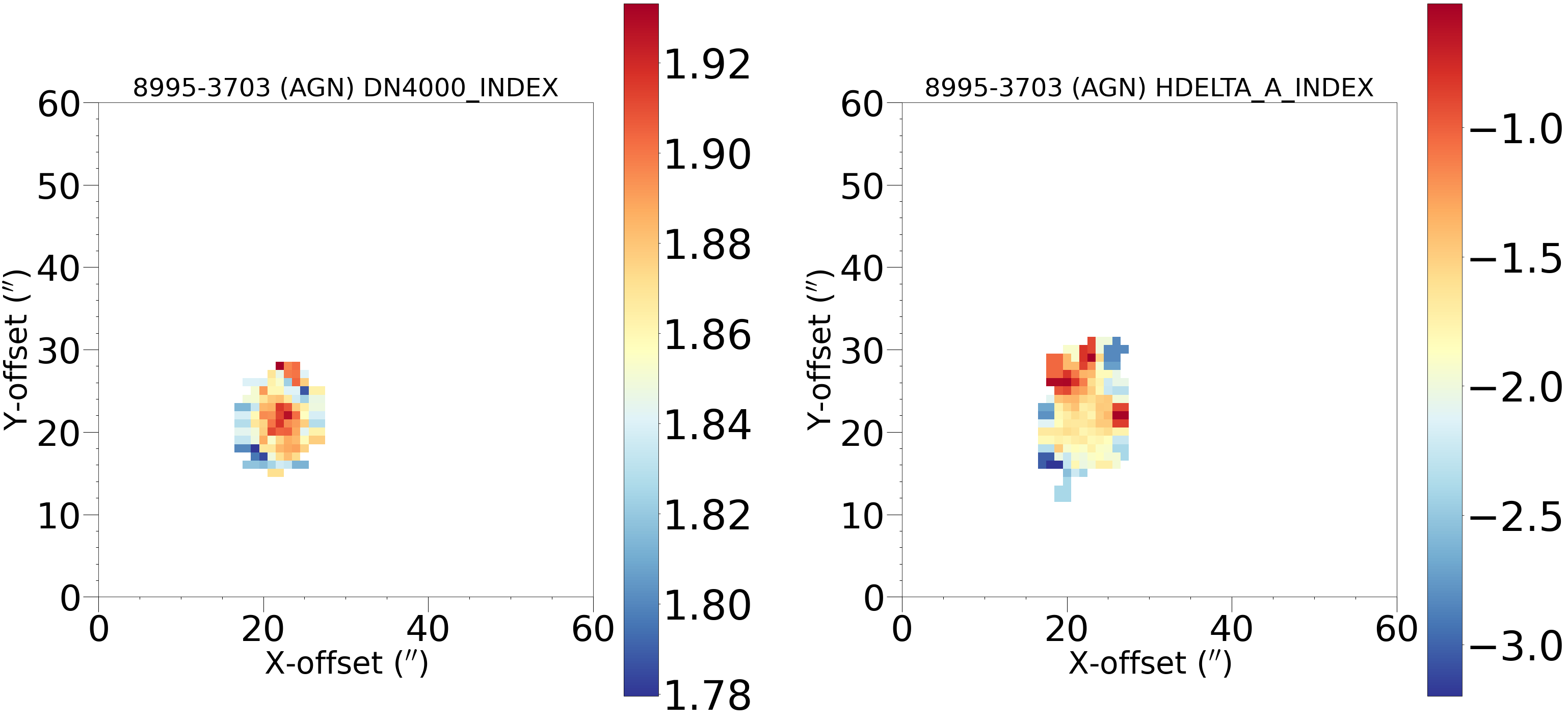}  &
\hspace{-.5cm}\includegraphics[height=0.19\textheight, keepaspectratio]{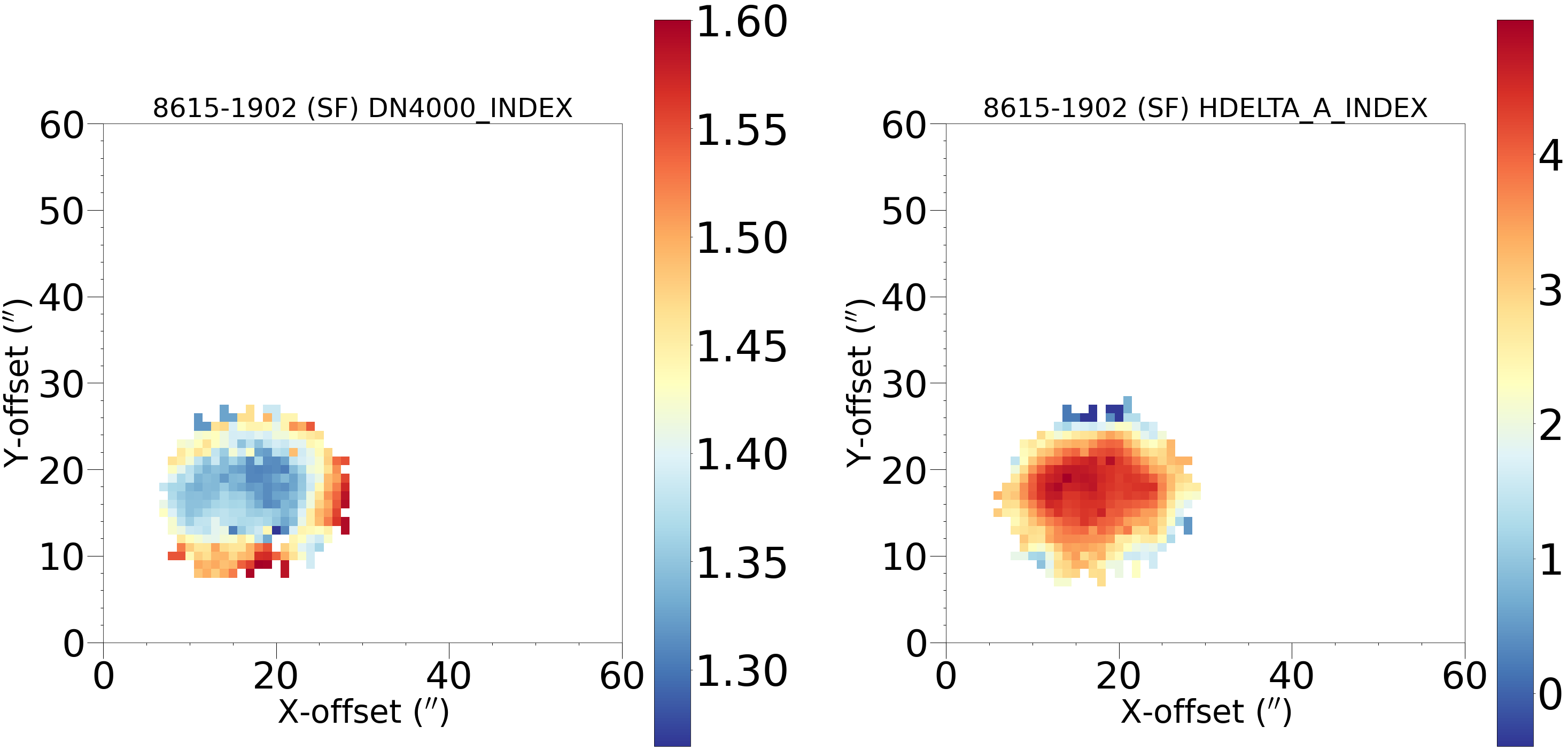}  \\
\hspace{-1.5cm}\includegraphics[height=0.19\textheight, keepaspectratio]{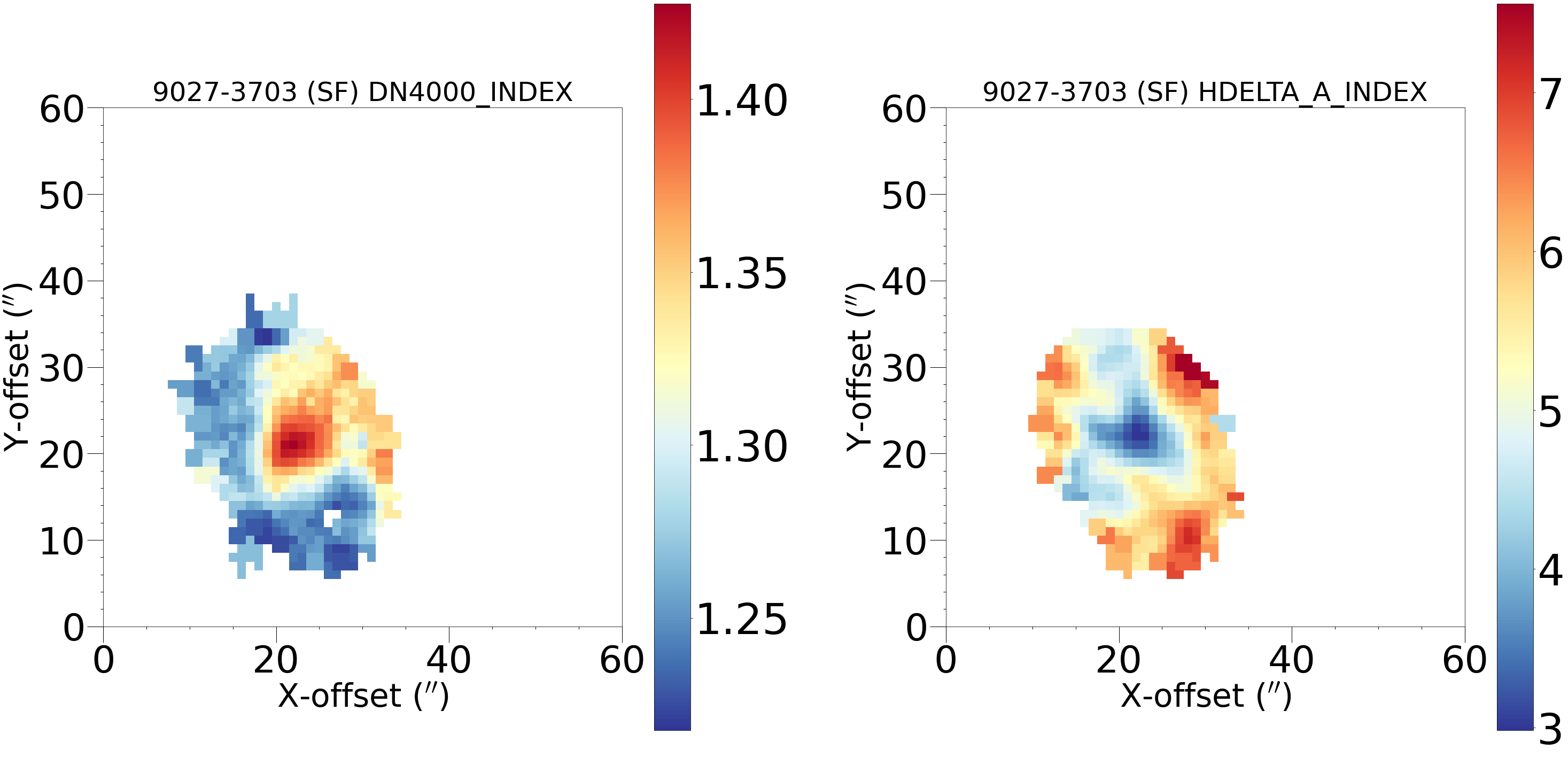}  &
\hspace{-0.5cm}\includegraphics[height=0.19\textheight, keepaspectratio]{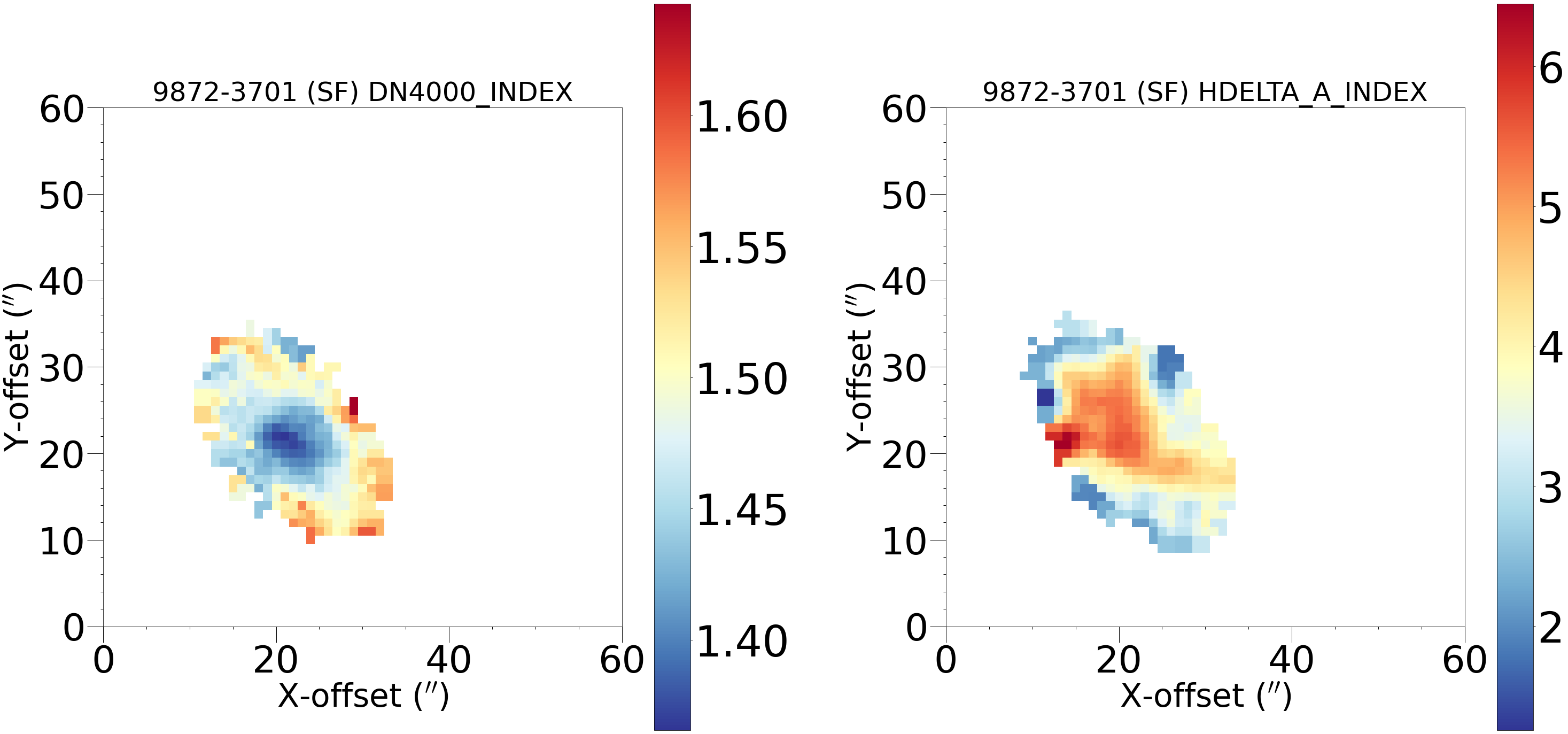}   \\
\end{tabularx}
\end{figure*}
\begin{figure*}
\ContinuedFloat
\centering
\begin{tabularx}{\textwidth}{X X}
\hspace{-1.5cm}\includegraphics[height=0.19\textheight, keepaspectratio]{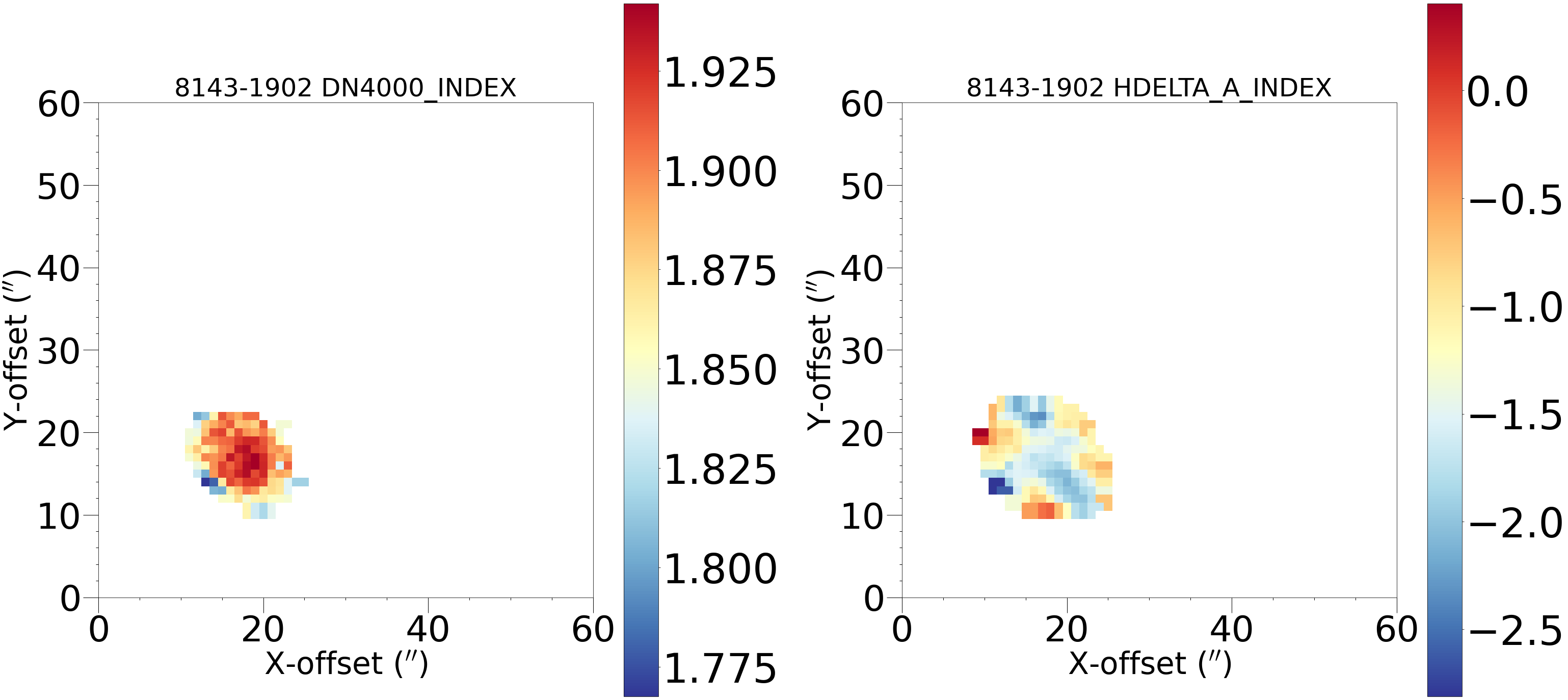}  &
\hspace{-0.1cm}\includegraphics[height=0.19\textheight, keepaspectratio]{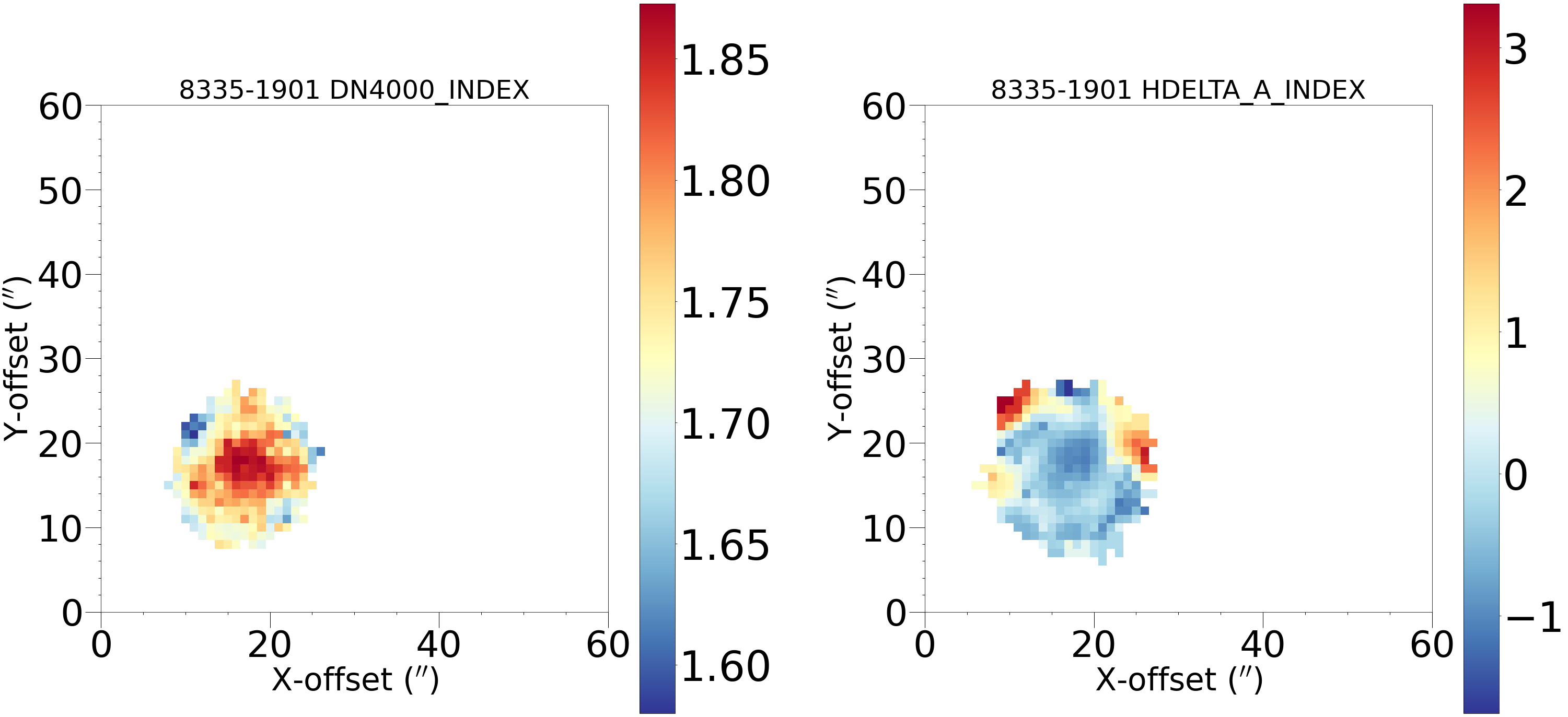}   \\
\hspace{-1.5cm}\includegraphics[height=0.19\textheight, keepaspectratio]{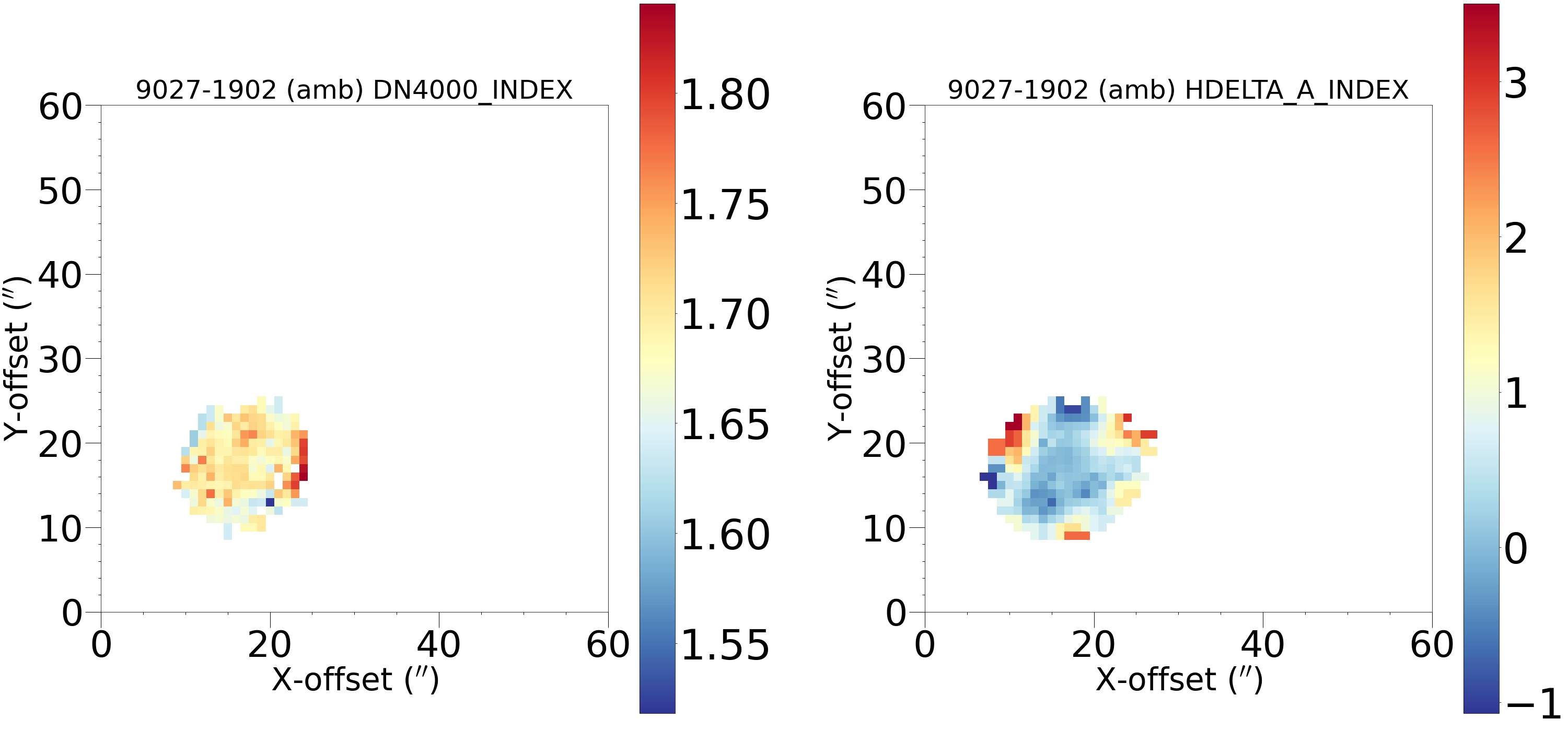}   \\
    \end{tabularx}
        \caption{The spatially resolved D$_{n}$4000 and H$\delta_{A}$ distributions of our sample. Two galaxies are shown per row. The first column shows the spatially resolved D$_{n}$4000 distribution, and the second the spatially resolved H$\delta_{A}$ distribution.}
    \label{fig:hdd}
\end{figure*}

\renewcommand{\thefigure}{6}
\begin{figure*}[htp]
    \centering
    \begin{tabularx}{\textwidth}{X X X X}

\hspace{-0.7cm}\rot{8143-3702 (AGN)} & \hspace{-5cm}\includegraphics[height=0.32\textheight,keepaspectratio]{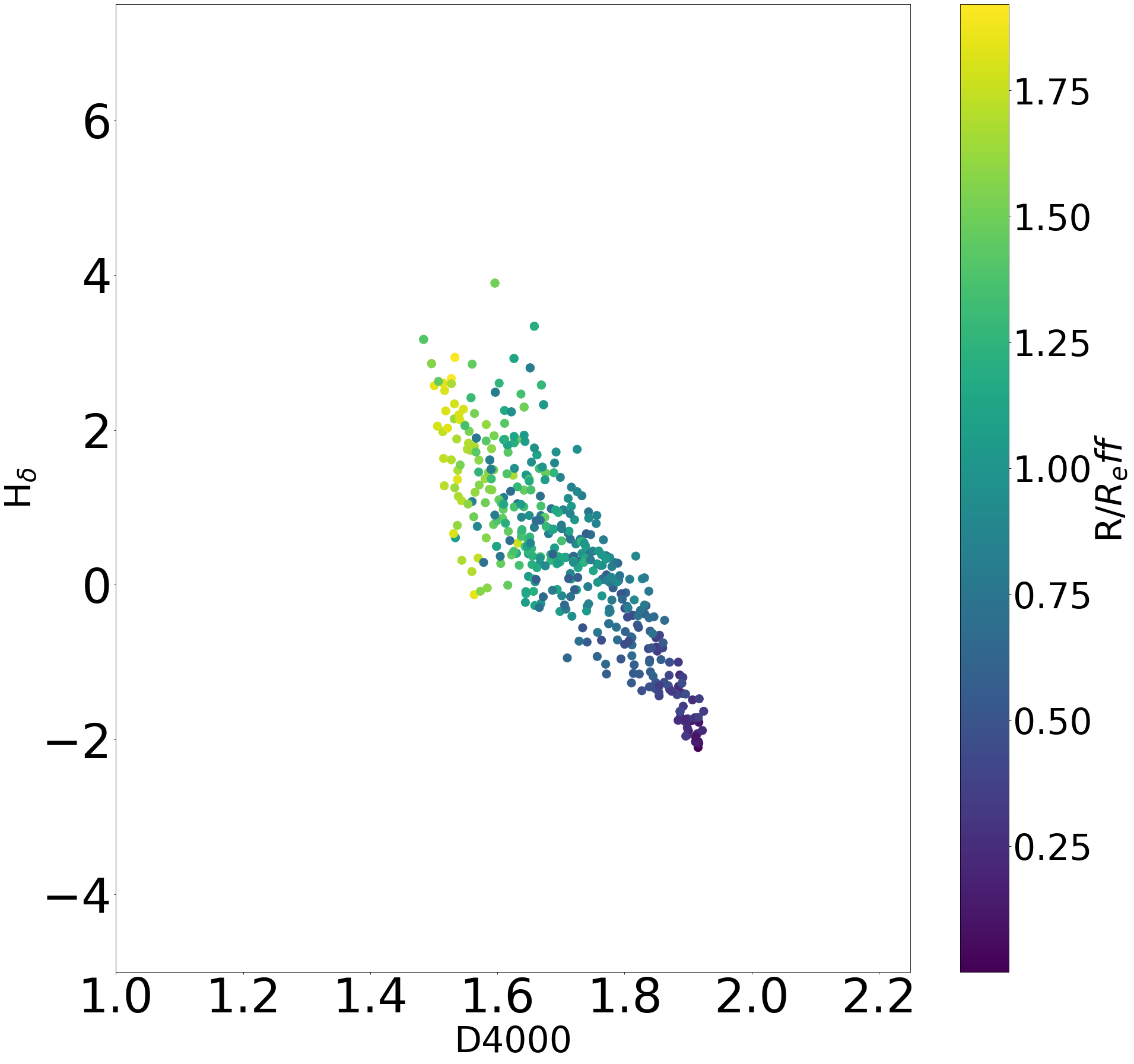} &
\hspace{0cm}\rot{8155-3702 (AGN)} &  \hspace{-4.3cm}\includegraphics[height=0.32\textheight,keepaspectratio]{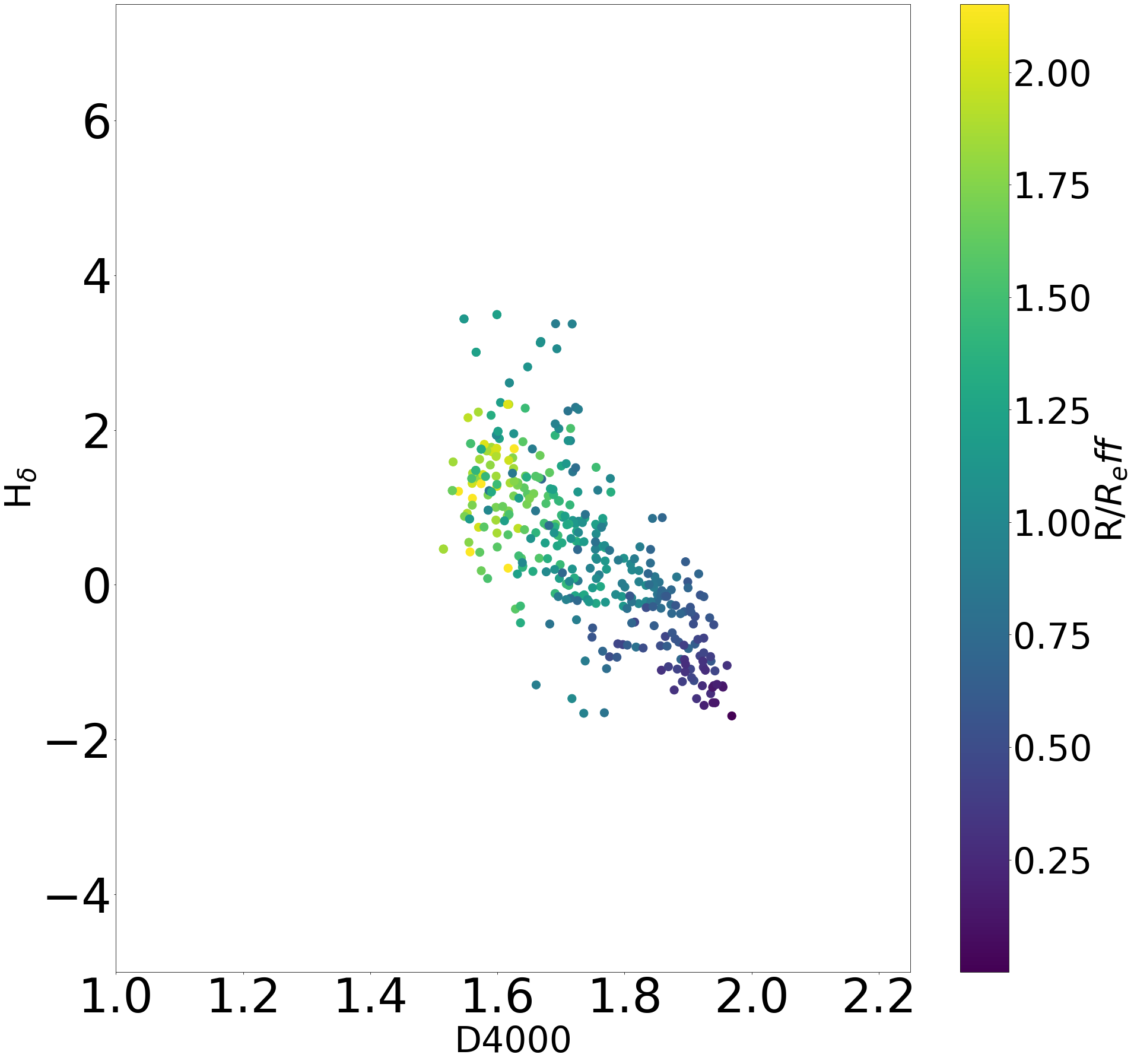} \\
\hspace{-0.7cm}\rot{8606-3702 (AGN)} &   \hspace{-5cm}\includegraphics[height=0.32\textheight,keepaspectratio]{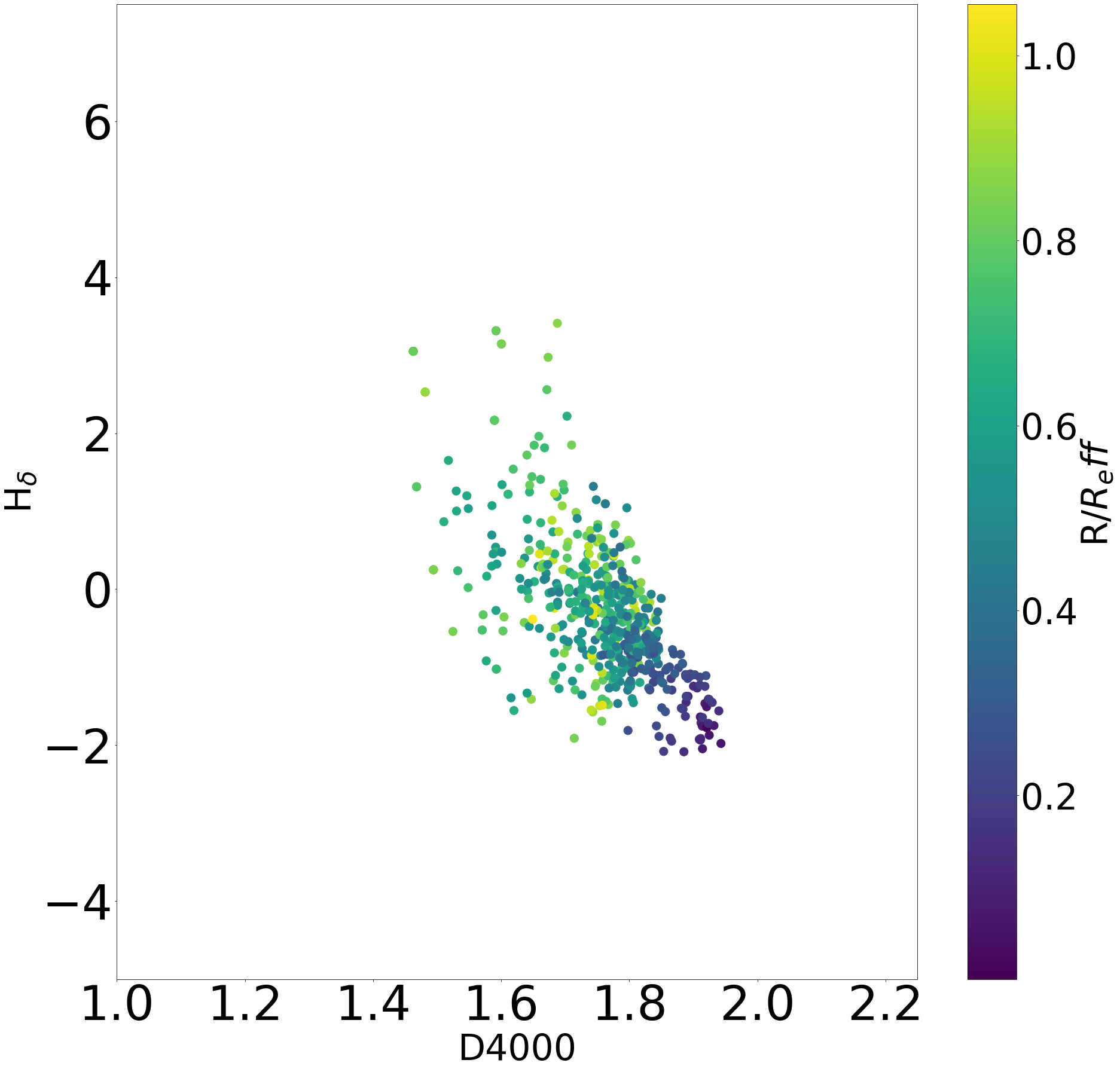} &
\hspace{0cm}\rot{8989-9101 (AGN)} &   \hspace{-4.3cm}\includegraphics[height=0.32\textheight,keepaspectratio]{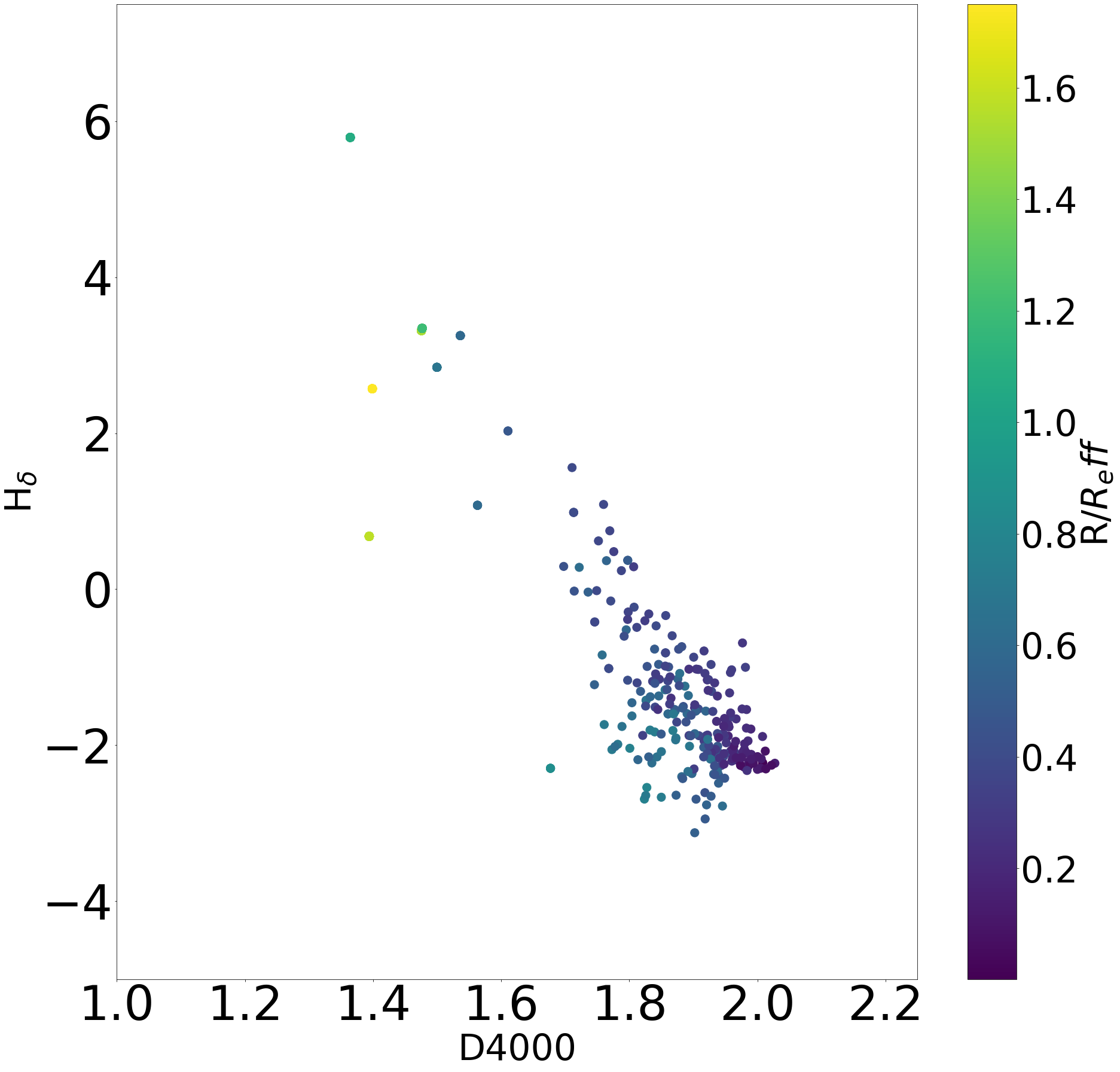} \\
\hspace{-0.7cm}\rot{8995-3703 (AGN)} &   \hspace{-5cm}\includegraphics[height=0.32\textheight,keepaspectratio]{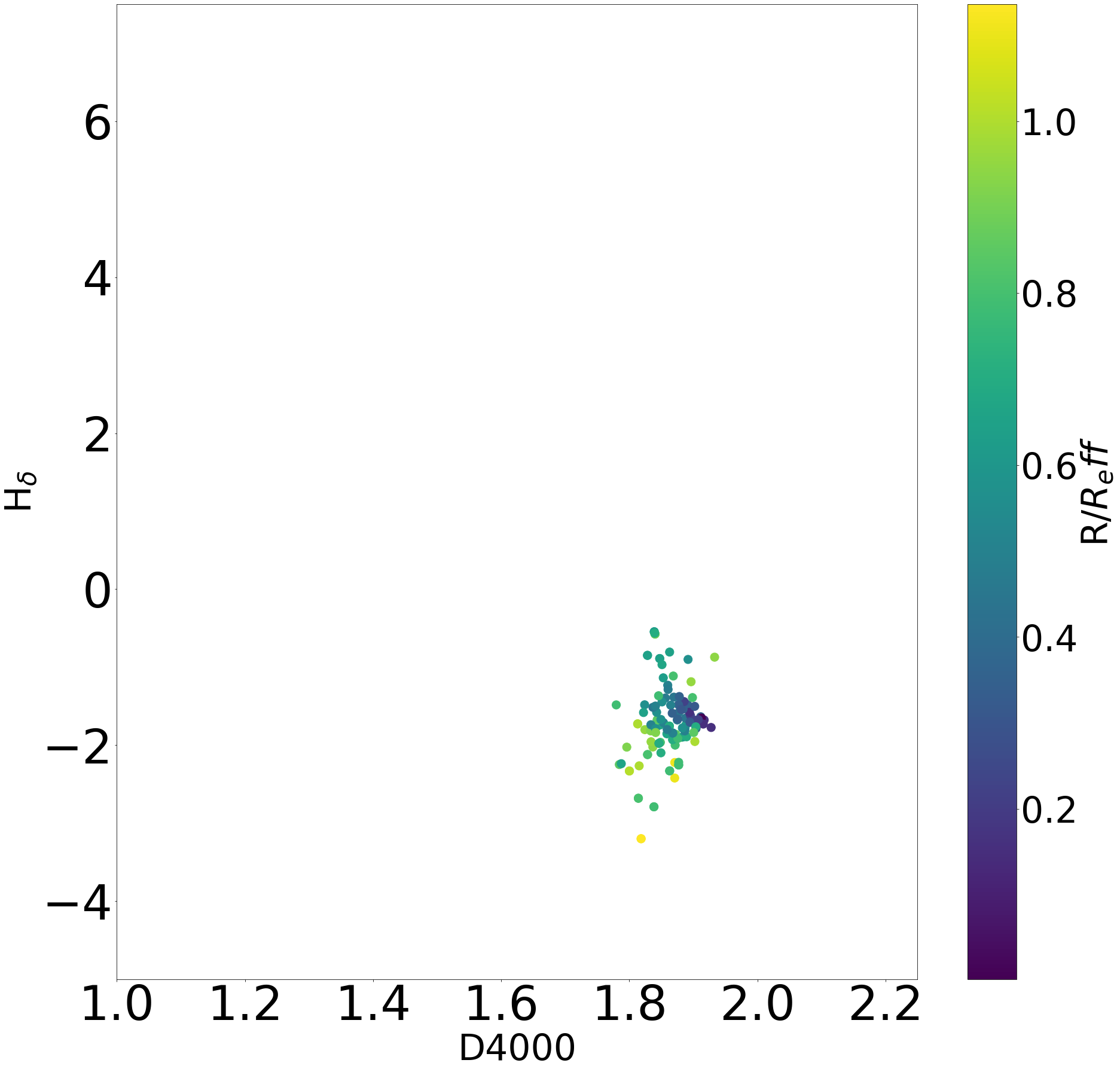} & \hspace{0cm}\rot{8615-1902 (SF)} &   \hspace{-4.3cm}\includegraphics[height=0.32\textheight,keepaspectratio]{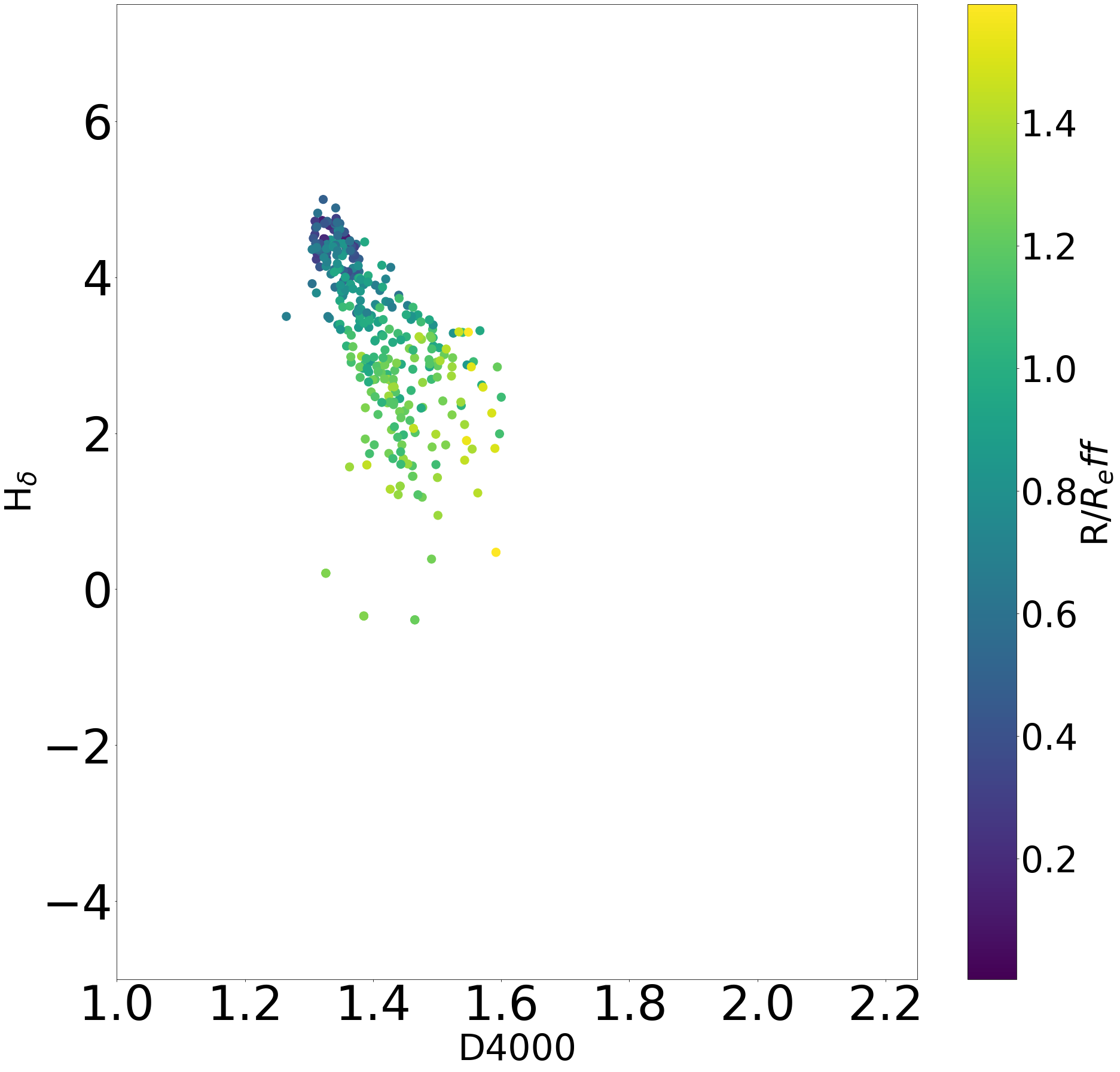} \\
     \end{tabularx}
\end{figure*}
\begin{figure*}
\ContinuedFloat
\centering
    \begin{tabularx}{\textwidth}{X X X X}
       \hspace{-0.7cm}\rot{9027-3703 (SF)} &   \hspace{-5cm}\includegraphics[height=0.32\textheight,keepaspectratio]{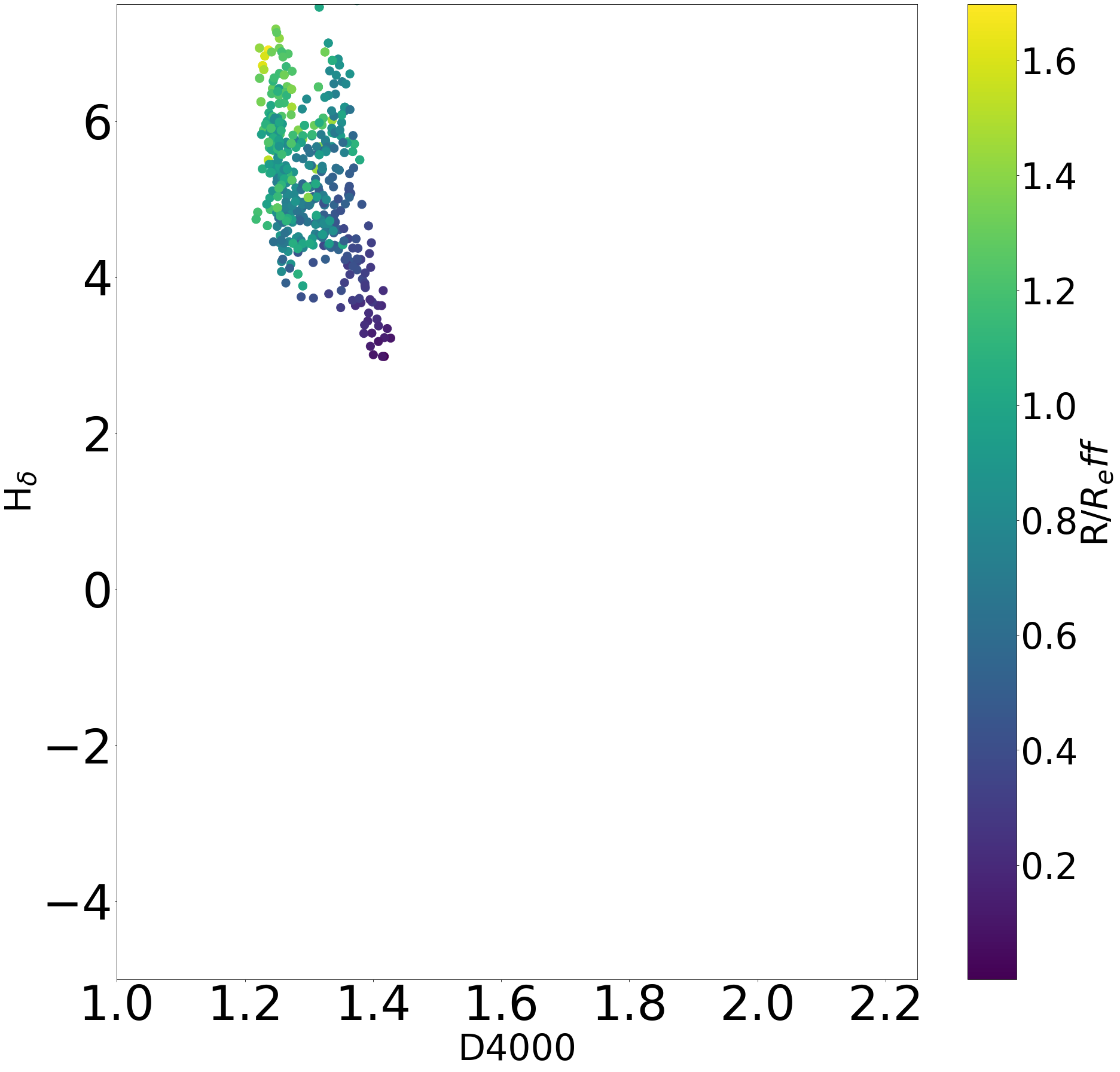} &
        \hspace{0cm}\rot{9872-3701 (SF)} &   \hspace{-4.3cm}\includegraphics[height=0.32\textheight,keepaspectratio]{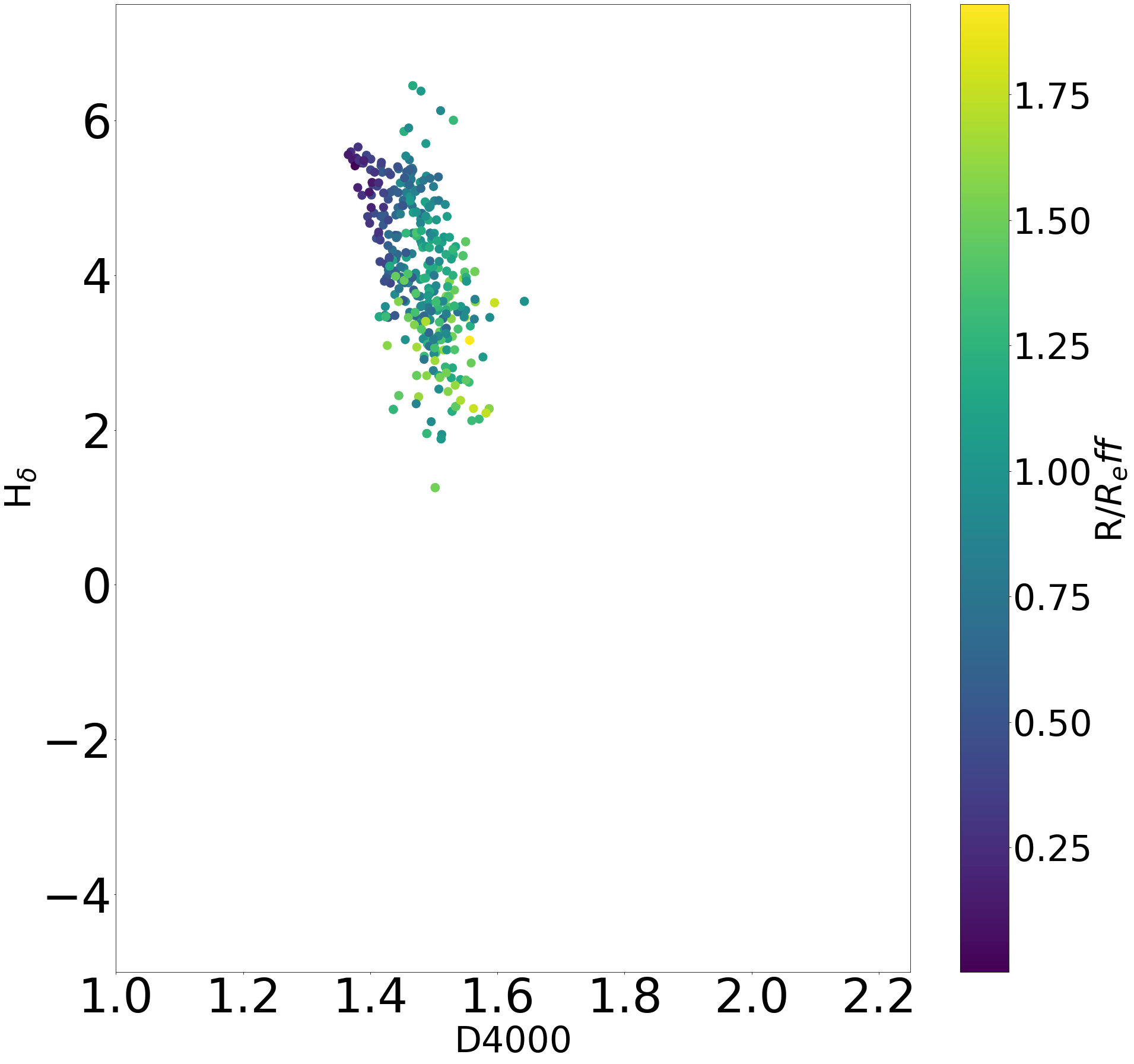} \\
         \hspace{-0.7cm}\rot{8143-1902 (unclassified)} &   \hspace{-5cm}\includegraphics[height=0.32\textheight,keepaspectratio]{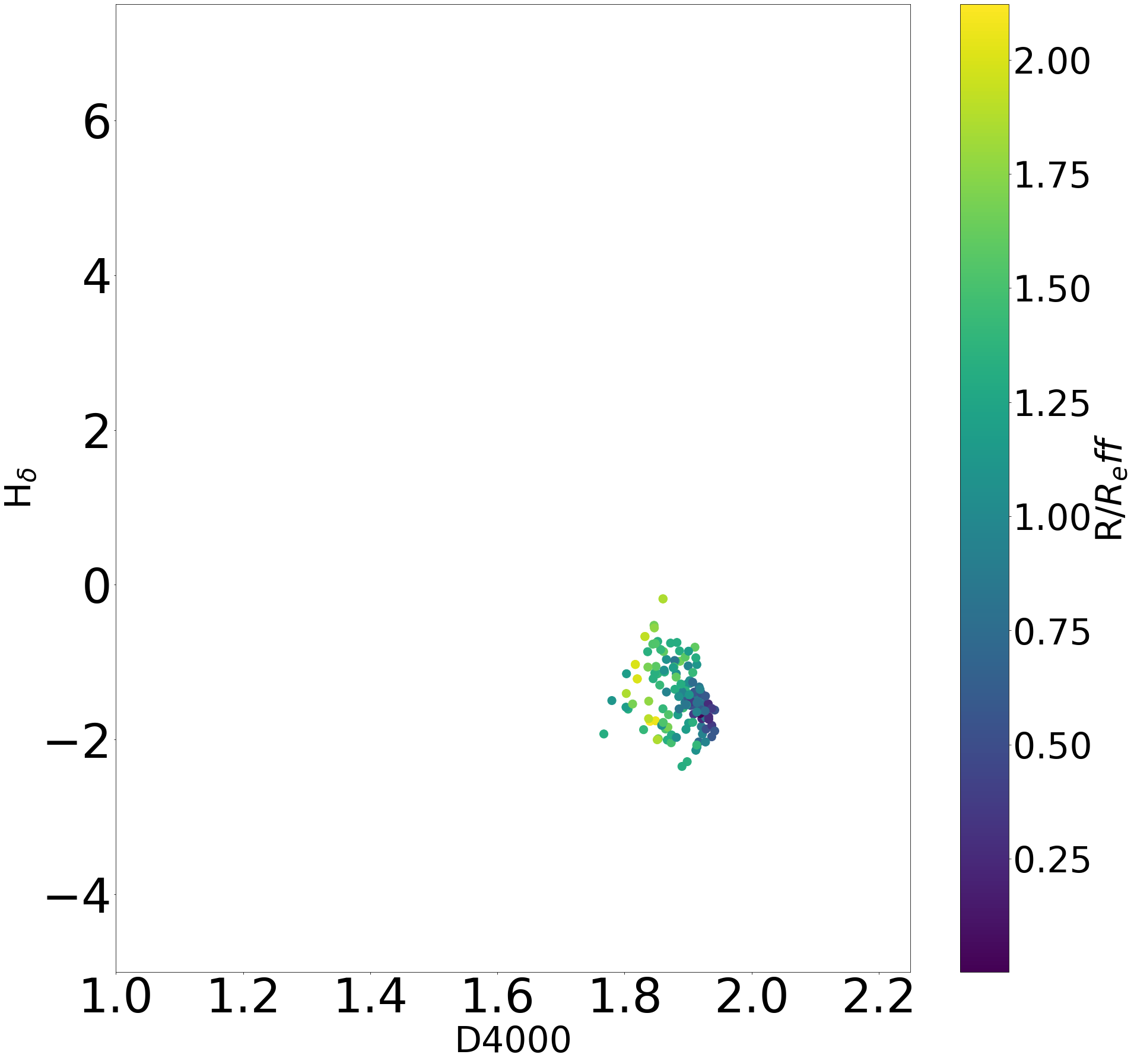} &
        \hspace{0cm}\rot{8335-1901 (unclassified)} &   \hspace{-4.3cm}\includegraphics[height=0.32\textheight,keepaspectratio]{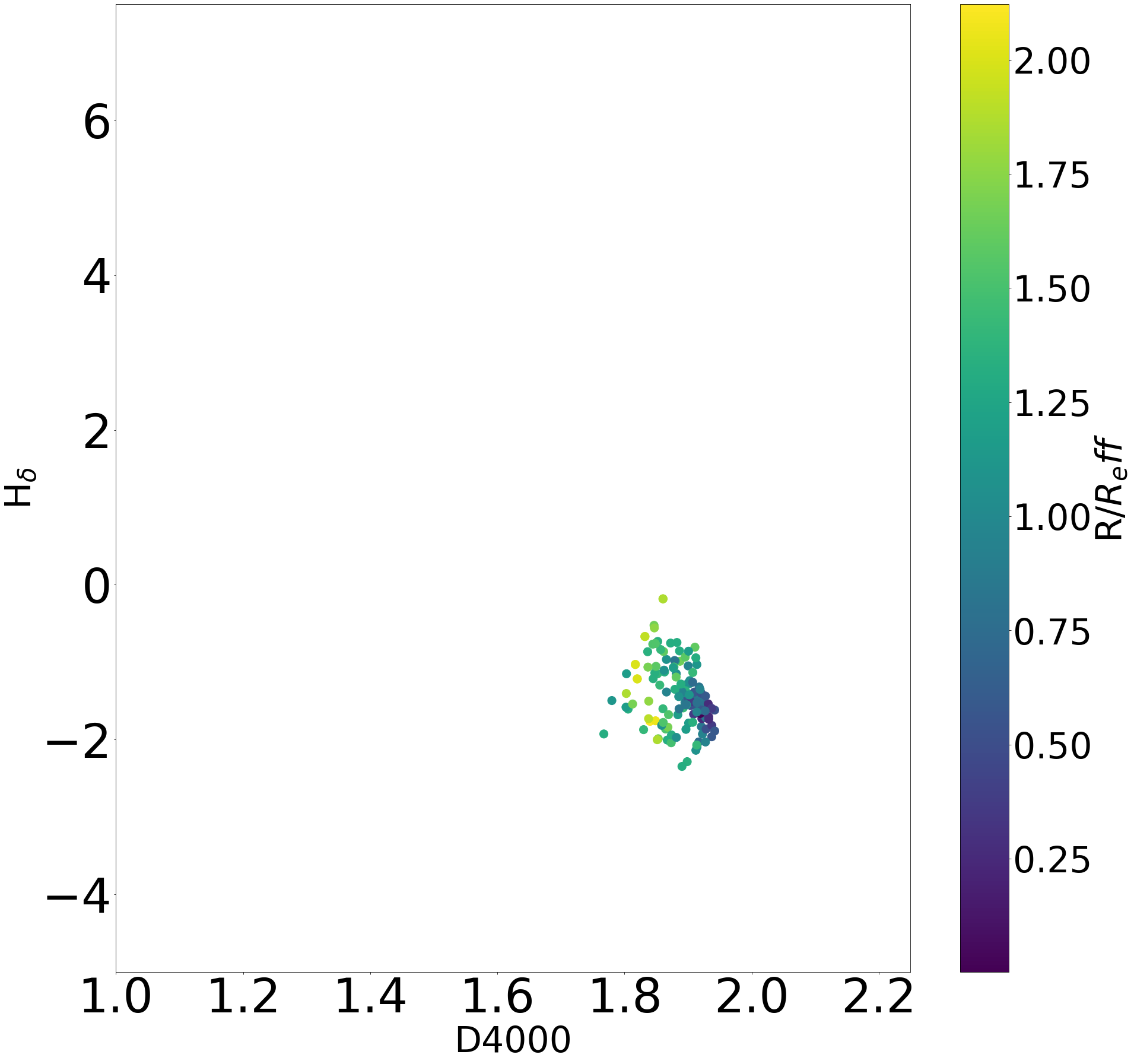} \\
        \hspace{-0.7cm}\rot{9027-1902 (ambiguous)} &   \hspace{-5cm}\includegraphics[height=0.32\textheight,keepaspectratio]{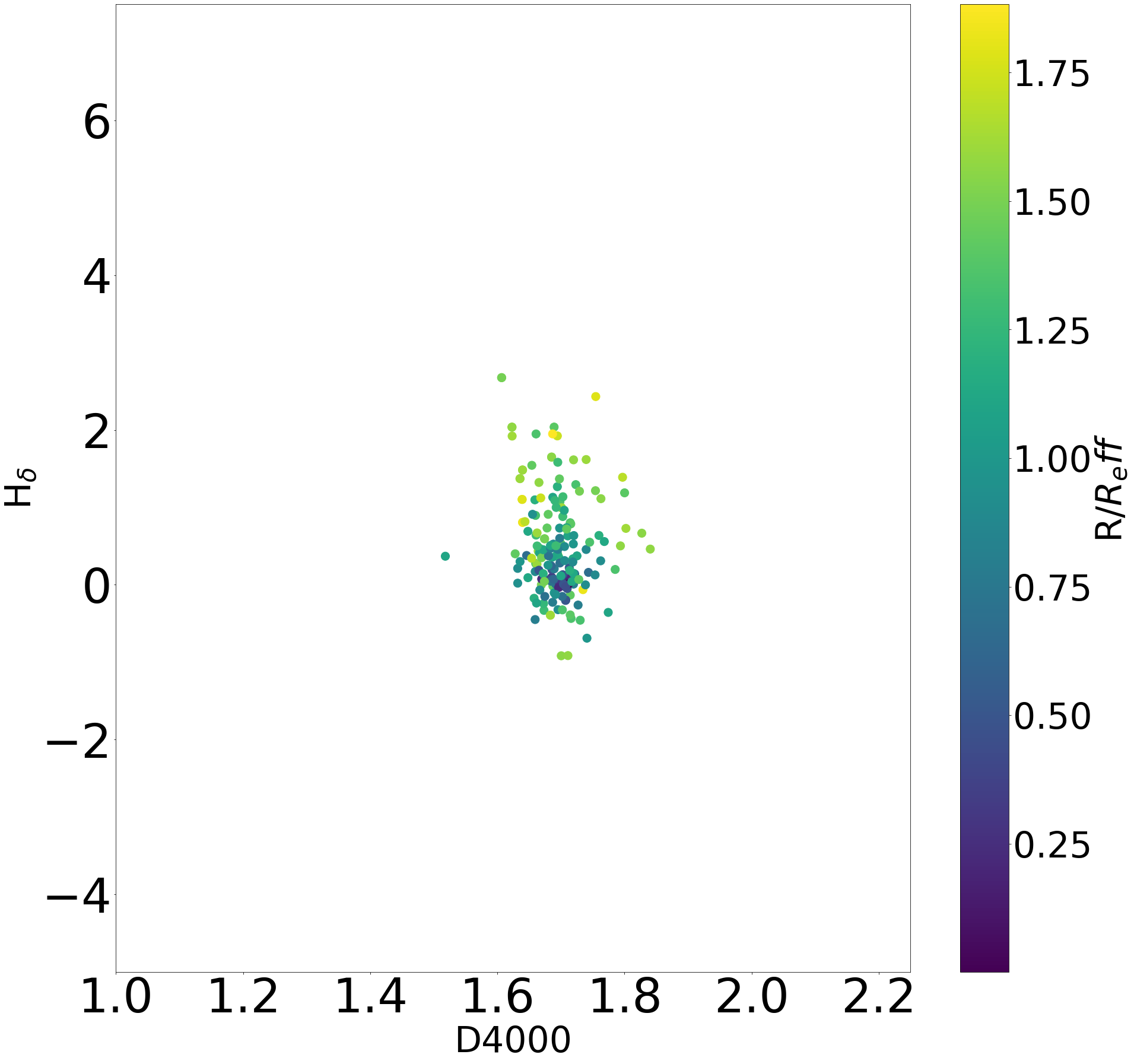} \\
    \end{tabularx}
         \caption{The \cite{Kauffmann_2003} diagnostic, or D$_{n}$4000 as a function of H$\delta_{A}$, colour coded by effective radius of our sample. Two galaxies are shown per row.}
    \label{fig:kauff}
\end{figure*}

Figure \ref{fig:hdd} shows the spatially resolved D$_{n}$4000 and H$\delta_{A}$ distributions of our sample. From left to right, the columns show the D$_{n}$4000 and H$\delta_{A}$ maps. Figure \ref{fig:kauff} shows the D$_{n}$4000 versus H$\delta_{A}$ plane plotted with their radial profiles. 
Our results show that two of the three star-forming galaxies, 9872\mbox{--}3701 and 8615\mbox{--}1902, have a distinctly different behaviour compared to the rest of the sample.

We see that 9872\mbox{--}3701 and 8615\mbox{--}1902 have increasing D$_{n}$4000 radially outward, with the core showing D$_{n}$4000 $ < 1.5$, indicating young stellar populations in the core compared to the outskirts. 
The rest of the sample show D$_{n}$4000 decreasing radially outward, with the core having ${\rm D}_{n}4000 >1.5$, denoting old stellar populations. 
The outskirts of the AGN-hosting galaxies show younger stellar populations. 

We also see that 9027--3703 and 8615-1902 have a high H$\delta_{A}$ in the core region, and lower values in the outskirts. This indicates that recent star formation occurred in the galaxy core.
The rest of the sample show an opposite behaviour, with the outskirts displaying a higher H$\delta_{A}$ than the core. However, the outskirts mostly display a lower value than the cores of the star-forming galaxies. 

Looking at H$\delta_{A}$ plotted as a function of D$_{n}$4000 in Fig. \ref{fig:kauff}, we can see that the radial gradient for 9027--3703 and 8615--1902 shows a decrease in H$\delta_{A}$/increase in D$_{n}$4000 radially outward, which is opposite to the rest of the sample.  
The unclassified and ambiguous galaxies do not show much of a gradient in their distributions.

\section{Discussion}
\label{section:Discussion}

The properties of kinematically misaligned or decoupled galaxies have been studied using MaNGA data in several works, such as \citet{Jin_2016} and \citet{2019arXiv191204522L}. However, these works do not focus specifically on the properties of galaxies with a KDC or galaxies with a counter-rotating core.
    
We investigate the spatially resolved kinematics, stellar population properties, and star formation histories of galaxies with a counter-rotating core. Specifically, these galaxies fall under 2-$\sigma$ galaxies.  We find that the stars and gas of these galaxies are decoupled, and the nature of decoupling differed depending on if the galaxy was identified as star-forming or AGN. We find that the recent star formation history of these galaxies also differed based on their identification.
Due to our small sample size of only five AGN galaxies, three star-forming, two unclassified and one ambiguous, these observed properties may not be a trend but simply individual special cases. We hope to conduct this analysis on a greater sample size in future works so that we can make definitive conclusions.



\subsection{Kinematic properties}

We find that galaxies with a counter-rotating stellar core do not show a similar property in their gas kinematics, i.e. the gaseous velocity maps show no kinematically distinct core.
There are multiple possible explanations as to why the gaseous kinematics are as such.

A large kinematic misalignment, such as KDCs, is said to be a relic of an external accretion event, such as a galaxy merger \citep[e.g.][]{1992A&A...258..250B, 1991ApJ...370L..65B}.
When such an event occurs, the accreted material flow towards the centre of the galaxy with an angular momentum different from the main body of the galaxy. 
The fate of this accreted material may explain why the gas does not show a distinct core. There are a number of explanations as to what occurred to the gas.
Much of the accreted gas associated to the KDC may be consumed by a merger-induced star formation event which has ceased at the time of observation, leaving no kinematically decoupled gas in the core \citep{Taylor_2018}.
Another possible explanation is the difference in collisionless/collisional nature of stellar and gaseous systems \citep{Crocker_2009} in interacting galaxies. In the timescale of galaxy interactions and mergers, stars are collisionless, so two systems of counter-rotating stars can co-exist in the same galaxy. However, because gaseous systems are collisional, one of the two systems must dominate over the other.

We also find that the rotational direction of the KDC differs depending on if the galaxy is classified as a star-forming galaxy or AGN. 
For the AGN-hosting galaxies, the stars of the KDC are counter-rotating with respect to the gaseous main body of the galaxy, and the gas and stars of the main body are co-rotating. 
The star-forming galaxies, on the other hand, have counter-rotating stars and gas, with the KDC co-rotating with the gaseous body. 
A possible explanation for the star-forming galaxies having a counter-rotating gaseous body is that the accreted counter-rotating gas is still present in the galaxy as the dominant system and is contributing to current star formation. Once the accreted counter-rotating gas has been consumed and the star formation ceased, the co-rotating gas dominates, leaving us with observations as the ones seen in the AGN-hosts and unclassified galaxy 8143\mbox{--}1902.
The ambiguous galaxy has a rotational behaviour similar to that of the star-forming galaxies, however we cannot make any definite statement because it is a single galaxy.

With the velocity dispersion ($\sigma$) maps, we find that the stellar and gaseous $\sigma$ maps show peaks in different locations, with the stellar $\sigma$ maps having two peaks, however such property is not seen in the gaseous $\sigma$ map.
This "2-$\sigma$" feature is known to appear in the remnants of major mergers, as shown in simulation works such as \citet{Tsatsi_2015}. 
The origin of these peaks are considered to be from gas accretion \citep[see][]{2011MNRAS.414.2923K} or from merger events \citep[see][]{2009MNRAS.393.1255C}.

The gas $\sigma$ maps shows differing properties, with four of the five AGN-hosts and the ambiguous galaxy seemingly showing central peaks, and all of the star-forming galaxies showing peaks perpendicular to the stellar velocity, with two of three showing no central peak.
The stellar-gaseous decoupling shown in the AGN galaxies can be explained through the same process as the velocity maps. 
Since gaseous systems are collisional, the gaseous kinematics of either the accreted gas or the galaxy experiencing accretion must dominate.

The gaseous velocity dispersion distribution in the star-forming galaxies is distinctly different from those in the galaxies classified as AGN-hosting and ambiguous. 
Among the three galaxies, two do not exhibit a central peak in the gaseous velocity maps. 
All three galaxies show symmetrical peaks in the outer regions, along the minor axis.
The lack of a peak in the central region differs from the distribution of gaseous velocity dispersions of star-forming galaxies investigated in previous works studying MaNGA galaxies such as \citet{Yu_2019}, even with effects such as beam smearing taken into account.
We also considered the possibility that these peaks coincide with local star formation. However, as shown from the H$\alpha$ flux diagrams in Fig. \ref{fig:haflux}, the locations of active star formation do not coincide with the peaks.
As the peaks lie in the outer regions, we have also looked into the local spectra of the areas where the peaks exist to check if the peaks are a result of errors or being a poor fit to intrinsic emission lines, which may be a result of noise or complex dynamics in the gas. However, the peaks did not coincide with regions of high error, which are masked in Fig. \ref{fig:galkin}. With regards to the fits to emission lines, as we can see from the spectra of the peak H$\alpha$ regions in Fig. \ref{fig:haspectra}, the H$\alpha$ emission lines do not seem to have poor Gaussian fits for two of the three galaxies. 9872\mbox{--}3701 may be a result of a poor fit.
The source of this lack of central peak is not well understood, and we plan to discuss it in our future work.

\subsection{Stellar population properties}

To understand more about the KDC in each galaxy in our sample, we investigated the spatially resolved stellar population properties, namely the spatially resolved ages and metallicities of the KDC and the surrounding main body of the galaxy, as well as the gradients of these properties. Similar to the kinematics, we found that the gradients differ depending on if the galaxy is classified as an AGN-host or star-forming.

\subsubsection{Age gradients}
The steepness of the age gradients differ based on the BPT classification of the galaxies. The galaxies classified as AGN and the one ambiguous galaxy have shallow light-weighted age gradients compared to the SF classified galaxies. These shallow gradients are similar with the average gradients of early-type galaxies surveyed in the MaNGA survey such as \citet{2017MNRAS.466.4731G}, which showed relatively shallow age gradients for early-type galaxies regardless of mass bin.

Two of the three star-forming galaxies, 9872\mbox{--}3701 and 8615\mbox{--}1902, show positive age gradients, whereas the third, 9027\mbox{--}3703, show a negative age gradient. These gradients are more significant than the ones found in the AGN-hosts. These gradients are not consistent with the age gradients found in \citet{2017MNRAS.466.4731G}. However, the SDSS morphological classification for 9027\mbox{--}3703 is a spiral galaxy, in such case the gradient is consistent with those of late-type galaxies in \citet{2017MNRAS.466.4731G}

The two unclassified galaxies both show positive gradients, however 8143\mbox{--}1902 shows a shallow gradient similar to that of the AGN-hosts, whereas 8335\mbox{--}1901 shows a steeper gradient, much like two of the three star-forming galaxies. Because there are only two samples, we hope future studies will give us more samples from which we can formulate a statement on.

\subsubsection{Metallicity gradients}

Similar to the age gradients, the metallicity gradients show different properties depending on its classification.

Four of the give AGN-host galaxies have greater than solar metallicity in the core region, and has a relatively sharp negative metallicity gradient compared to the star-forming galaxies.  This demonstrated gradient is consistent with those shown in previous MaNGA works such as \citet{2017MNRAS.466.4731G}. These works suggest an "outside-in" formation scenario, where the outer region forms stars first then the core forms younger, metal rich stars later. However, this scenario contradicts our findings in the age gradients.
The three SF galaxies show no distinct metallicity gradient.
The unclassified galaxies both show negative gradients, however much like their age gradients 8143\mbox{--}1902 shows a shallower gradient compared to 8335\mbox{--}1901. Whereas 8143\mbox{--}1902 shows a similar age gradient to that of AGN-hosts, the metallicity gradient is shallower compared to them, and are also inconsistent with those shown in works such as \citet{2017MNRAS.466.4731G}. On the other hand, 8335\mbox{--}1901 shows a steep age gradient similar to that of star-forming galaxies in our sample, however the metallicity gradient is not consistent with them. Again, these are only two samples, so we cannot make any statements whether or not there is a trend.
There is a clear difference in the age and metallicity gradients between the AGN-host galaxies and SF galaxies, but the gradients of individual galaxies do not give us a complete picture of their formation scenarios. This discussion will be for a later paper. 

\subsubsection{Star formation histories}

We discuss here the spatially resolved star formation histories of our sample using the using the D$_{n}$4000 versus H$\delta_{A}$ diagnostic diagram developed by \citet{Kauffmann_2003}.

The AGN-hosting galaxies show a gradient where the galaxy is dominated by older stellar populations as indicated by their D$_{n}$4000 values, and the stellar population tend to get younger radially outwards. 
The lack of young stellar populations in the central regions is consistent with our findings in the kinematic maps, because they indicate no recent star formation. 
These galaxies exhibit an inside-out quenching star formation history, with the outer regions showing the most recent star formation.
Star-forming galaxy 9072\mbox{--}3702 also shows a gradient where the stellar population where the stellar populations get younger radially outward. However, the D$_{n}$4000 values indicate that the galaxy has experienced recent star formation, which is consistent with our findings in the kinematic maps.

All three star-forming galaxies are dominated by young stellar populations, as indicated by the galaxies being dominated by spaxels with D$_n$4000 $> 1.5$. 
The radial gradients for two of the three, 9027\mbox{--}3703 and 8615\mbox{--}1902 indicate that the most recent star formation occurred in the core. 
The gaseous velocity maps show that the gas is counter-rotating with the main stellar body, it is possible that the accreted counter-rotating gas, which is the dominating system for star-forming galaxies, is still present and forming stars.
We will require a larger number of samples to investigate whether or not the star formation history radial gradients differ depending on if the galaxy is star-forming or an AGN-host. We hope to find more star-forming galaxies with a counter-rotating core in future studies so that we can make a definitive statement.

The two unclassified galaxies show stellar populations getting younger radially outwards, with the galaxy being dominated by older stellar populations, exhibiting inside-out quenching similar to the AGN-host galaxies.

The ambiguous galaxy does not show a gradient that follow outside-in or inside-out scenarios.
\subsubsection{Comparison with previous studies of $2-\sigma$ galaxies}

We have compared our findings in stellar population properties with previous works studying KDCs and $2-\sigma$ galaxies. The AGN-hosting galaxies show shallow age gradients, and four of the five galaxies show relatively sharp negative metallicity gradients. These results are similar to that found in works studying NGC 448 \citep[e.g.]{Katkov_2016, Nedelchev_2019}, which suggest rapid outside-in star formation. However, the diagnostic diagram indicates an opposite scenario, because the indicators for most recent star formation lie in the outer region.

For the star-forming galaxies, two of the three galaxies, 9872\mbox{--}3701 and 8615\mbox{--}1902, show a relatively sharp positive age gradient, with the youngest stars in the core regions, and a shallower metallicity gradient compared to the AGN-hosts. The stellar age gradients are consistent with the radial profiles obtained using the Kauffmann diagnostic, where the same two galaxies show the youngest stars in the inner regions  and 9027\mbox{--}3703 showing the youngest stars in the outer regions. The positive gradients are similar to what was seen in previous studies such as \citet{Coccato_2011} and \citet{Coccato_2012}, however the shallow metallicity gradients do not seem to match what was found in these studies.

Spectral decomposition of the main and counter-rotating components may tell us more about the radial profiles of each component, as well as their star formation histories. This discussion will be for a later paper.

\subsubsection{Comparison with galaxies without a counter-rotating core}

We compared the results of our sample to AGN-hosting and star-forming galaxies without a counter-rotating core. 
The star formation histories of both AGN-hosting and star-forming galaxies without a counter-rotating core show a radial gradient where the core has an older stellar population compared to the outside regions. 
As both AGN-hosting and star-forming galaxies without a counter-rotating core, as well as AGN-hosting galaxies with a counter-rotating core exhibit similar radial gradients that support inside-out quenching of star formation, it may be feasible to say that the star-forming galaxies with a counter-rotating core have a distinct star formation pathway, and the star formation history on its own cannot distinguish a galaxy with a counter-rotating core from a galaxy without one. 
These results are consistent with previous studies such as \citet{Davies_2001} and \citet{McDermid_2015}, with the former finding that KDCs have a similar formation history as its main body (i.e. indistinguishable from non-KDC galaxies), and the latter finding that kinematics alone do not place a large constraint on evolutionary pathways.
Galaxies observed in the MaNGA survey have a higher percentage of inside-out quenching compared to outside-in quenching \citep{Lin_2019}, so the distribution of star formation histories of our sample may be due to sampling bias.

\renewcommand{\thefigure}{7}
\begin{figure*}
    \centering
    \begin{tabularx}{\textwidth}{X X X}
 \includegraphics[width=0.3\textwidth, keepaspectratio]{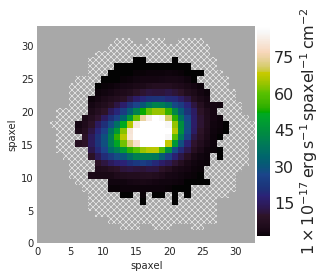} & \includegraphics[width=0.3\textwidth, keepaspectratio]{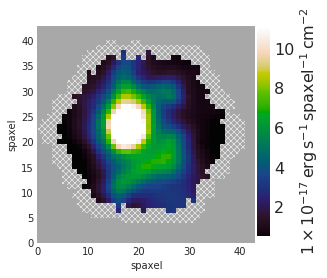} &
    \includegraphics[width=0.3\textwidth, keepaspectratio]{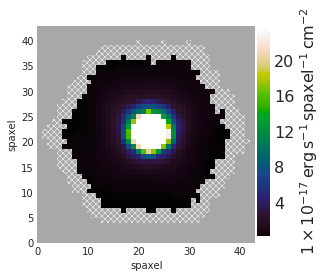}\\
    \end{tabularx}
    \caption{The H$\alpha$ flux diagrams for the three star-forming galaxies. From left to right, 8615\mbox{--}1902, 9027\mbox{--}3703, 9872\mbox{--}3701.}
    \label{fig:haflux}
\end{figure*}
\renewcommand{\thefigure}{8}
\begin{figure*}
    \centering
    \begin{tabularx}{\textwidth}{X}
 \includegraphics[height=0.3\textheight, keepaspectratio]{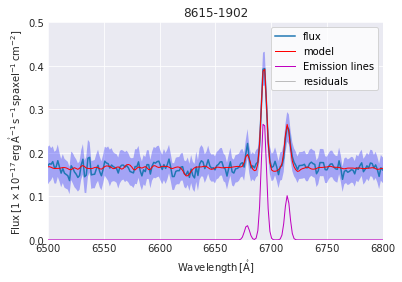} \\ \includegraphics[height=0.3\textheight, keepaspectratio]{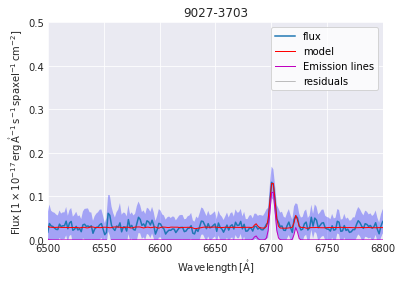} \\
    \includegraphics[height=0.3\textheight, keepaspectratio]{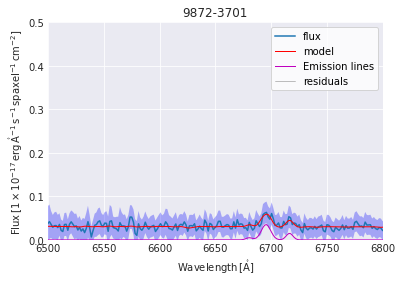}\\
    \end{tabularx}
    \caption{The H$\alpha$ emission line region spectra of the H$\alpha$ emission line peak spaxels for the three star-forming galaxies. From top to bottom, 8615\mbox{--}1902, 9027\mbox{--}3703, 9872\mbox{--}3701.}
    \label{fig:haspectra}
\end{figure*}



\section{Conclusion}
\label{section:Conclusion}
We have visually selected galaxies with a kinematically distinct stellar core (KDC) that are also counter-rotating using two-dimensional galaxy kinematic maps, and analyzed their spatially resolved kinematic properties, stellar population properties and star formation histories. We discovered that these properties may differ depending on if the galaxies were classified star-forming or AGN by their spatially resolved BPT diagrams.

\begin{itemize}
\item The rotational orientation of the galaxy main body and gas differed for star-forming and AGN galaxies. star-forming galaxies had counter-rotating stellar main body and gas, meaning the counter-rotating core was co-rotating with gas. AGN galaxies had co-rotating stellar main body and gas, so the core was counter-rotating with the main body and gas.

\item The stellar velocity dispersion maps indicated that galaxies with a counter-rotating core were "2-$\sigma$" galaxies, which show two off-centred symmetrical peaks aligned with the major axis. The stellar $\sigma$ maps and gaseous $\sigma$ maps were decoupled for both star-forming and AGN galaxies. The star-forming galaxies showed peaks along the minor axis in the gaseous $\sigma$ maps, and AGN galaxies showed a central peak.

\item The age and metallicity gradients of the AGN-host galaxies were similar to those of the average of the field sample of MaNGA galaxies. However, the gradients observed in the star-forming galaxies showed gradients that were not similar to the average. The AGN galaxies showed similar gradients to that of other $2-\sigma$ galaxies studied in previous works, whereas the star-forming galaxies did not.

\item Stellar populations of the sample indicated an inside-out quenching star formation history for most of the galaxies. 
Two of the the star-forming galaxies showed different behaviour, with the core having a younger population compared to the outskirts. 

\end{itemize}
 Based on our results, we can see that even though all of our galaxies can be seen as ones with counter-rotating cores, there is a distinct difference in their properties, which may depend on their ionisation source (AGN-host or star-forming). Whereas the AGN-hosts showed properties similar to those found in previous works, the star-forming galaxies showed properties not necessarily consistent with previous works. However, due to a small sample size, we cannot make any definitive statements.
This difference may be an indicator of the evolutionary stage the galaxy is in of a merger/accretion process.
The star-forming galaxies are in an earlier stage of merger/accretion, where the galaxy has accreted gas with a distinct angular momentum and is forming stars, most recently in the core region, with the same angular momentum as the gas.
The AGN-host galaxies are in a later stage of merger/accretion, where the accreted external gas has been consumed by star formation activity, leaving only the gas from the initial galaxy, but the stars with the same angular momentum as the gas have not yet relaxed and have retained their kinematic information. The galaxy as undergone inside-out quenching, possibly as a result of AGN feedback.

Due to the above reasons, in future studies, instead of investigating these galaxies with a counter-rotating core as "star-forming" or "AGN-host", it may be more feasible to conduct investigations with the consideration that galaxies with a counter-rotating core fall under a single evolutionary timeline, and the BPT classifications are an indicator of evolutionary stage. However, due to the small sample size, we cannot make a solid conclusion, and an increased sample size and further studies will be required to solidify this statement. 
Future SDSS data releases expect over 10,000 unique MaNGA targets, so we expect to have a greater number of galaxies with a counter-rotating core to conduct this investigation on, allowing us to draw stronger conclusions.



\begin{acknowledgements}
We would like to thank the anonymous referees for their useful and helpful feedback and comments on the previous version of this manuscript.
This work has been supported by JSPS Grants-in-Aid for Scientific Research (17H01110 and 19H05076). 
This work has also been supported in part by the Sumitomo Foundation Fiscal
2018 Grant for Basic Science Research Projects (180923), and the Collaboration Funding of the Institute of Statistical Mathematics ``New Development of the Studies on Galaxy Evolution with a Method of Data Science''.
\end{acknowledgements}

%
%

\bibliography{References}
\bibliographystyle{aa}
\begin{appendix}
	\raggedleft
    \section{BPT diagrams of sample}
    \vspace{0.5cm}
    \begin{tabular} {m{0cm} m{5.1cm} m{0cm} m{5.1cm} m{0cm} m{5.1cm}}
     \rotatebox[origin=c]{90}{8155-3702 (AGN)}  & 
        \includegraphics[width=0.25\textwidth]{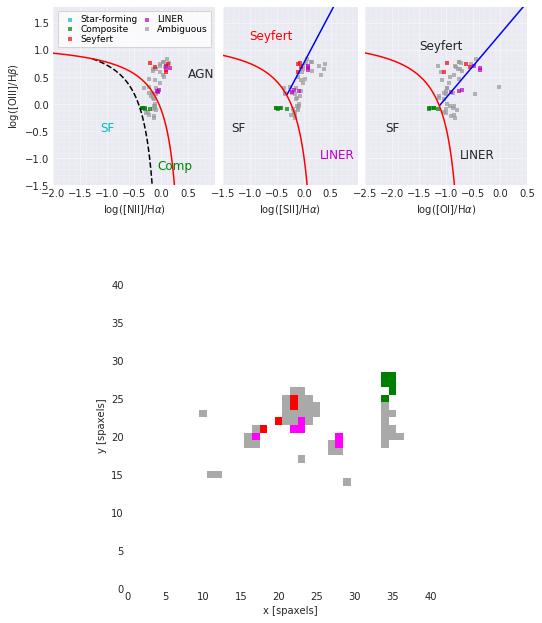} &
      \rotatebox[origin=c]{90}{8143-3702 (AGN)}  &  
        \includegraphics[width=0.25\textwidth]{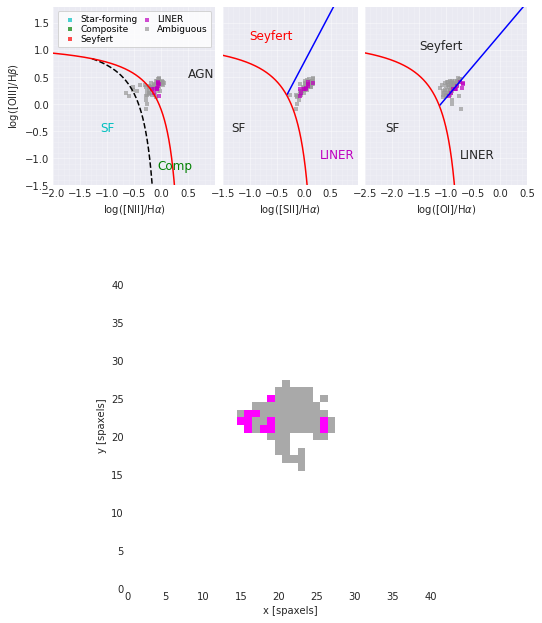} &
      \rotatebox[origin=c]{90}{8606-3702 (AGN)}  &  
        \includegraphics[width=0.25\textwidth]{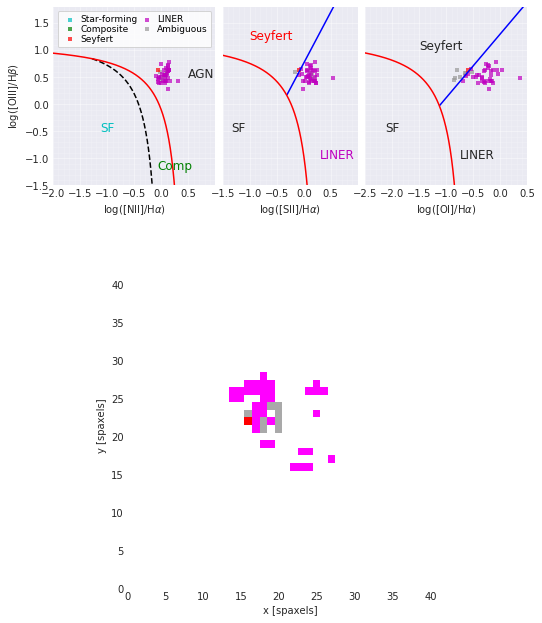} \\
     \rotatebox[origin=c]{90}{8995-3703 (AGN)}  &  
        \includegraphics[width=0.25\textwidth]{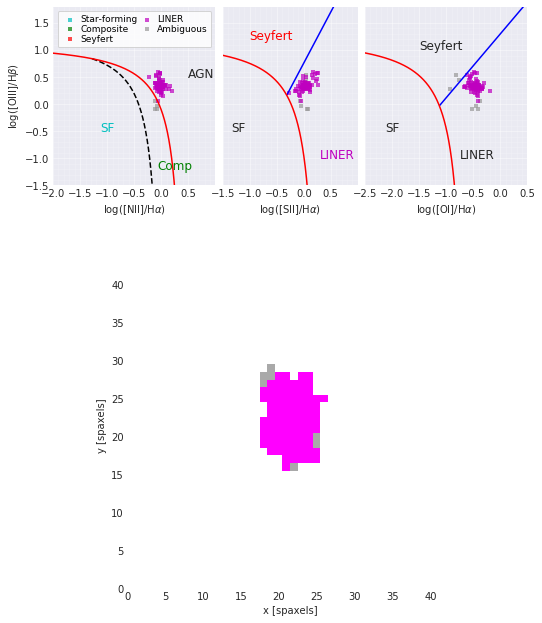} &
     \rotatebox[origin=c]{90}{8989-9101 (AGN)}  &  
        \includegraphics[width=0.25\textwidth]{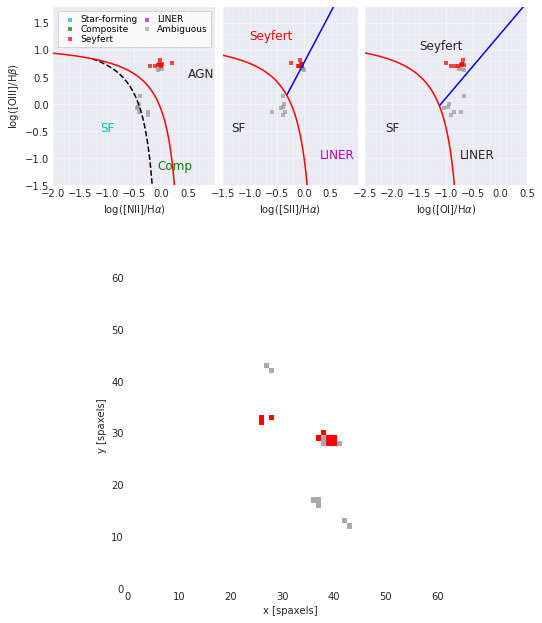} &
       \rotatebox[origin=c]{90}{8615-1902 (SF)}  &  
        \includegraphics[width=0.25\textwidth]{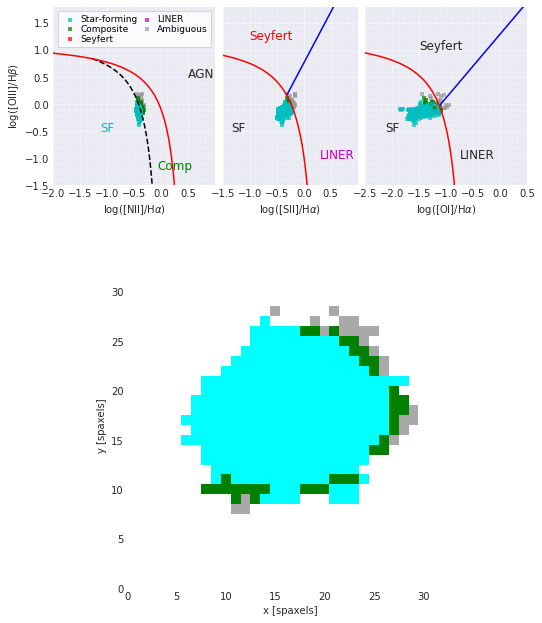} \\
        \rotatebox[origin=c]{90}{9027-3703 (SF)}  &  
\includegraphics[width=0.25\textwidth]{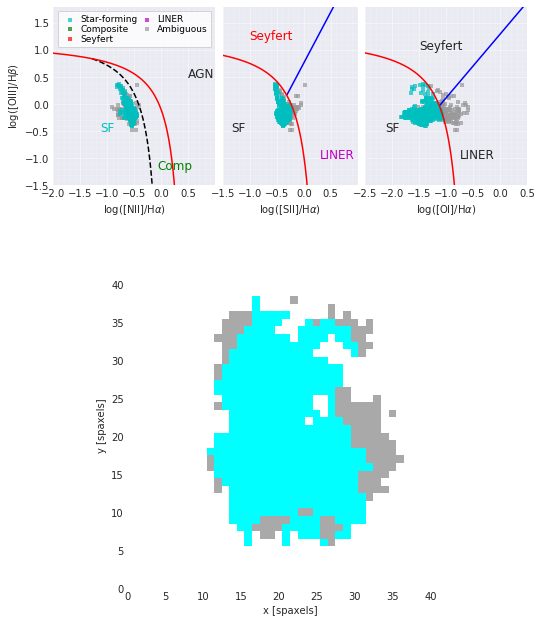} &
         \rotatebox[origin=c]{90}{9872-3701 (SF)}  &  
        \includegraphics[width=0.25\textwidth]{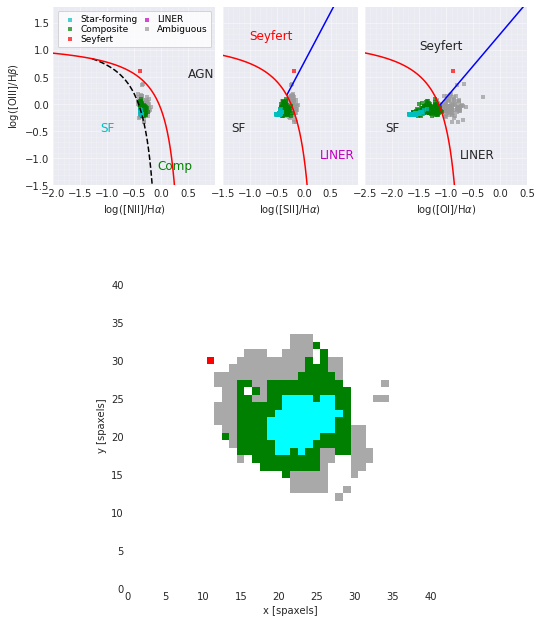} &
      \rotatebox[origin=c]{90}{8143-1902}  &  
        \includegraphics[width=0.25\textwidth]{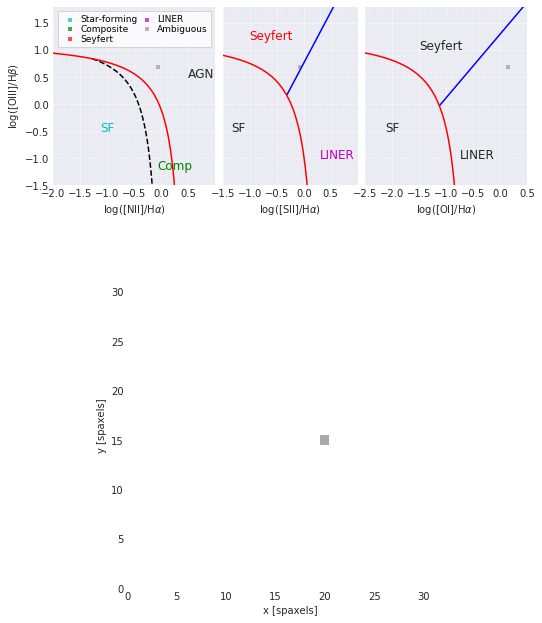} \\
      \rotatebox[origin=c]{90}{8335-1901}  &  
        \includegraphics[width=0.25\textwidth]{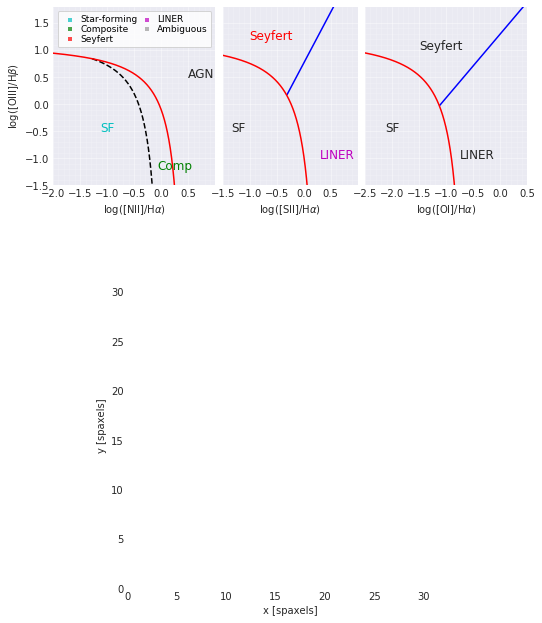} &
          \rotatebox[origin=c]{90}{9027-1902 (ambiguous)}  &  
        \includegraphics[width=0.25\textwidth]{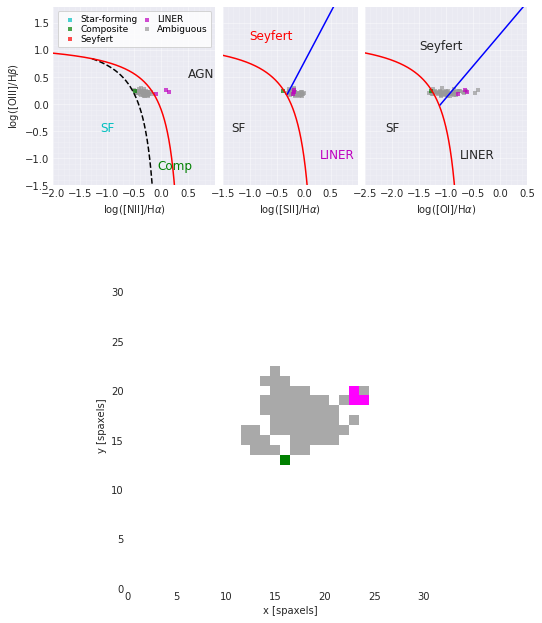} \\
   
    \end{tabular}
 
Spatially resolved BPT diagrams of our sample. Blue spaxels represent star-forming, pink and red spaxels represent AGN, green spaxel represent composite, grey spaxels represent ambiguous.
 \end{appendix}
\end{document}